\documentclass[12pt]{iopart}
\usepackage{iopams}  
\usepackage{graphicx}
\begin{document}

\title[Hamiltonian Hopf bifurcations in the discrete nonlinear Schr\"odinger 
trimer]{Hamiltonian Hopf bifurcations in the discrete nonlinear Schr\"odinger 
trimer: oscillatory instabilities, quasiperiodic solutions and a 'new' type
of self-trapping transition}

\author{Magnus Johansson}

\address{Department of Physics and Measurement Technology, 
Link\"oping University, S-581~83 Link\"oping, Sweden}

\ead{mjn@ifm.liu.se}

\begin{abstract}
Oscillatory instabilities in Hamiltonian anharmonic lattices 
are known to appear through 
Hamiltonian Hopf bifurcations of certain time-periodic solutions of 
multibreather type. Here, we analyze the basic mechanisms for this scenario 
by considering the simplest possible model system of this kind where they 
appear: the three-site 
discrete nonlinear Schr\"odinger model with periodic boundary conditions. 
The stationary solution having equal amplitude and opposite phases on two 
sites and zero amplitude on the third is known to be unstable for an 
interval of 
intermediate amplitudes. We numerically 
analyze the nature of the two bifurcations 
leading to this instability and find them to be of two different types. 
Close to the lower-amplitude threshold 
stable two-frequency quasiperiodic solutions exist 
surrounding the unstable stationary solution, and the dynamics remains 
trapped around the latter so that in particular the amplitude of the 
originally unexcited site remains small. By contrast, 
close to the higher-amplitude 
threshold all two-frequency quasiperiodic solutions are detached from the 
unstable stationary solution, and the resulting dynamics is of 
'population-inversion' type involving also the originally unexcited site. 

\end{abstract}


\submitto{\JPA}


\maketitle

\section{Introduction}
As indicated by a number of recent results 
(e.g.\ \cite{MAF98,MA98,JK99,MJKA00,KBR01,MK01,MJKA02,JMAK02,AACR02}), 
oscillatory instabilities of 
time-periodic solutions in Hamiltonian anharmonic lattice models are quite 
ubiquitous, and likely to play an important role in the dynamics 
of any physical system which, in some approximation, can be described by 
such a model. Typically they are associated with an inhomogeneous 
amplitude distribution of the solution dividing the lattice into two 
sublattices of small- and large-amplitude sites, respectively, and 
oscillatory instabilities appear when internal oscillations of the 
sublattices resonate. 

As long as the harmonic coupling is not too strong, time-periodic solutions 
for coupled anharmonic oscillators can be conveniently described using the 
multibreather formalism 
\cite{MA94, Aub97}, utilizing the ('anticontinuous') limit of zero 
coupling  to identify 
each solution with a coding sequence. In the 
simplest case of time-reversible solutions with one single determination 
of the oscillation pattern for the individual uncoupled oscillators at a 
given 
period, the code is defined as +1 for each site oscillating with 
a certain reference phase, -1 for each site oscillating 
with the opposite phase, and 0 for each site at rest.  
(This scheme can be readily generalized to include 
e.g.\ higher-harmonic oscillations by allowing codes $\pm 2, \pm 3, ...$, or 
non-time-reversibility by allowing arbitrary phase torsions \cite{Aub97}). 
This 
identification is unique for sufficiently weak coupling \cite{MA94, Aub97}, 
and can be obtained by numerical continuation using Newton-type 
schemes (e.g.\ \cite{MA96}). 

In particular, for Klein-Gordon (KG) chains of harmonically coupled soft 
anharmonic oscillators, most known examples of  oscillatory instabilities 
are associated with solutions whose 
coding sequences contain blocks of the kind $..., 1, 0^p, -1, 0, ...$, with 
integer $p\geq 0$ ($0^p$ means 0 
repeated $p$ times). 
(For hard oscillators, a corresponding statement is true 
provided that the code sequence is multiplied with $(-1)^n$.) Thus, 
large-amplitude sites of 
alternating phases (codes $\pm 1$) and small-amplitude sites 
(code $0$) provide a two-sublattice structure. 
Generically, such solutions are linearly stable for weak coupling 
when the  internal oscillation frequencies for the two sublattices are 
separated, but for larger coupling the frequencies may
overlap causing resonances and instabilities. By contrast, blocks 
of the kind  $..., 1, 0^p, 1, 0, ...$
typically yield nonoscillatory instabilities already for weak couplings. 
This difference can be related to the different natures of extrema for 
a corresponding effective 
action, defined in \cite{Aub97} as a function of the individual phases at 
each site (see also \cite{CA97, JA97, JK99}, and \cite{MK99} for further 
developments).

Oscillatory instabilities were
explicitly demonstrated e.g.\ for:  
(i) two-site 
out-of-phase breathers \cite{MAF98, JA97} 
('twisted localized modes'  \cite{DKL98, KBR01})
 with code $...0, 0, 1, -1, 0, 0...$; 
(ii) discrete dark solitons \cite{JK99,KKC94,MJKA02}
 ('dark breathers' \cite{AACR02}) with 
a hole and a phase shift added to a stable 
background 
wave, code 
$...1, -1, 1, -1, 0, 1, -1, 1, -1...$; (iii) nonlinear standing waves 
\cite{MJKA00, MJKA02, JMAK02}, corresponding to periodic or 
quasiperiodic repetitions of blocks of codes of type $\{1, 0^p, -1,0^q\}$ 
with particular combinations of integers $p,q\geq0$. 
These 
instabilities may play an important role in the processes of 
breather formation and evolution to energy equipartition 
\cite{MJKA02, JMAK02}. Oscillatory instabilities 
are found also e.g.\ for 
two-dimensional vortex-breathers \cite{CA97, JAGCR98, MK01}.

Mathematically, oscillatory instabilities appear through 
'Krein collisions' (e.g.\ \cite{HM87})
of eigenvalues to the corresponding linearized equations of
opposite Krein (symplectic) signatures\footnote{We here conform to standard 
terminology, although historically the concept of symplectic 
signature was known before Krein, see e.g.\ \cite{Bridges97}.}. 
Although well-known from  other 
branches of physics, their relevance  for the dynamics of 
anharmonic 
lattices was realized mainly after  \cite{Aub97}. 
Generally linear stability of time-periodic solutions is studied through 
the linearization of a Poincar{\'e} return map with the exact solution as a 
fixed point (Floquet map), and its symplectic nature  implies 
that 
all eigenvalues (Floquet multipliers) must lie on the unit circle for 
stability. In a Krein collision, four eigenvalues leave 
the unit circle as a complex quadruplet. 

However, for weakly coupled KG lattices the small-amplitude dynamics can be 
approximated, through a multi-scale expansion  (e.g.\ \cite{KP92, MJKA02}), 
by a discrete nonlinear Schr{\"o}dinger (DNLS) equation. 
A time-periodic KG solution then becomes a
purely harmonic (stationary) DNLS solution, 
which can be made
time-independent by studying the dynamics 
in a frame rotating with this frequency. 
Thus, performing also the stability analysis in the 
rotating frame reduces the problem to studying
stability of (relative) 
equilibria, as no explicit time-dependence appears in the linearized 
equations \cite{CE85,ELS85}. Linear stability demands all eigenvalues 
of the linearized flow to lie on the imaginary axis, and a Krein collision 
manifests itself as two pairs of complex conjugated imaginary eigenvalues 
with opposite Krein signature colliding and going out in 
the complex plane as a quadruplet (e.g.\ \cite{MacKay86}). This collision is 
also (e.g.\ \cite{Bridges90, Bridges91}) referred to as a 1:-1 
resonance (equal frequencies, opposite Krein signatures), and corresponds 
to a bifurcation of two (relative) time-periodic solutions which, due to 
its 
similarity to a Hopf bifurcation in dissipative systems, has been 
termed \cite{vdM85} a 
Hamiltonian Hopf (HH) bifurcation. 

Following a recent description of the essential aspects of the 
HH bifurcation \cite{LR01}, we briefly recall its 
phenomenology (see \cite{Bridges90,Bridges91} for a rigorous 
description). HH bifurcations may be of two different types 
 called \cite{LR01}
'type I' and 'type II', respectively (in a related context, these 
bifurcation classes were called \cite{Sevryuk} 'hyperbolic regime' and 
'elliptic regime', respectively, referring to the nature of 
the corresponding normal forms). In the type I bifurcation, 
two ('Lyapunov') families of stable periodic solutions  
surround the equilibrium in its 
stable regime, merge at the bifurcation point, and form a single family 
detached from the equilibrium in its unstable regime. For type II, 
the two periodic solutions belong to a single connected family on the stable 
side of the bifurcation, which shrinks to a fixed point and disappears 
on the unstable side. Thus, while for type I
stable periodic solutions exist close to (but away from) 
the equilibrium also in its unstable regime, for type II 
no periodic 
solutions exist close to the unstable equilibrium. 
Since for the DNLS equation this description refers to the 
rotating frame, the (relative) equilibria correspond to time-periodic 
stationary solutions and the (relative) periodic solutions
generally to quasiperiodic two-frequency solutions in the 
original 
frame. For the DNLS equation (and other equations 
with particular symmetries \cite{BV02}) such quasiperiodic solutions 
may persist,  and 
even be spatially localized, also for infinite lattices
\cite{MA94,JA97,KBR01}.

It is our purpose to describe, mainly by
numerical means, the nature of these 
bifurcations, aiming at an increased understanding for the 
dynamics in the oscillatory instability regime.
We focus here on the simplest possible model where the above described 
mechanism for oscillatory instabilities can occur: the 3-site DNLS-model 
with periodic boundary conditions and the stationary solution with code 
$+1, -1, 0$. The 
fact that this solution exhibits a complex 
instability is known already since the pioneering works of Eilbeck et.al.\ 
\cite{CE85,ELS85}; however, in spite of the 
wealth of existing literature on this model 
(e.g.\ \cite{CE85,ELS85,FFS89,Cruzeiro90,Hennig,AK93,MT93,NHMM01,FP01} and 
references 
therein), apparently the unstable dynamics of these 
stationary states and the 
properties of the surrounding quasiperiodic solutions were never before 
analyzed. Future work will extend our results for the DNLS trimer 
to the 
above-mentioned examples for infinite lattices. 

In addition to its importance for
general anharmonic lattice dynamics, 
the DNLS model has many other particular 
applications (see e.g.\ \cite{KRB01,EJ02} for recent reviews).
In particular, the recent 
experimental developments in the fields
of coupled optical waveguide arrays 
(e.g.\ \cite{CJ88, waveguides}) and  Bose-Einstein condensates 
(e.g.\ \cite{Smerzi,NHMM01,FP01,ABDKS01}) provide 
very exciting possibilities for 
practical utilization of the theoretical results. 

The outline is as follow. In section \ref{sec:model} we 
define the DNLS model and describe the stationary $\{1,-1,0\}$ solution 
and its linear stability properties, partly recapitulating material 
from \cite{CE85,ELS85}. In section \ref{sec:2freq} we use the 
technique developed in \cite{JA97} to calculate 
numerically families of  exact quasiperiodic two-frequency solutions 
surrounding the 
stationary  solution, and  analyze the nature of 
the two HH bifurcations associated with its  Krein collisions. 
We find them
to be of different types, with a type I bifurcation
at the low-amplitude instability threshold for the stationary 
solution and a type II 
bifurcation at the high-amplitude instability limit. 
In section \ref{sec:dyn} we show that, as a consequence,
the instability-induced dynamics of the $\{1,-1,0\}$ solution 
exhibits a new type of self-trapping transition: 
in the low-amplitude part of the unstable regime the dynamics remains
trapped 
around the initial solution, while in the high-amplitude part the dynamics 
also to a significant extent involves the initially 
unexcited site in a 'population-inversion' \cite{FP02} type dynamics. 
Section \ref{Conclusions} concludes the paper.

While this work was in progress, we became 
aware of a preprint of \cite{FP02} (published after our  
original submission) which, within the framework of 
describing 
the dynamics of three 
coupled Bose-Einstein condensates, reports numerical simulations exhibiting 
self-trapping effects and 'population-inversion' dynamics similarly to our 
observations. Whenever appropriate we will discuss the relation between 
our results and those of \cite{FP02}. 

\section{The DNLS model, stationary solution and linear stability}
\label{sec:model}

We consider the following form of the DNLS equation
\begin{equation}
 \rmi \dot{\psi}_{n}+C(\psi_{n+1}+\psi_{n-1}-2 \psi_{n}) 
+ |\psi_{n}|^{2}\psi_{n} = 0 ,
 	\label{DNLS}
 \end{equation}
where we assume $C\geq0$. It is the equation of 
motion for the Hamiltonian 
\begin{equation} 
       H\left(\{\rmi \psi_n\},\{\psi_n^\ast\}\right)= 
\sum_{n=1}^N \left( C |\psi_{n+1}-\psi_n | ^ 2 
-\frac{1}{2}|\psi_n|^4 \right),
      \label{hamil} 
\end{equation} 
with canonical conjugated variables $\{\rmi\psi_n\},\{\psi_n^\ast\}$. 
The 
DNLS equation has also a second conserved quantity, the 
norm\footnote{We here conform to established nomenclature in the physics 
literature, although mathematically ${\mathcal N}$ is rather the square of 
a norm.}
(excitation number)
${\mathcal N } = \sum_{n=1}^N |\psi_n|^2$. 
We consider the trimer, $N=3$, and impose periodic boundary conditions 
$\psi_0=\psi_3$, $\psi_{4}=\psi_1$ describing a triangular 
configuration with identical coupling between all sites. 
Note that we consider the case of positive intersite 
coupling and negative (attractive) nonlinearity in \eref{hamil}, but 
 the general case can be 
recovered since changing the sign of $C$ in \eref{DNLS} is 
equivalent to making the transformation
$\psi_n \rightarrow (-1)^n \rme ^ {-\rmi 4Ct}\psi_n$, while the same 
transformation followed by a time reversal $t\rightarrow -t$ is equivalent 
to changing the sign of the nonlinearity. However, as 
$N$ here is odd, these transformations must be 
accompanied by 
a change of boundary conditions from periodic to antiperiodic. Thus, our 
model is not equivalent to the periodic three-site configuration with 
positive coupling and positive 
(repulsive) 
nonlinearity considered in \cite{FP02}. 
In the latter case, the obtained solutions are 
equivalent to a subclass (antisymmetric under transformation 
$n\rightarrow n+3$) of 
the solutions to \eref{DNLS} with $C>0$ for a hexagonal geometry ($N=6$) 
analyzed in \cite{Eilbeck87}. We also note that when $C\neq0$, it can 
be normalized to $C=1$ by the transformation 
$\psi_n \rightarrow \sqrt{C}\psi_n$, $t\rightarrow t/C$. 

A time-periodic, stationary solution to \eref{DNLS} with frequency 
$\Lambda$ has the form 
\begin{equation}
\psi_n^{(\Lambda)}(t)=\phi_n^{(\Lambda)} \rme^{\rmi \Lambda t} 
\label{stationary}
\end{equation}
with time-independent $\phi_n^{(\Lambda)}$. 
We consider the particular family of solutions
\begin{equation}
\{\phi_n^{(\Lambda)}\} = \{A, -A, 0\}, \; \mbox{with}\;
A= \sqrt{\Lambda + 3 C}\;  ; \; \Lambda > -3C ,
\label{ALambda}
\end{equation}
for which the norm and Hamiltonian can be expressed as
\begin{equation}
{\mathcal N } = 2 A^2 = 2 (\Lambda+3C),
\label{NLambda}
\end{equation}
\begin{equation}
H= A^2 (6C - A^2) = 9C^2 - \Lambda ^2 = 
\frac{{\mathcal N }}{2} \left(6 C - \frac{{\mathcal N }}{2}\right) .
\label{HLambda}
\end{equation}
Linearizing \eref{DNLS}
around a stationary solution by 
writing $\psi_n(t) = \left(\phi_n^{(\Lambda)} + \epsilon_n(t)\right) \rme ^ 
{\rmi\Lambda t}$ yields: 
\begin{equation}
\rmi \dot{\epsilon}_{n} +C(\epsilon_{n+1}+\epsilon_{n-1}-2\epsilon_{n}) 
+ 2|\phi_{n}^{(\Lambda)}|^{2} \epsilon_{n}
+ (\phi_{n}^{(\Lambda)})^{2} \epsilon_{n}^{*} - \Lambda \epsilon_n = 0 .
        \label{linear}
\end{equation}
For real $\phi_n^{(\Lambda)}$ 
a linear eigenvalue problem is obtained by writing 
$\epsilon_n = a_n \rme^{-\rmi \omega_{\rm l} t} + 
b_n^\ast \rme^{\rmi\omega_{\rm l}^\ast t}$: 
\begin{eqnarray}
(2C+\Lambda - 2 (\phi_{n}^{(\Lambda)})^{2}) a_{n} - C(a_{n+1}+a_{n-1}) -
(\phi_{n}^{(\Lambda)})^{2} b_{n} &=& \omega_{\rm l} a_{n} \nonumber \\
(\phi_{n}^{(\Lambda)})^{2} a_{n} + C(b_{n+1} + b_{n-1}) - (2C +\Lambda -2
(\phi_{n}^{(\Lambda)})^{2}) b_{n} &=& \omega_{\rm l} b_{n} .
\label{eigeneqab}
\end{eqnarray}
Then, linear stability is equivalent to all eigenfrequencies 
$\omega_{\rm l}$ being real. 
Alternatively, one may use the linear combinations 
$U_{n}=a_{n}+b_{n}$, $W_{n}=a_{n}-b_{n}$ so that \eref{eigeneqab} becomes 
\begin{eqnarray}
{\mathcal{L}}_0 W_n \equiv 
(2C+\Lambda - (\phi_{n}^{(\Lambda)})^{2}) W_{n}-C(W_{n+1}+W_{n-1})
&=&  \omega_{\rm l} U_{n} \nonumber \\
{\mathcal{L}}_1 U_n \equiv(2C+\Lambda -3 (\phi_{n}^{(\Lambda)})^{2}) U_{n}
-C(U_{n+1}+U_{n-1})&=& \omega_{\rm l} W_{n} .
\label{eigeneqUW}
\end{eqnarray}
The usefulness of this formulation becomes 
clear by noting that, if $\epsilon_n$ is expanded in its real and imaginary 
parts $\epsilon_n=\epsilon_n^{(\rm r)} + \rmi 
\epsilon_n^{(\rm i)}$, \eref{linear} can 
be written in the Hamiltonian form
\begin{equation}
\left ( \begin{array}{c}\{\dot{\epsilon}_n^{(\rm r)}\} \\ 
\{\dot{\epsilon}_n^{(\rm i)}\}
\end{array} \right) = 
\left ( \begin{array}{c}\; 0\;\;\;\;\;{\mathcal{L}}_0 \\ 
{-\mathcal{L}}_1\;\;\;0  \end{array} \right)
\left ( \begin{array}{c}\{{\epsilon}_n^{(\rm r)}\} \\
\{{\epsilon}_n^{(\rm i)}\}
\end{array} \right)
\label{matrix}
\end{equation}
where ${\mathcal{L}}_0$ and ${\mathcal{L}}_1$ defined by \eref{eigeneqUW} 
are symmetric. Then, if $\rmi \omega_{\rm l}$ is a purely imaginary 
eigenvalue of 
\eref{matrix} with $\omega_{\rm l} >0$, the corresponding eigenvector is 
$\left ( \begin{array}{c}\{U_n\} \\ \{\rmi W_n\} \end{array} \right)$
where $\left(U_n,W_n\right)$ is the (real) solution of \eref{eigeneqUW} with 
eigenfrequency $\omega_{\rm l}$. Thus, to each real eigenfrequency 
$\omega_{\rm l} > 0$ of 
\eref{eigeneqab} and \eref{eigeneqUW} we can associate a Krein signature 
${\mathcal K}(\omega_{\rm l})$
from the symplectic product of the corresponding eigenvector 
with eigenvalue $\rmi \omega_{\rm l}$ of \eref{matrix} 
with itself, which, using the standard sign convention as in 
e.g.\ \cite{Bridges97}, reads
\begin{equation}
{\mathcal K}(\omega_{\rm l}) = 
 {\rm sign}  \sum_n  U_n W_n
 =  {\rm sign} \sum_n \left[|a_n|^2-|b_n|^2\right] .
\label{Krein_sig}
\end{equation}
The Krein signature is then the sign 
of the (Hamiltonian) energy carried by the linear eigenmode 
\cite{Bridges97}. 
(In an earlier publication \cite{JA00}, 
an opposite sign convention was used.)

Since \eref{DNLS} is invariant under global phase rotations, 
$\omega_{\rm l}=0$ is always an eigenvalue of algebraic multiplicity 2 
and geometrical multiplicity 1, corresponding to 
 $U_n=0, W_n=\phi_n^{(\Lambda)}$ ('phase mode'). For the  
solution \eref{ALambda}, 
the remaining eigenfrequencies are \cite{CE85} 
\begin{equation}
\fl
\left(\frac{\omega_{\rm l}}{C}\right)^2 = \frac{1}{2} 
\left( \frac{\Lambda}{C}+4 \right)^2 +4 
\pm \frac{1}{2} \sqrt{\left(\frac{\Lambda}{C}-6 \right)
\left(\left( \frac{\Lambda}{C}\right)^3 + 6  
\left(\frac{\Lambda}{C}\right)^2 
+4 \frac{\Lambda}{C} -24 \right)} .
\label{roots}
\end{equation}
\begin{figure}
\centerline{\includegraphics[height=0.8\textwidth,angle=270]{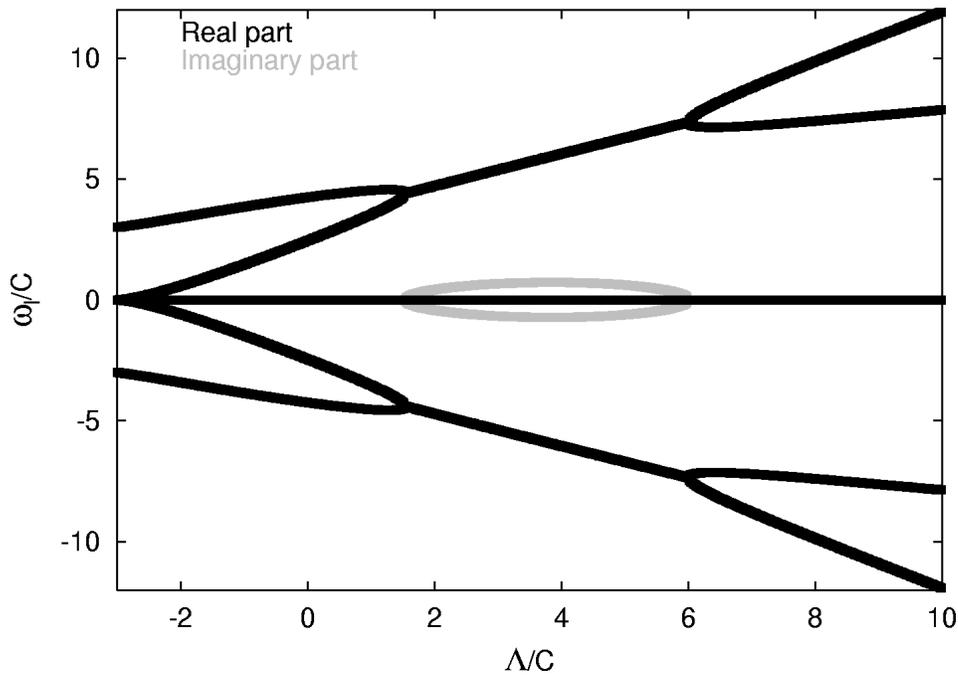}}
\caption{\label{fig_wL} 
Scaled eigenfrequencies $\omega_{\rm l}/C$ of 
\eref{eigeneqab}, \eref{eigeneqUW} versus scaled frequency 
$\Lambda/C$ of the stationary solution 
\eref{ALambda}. Dark (grey) lines represent real 
(imaginary) parts of $\omega_{\rm l}/C$. 
}
\end{figure} 
These 
are real
for small or large $\Lambda/C$,
 but complex for 
$x_\ast < \Lambda/C<6$ where $x_\ast \approx 1.5385$ is the only 
real solution to $x^3+6x^2+4x-24=0$ \cite{CE85,ELS85} 
(figure \ref{fig_wL}.) Close to the linear 
limit $\Lambda/C\rightarrow -3$ we can expand \eref{roots} using 
$\lambda=A^2/C=\Lambda/C+3 = {\mathcal N }/2C $ as a 
small parameter, and obtain
$|\omega_{\rm l +}|/C\approx 3+\lambda/3+4\lambda^2/27$ and 
$|\omega_{\rm l -}|/C\approx 2\lambda^{3/2}/3$. For $\omega_{\rm l +}$ the 
corresponding eigenmode can be expressed as 
\begin{equation}
\fl
\left\{\left ( \begin{array}{c}{a}_n \\ {b}_n \end{array} \right)\right\}=
\left\{\left ( \begin{array}{c} -2\lambda/9 \\ 1+2\lambda/3 \end{array} 
\right),
\left ( \begin{array}{c} -2\lambda/9 \\ 1+2\lambda/3 \end{array} 
\right),
\left ( \begin{array}{c} \lambda/9 \\ 1 \end{array} 
\right)\right\} +{\mathcal O} (\lambda^2) 
\label{lin+}
\end{equation}
with
${\mathcal K}(\omega_{\rm l +}) = -1$, while for $\omega_{\rm l -}$ 
the eigenmode is most conveniently described as 
\begin{equation}
\left\{\left ( \begin{array}{c}{U}_n \\ {W}_n \end{array} \right)\right\}=
\left\{\left ( \begin{array}{c} 1 \\ \sqrt{\lambda} \end{array} 
\right),
\left ( \begin{array}{c} 1 \\ \sqrt{\lambda} \end{array} 
\right),
\left ( \begin{array}{c} -2 -2 \lambda \\ -2 \sqrt{\lambda} \end{array} 
\right)\right\} +{\mathcal O} (\lambda^{3/2})
\label{lin-}
\end{equation}
with ${\mathcal K}(\omega_{\rm l -}) = +1$. On the other hand, close to the 
anticontinuous limit $\Lambda/C\rightarrow \infty$, we obtain from 
\eref{roots}
the eigenfrequencies 
$|\omega_{\rm l +}|/C\approx \Lambda/C + 2$ and 
$|\omega_{\rm l -}|/C \approx 2 \sqrt{\Lambda/C} + 5 \sqrt{C/\Lambda}$. 
Then, we can write the corresponding eigenmode for $\omega_{\rm l +}$ as
\begin{equation}
\left\{\left ( \begin{array}{c}{a}_n \\ {b}_n \end{array} \right)\right\}=
\left\{\left ( \begin{array}{c} 0 \\ C/\Lambda \end{array} 
\right),
\left ( \begin{array}{c} 0 \\ C/\Lambda \end{array} 
\right),
\left ( \begin{array}{c} 1 \\ 0 \end{array} 
\right)\right\} +{\mathcal O} ((C/\Lambda)^2) 
\label{ac+}
\end{equation}
 with 
${\mathcal K}(\omega_{\rm l +}) = +1$, and the eigenmode corresponding to 
$\omega_{\rm l -}$ as 
\begin{equation}
\fl
\left\{\left ( \begin{array}{c}{U}_n \\ {W}_n \end{array} \right)\right\}=
\left\{\left ( \begin{array}{c} -\sqrt{C/\Lambda} \\ 1 \end{array} 
\right),
\left ( \begin{array}{c} -\sqrt{C/\Lambda} \\ 1 \end{array} 
\right),
\left ( \begin{array}{c} 0 \\ 2C/ \Lambda \end{array} 
\right)\right\} +{\mathcal O} ((C/\Lambda)^{3/2})
\label{ac-}
\end{equation}
with ${\mathcal K}(\omega_{\rm l -}) = -1$. Thus, the Krein signatures of 
the frequencies $\omega_{\rm l +}$ and $\omega_{\rm l -}$ 
are interchanged by the two Krein 
collisions at the boundaries of the complex regime, where the stationary 
solution is unstable (and the Krein signature undefined). Note also that 
the eigenmodes corresponding to nonzero $\omega_{\rm l}$ 
are symmetric around 
 $n=3$, and thus their excitation breaks the spatial antisymmetry 
of the stationary solution around this site. 

By considering the 
solution \eref{ALambda} as 
periodically repeated in an infinite lattice, it becomes,
in the terminology of 
\cite{MJKA00,MJKA02,JMAK02}, 
a nonlinear
standing wave 
with wave vector $Q=2 \pi/3$ of 'type H',  for which \eref{roots} gives  
the subset of all eigenfrequencies 
corresponding to 
eigenmodes with spatial period 3. Close to the linear limit,  
\eref{lin+} can then be interpreted as a phonon mode with wave vector 
$q = 0$, while  \eref{lin-} represents a translational mode 
of wave vector $q=2\pi/3$ corresponding to a sliding 
towards the 'type E' $Q=2\pi/3$ standing wave having a periodic 
repetition of the codes +1,-1,-1. Close to the anticontinuous limit, 
\eref{ac+} represents a 'hole mode' localized at the zero-amplitude 
site, while  \eref{ac-} corresponds to 
an internal oscillation at the  non-zero 
amplitude sites. 
In the intermediate, unstable, regime, these characters of the 
eigenmodes become mixed.

\section{Two-frequency solutions: bifurcations and linear stability}
\label{sec:2freq}

As shown in  \cite{JA97}, exact, generally quasiperiodic, 
two-frequency solutions to the 
DNLS equation \eref{DNLS} exist and can be obtained 
from the ansatz 
\begin{equation}
\psi_n(t)=\phi_n(t) \rme^{\rmi \omega_0 t},  
\label{2freq}
\end{equation}
so that in the variables $\phi_n$ the DNLS equation \eref{DNLS} becomes
\begin{equation}
 \rmi \dot{\phi}_{n}+C(\phi_{n+1}+\phi_{n-1}-2 \phi_{n}) 
+ |\phi_{n}|^{2}\phi_{n} -\omega_0 \phi_n = 0 .
 	\label{DNLSphi}
 \end{equation}
Thus, the frequency $\omega_0$, corresponding to a pure phase rotation, 
becomes a parameter in \eref{DNLSphi}, and for a given 
$\omega_0$ we may search for time-periodic solutions to \eref{DNLSphi} 
with frequency 
$\omega_{\rm b}$ fulfilling $\phi_n(t+2\pi/\omega_{\rm b})=\phi_n(t)$. 
In particular, any stationary solution to \eref{DNLS} of the form 
\eref{stationary} is trivially a time-periodic solution to \eref{DNLSphi} 
of the form $\phi_n(t)=\phi_n^{(\Lambda)} \rme^{\rmi \omega_{\rm b} t}$ 
with frequency $\omega_{\rm b} = \Lambda - \omega_0$.
However,  when the linear eigenfrequencies $\omega_{\rm l}$ \eref{roots} 
are real and nonzero, in general (Lyapunov) families of genuine 
two-frequency 
solutions 
(generally quasiperiodic in the original frame)
bifurcate from the stationary solutions 
when $\omega_{\rm l}(\Lambda)+\omega_0 = \Lambda$. 
In the limit 
of small-amplitude oscillations around the stationary solution, these   
have frequencies $\omega_{\rm b}=\omega_{\rm l}$ in the frame 
rotating with the stationary 
frequency $\Lambda$, and 
coincide with the linear eigenmodes.  

Using standard numerical continuation techniques 
described in \cite{JA97}, 
these two-frequency solutions can be continued versus 
$\omega_{\rm b}$ 
towards larger oscillation amplitudes away from the 
stationary solutions, and explicitly calculated to computer precision in 
their whole domain of existence. Using standard fortran routines for 
numerical integration and matrix inversion, when necessary in quadruple 
precision, we may generally obtain solutions fulfilling 
$||\phi_n(t+2\pi/\omega_{\rm b})-\phi_n(t)|| < \epsilon$ with 
$\epsilon \sim 10^{-30}$, although 
for strongly unstable solutions inaccuracies in the numerical integration 
might limit the attainable precision to larger $\epsilon$. 
All results presented here have been obtained by first 
demanding a precision $\epsilon\sim 10^{-12}$, and then checked 
to be invariant when decreasing $\epsilon$.

The two-frequency solutions to \eref{DNLS} belong generally 
to two-parameter families, and there is some freedom in choosing independent 
parameters for the numerical continuation. 
(Since $C$ only yields a rescaling, 
we put $C=1$ in all numerical calculations.)
One may choose 
e.g.\ 
frequencies $\omega_0, \omega_{\rm b}$, or
norm  $\mathcal N$ and action 
('area') integral $I$ of 
the time-periodic solution to \eref{DNLSphi}, i.e.\ in the frame rotating 
with frequency $\omega_0$. Using in this frame the canonical variables 
 $\{\rmi\phi_n\},\{\phi_n^\ast\}$, corresponding to a transformation of 
the original Hamiltonian into $H'=H+\omega_0 {\mathcal N}$, $I$ becomes 
\begin{equation}
I=\frac{1}{2\pi}\int_0^{2\pi/\omega_{\rm b}} \sum_{n=1}^N \phi_n^\ast
\rmi \dot{\phi}_{n} \rmd t = H' / \omega_{\rm b} 
- \frac{1}{4\pi}\int_0^{2\pi/\omega_{\rm b}} \sum_{n=1}^N |\phi_n|^4 \rmd t .
\label{action}
\end{equation}
For a stationary solution with 
$\rmi\dot{\phi}_{n}= - \omega_{\rm b} \phi_n$, $I$ is simply proportional 
to the 
norm ($I=-{\mathcal N}$)\footnote{The 
minus-sign arises only because in terms of action/angle 
variables, frequencies are defined 
with a sign opposite to that of \eref{stationary} and \eref{2freq}, 
see e.g.\ \cite{JA97} and section \ref{sec:dyn}.}, while for 
genuine two-frequency solutions $I$ and ${\mathcal N}$ are independent. 
In the following, we choose as 
independent parameters the frequency $\omega_{\rm b}$ in the rotating 
frame and 
the norm $\mathcal N$, and perform the continuation versus 
$\omega_{\rm b}$ for 
fixed norm by, at each value of $\omega_{\rm b}$, vary the 
parameter $\omega_0$ in 
\eref{DNLSphi} so 
to obtain a solution with the desired value of $\mathcal N$ 
(this procedure may fail at particular points if the dependence 
${\mathcal N} (\omega_0)$ for solutions at fixed $\omega_{\rm b}$ is 
nonmonotonous or multivalued; examples of this can be found below).

Linear stability of two-frequency solutions is obtained by
linearizing  \eref{DNLSphi} around an exact time-periodic 
$\phi_n(t) = \phi_n^{(\rm r)}(t) + \rmi \phi_n^{(\rm i)}(t)$ 
($\phi_n^{(\rm r)}$, $\phi_n^{(\rm i)}$ real). 
The linearized equation for the small 
perturbation 
$\epsilon_n(t)$ will then have the same form as \eref{linear} with $\Lambda$ 
replaced by $\omega_0$, and the real and constant $\phi_n^{(\Lambda)}$ 
replaced by a generally complex and time-periodic 
$\phi_n(t)$. Decomposing
$\epsilon_n(t) = \epsilon_n^{(\rm r)}(t) + \rmi \epsilon_n^{(\rm i)}(t)$
as in \eref{matrix}, 
the 
linearized equations can be written in Hamiltonian form as \cite{CE85,ELS85}
\begin{eqnarray}
\fl
\left ( \begin{array}{c}\{\dot{\epsilon}_n^{(\rm r)}\} \\ 
\{\dot{\epsilon}_n^{(\rm i)}\}
\end{array} \right) = 
\left ( \begin{array}{cc} 0 & {\bf{1}}\\ 
{-\bf{1}} & 0  \end{array} \right) \nonumber \\
\fl\times
\left ( \begin{array}{cc} \left(\omega_0 -3 (\phi_n^{(\rm r)})^2 
-(\phi_n^{(\rm i)})^2\right) {\bf 1}  - C \Delta 
& -2 \phi_n^{(\rm r)} \phi_n^{(\rm i)} {\bf {1}}\\ 
-2 \phi_n^{(\rm r)} \phi_n^{(\rm i)} {\bf 1} &  
\left (\omega_0 - (\phi_n^{(\rm r)})^2 -3 (\phi_n^{(\rm i)})^2 
\right) {\bf 1}  - C \Delta 
\end{array} \right) 
\left ( \begin{array}{c}\{{\epsilon}_n^{(\rm r)}\} \\ 
\{{\epsilon}_n^{(\rm i)}\}\\
\end{array} \right)\nonumber\\
\label{matrix2}
\end{eqnarray}
where $\bf{1}$ denotes the $N$-dimensional unit matrix and $\Delta$ the 
discrete Laplacian, $\Delta \epsilon_n = \epsilon_{n+1}+\epsilon_{n-1}-
2\epsilon_n$ (note that the second matrix in \eref{matrix2} is symmetric). 
Integrating \eref{matrix2} numerically over the period 
$T_{\rm b} = 2\pi/\omega_{\rm b}$ of $\phi_n$ yields the symplectic 
$2N \times 2N$-dimensional Floquet matrix $\mathbf{F_{T_{\rm b}}}$ 
defined by 
\begin{equation}
\left(\begin{array}{c}\{\epsilon_n^{(\rm r)}(T_{\rm b})\} \\ 
\{\epsilon_n^{(\rm i)}(T_{\rm b})\} 
\end{array} \right) = \mathbf{F_{T_{\rm b}}}
\left(\begin{array}{c}\{\epsilon_n^{(\rm r)}(0)\} \\ 
\{\epsilon_n^{(\rm i)}(0)\} 
\end{array} \right) , 
\label{Floquet}
\end{equation}
and the solution $\phi_n(t)$ is linearly stable if and only if all 
eigenvalues $\alpha$ of $\mathbf{F_{T_{\rm b}}}$ lie on the unit circle. 
In the particular case of a stationary solution, the eigenvalues 
$\alpha$ of $\mathbf{F_{T_{\rm b}}}$ are related 
to the linear eigenfrequencies $\omega_{\rm l}$ of 
\eref{eigeneqab}, \eref{eigeneqUW} as 
$\alpha=\rme^{\rmi 2\pi \omega_{\rm l}/\omega_{\rm b}}$ with 
$\omega_{\rm b}=\Lambda - \omega_0$. 

For a genuine two-frequency solution there are two arbitrary phases, one 
corresponding to global phase rotations which is put to zero 
at $t=0$ in the ansatz \eref{2freq}, and another corresponding to the 
phase of the time-periodic solution in the rotating frame 
which we fix by imposing time reversibility, 
$\phi_n^\ast(-t)=\phi_n(t)$ (i.e.\ $\phi_n^{(\rm i)}(0) = 0$). 
To each of these phase 
degeneracies there is a corresponding solution to the linearized equations 
\eref{matrix2} with period $T_{\rm b}$, and thus a pair of eigenvalues 
$\alpha=+1$ 
of $\mathbf{F_{T_{\rm b}}}$ \cite{JA97}. 
The solution corresponding to an infinitesimal global phase 
rotation is $\epsilon_n(t)=\rmi \phi_n(t)$ 
(i.e.\ $\left (\{\epsilon_n^{(\rm r)}\},\{\epsilon_n^{(\rm i)}\} \right) 
= \left (\{-\phi_n^{(\rm i)}\},\{\phi_n^{(\rm r)}\} \right) $), 
while the solution corresponding to an infinitesimal translation of 
the phase in the rotating frame is 
$\epsilon_n(t)=\dot{\phi}_n(t)$
(i.e. $\left (\{\epsilon_n^{(\rm r)}\},\{\epsilon_n^{(\rm i)}\} \right) 
= \left (\{\dot{\phi}_n^{(\rm r)}\},\{\dot{\phi}_n^{(\rm i)}\} \right) $).
(Note that for stationary solutions, these become linearly dependent 
and there is only one pair of eigenvalues at +1.)
Thus, for the particular case $N=3$ there is for 
a genuine two-frequency solution at most 
one pair of Floquet eigenvalues $\alpha$ not identical to +1, so 
the only way such a solution can change its stability is through 
collisions of the members of this pair either at +1 or at -1. 

Furthermore, due to the invariance of the DNLS equation and its 
linearization under the transformation $\phi_n \rightarrow -\phi_n$ and 
the antisymmetry ($\phi_2^{(\Lambda)}=-\phi_1^{(\Lambda)}; 
\phi_3^{(\Lambda)}=0$)
of the particular stationary solutions, 
the families of two-frequency solutions bifurcating from them  
will have the additional space-time symmetry 
$\phi_1(t+T_{\rm b}/2)=\phi_2(t), \phi_3(t+T_{\rm b}/2)=\phi_3(t)$. 
This implies that the time-periodic solution $\phi_n$ 
with $\phi_1(0)\neq\phi_2(0)$
contains only odd harmonics of its fundamental frequency $\omega_{\rm b}$. 
As a conseqeunce, collisions of Floquet eigenvalues at -1 
corresponding to period-doubling bifurcations will not yield instabilities, 
which can be seen by decomposing the Floquet matrix as 
$\mathbf{F_{T_{\rm b}}}={\mathbf P_{1\leftrightarrow 2}}
\mathbf{F_{T_{\rm b}/2}}
{\mathbf P_{1\leftrightarrow 2}}
\mathbf{F_{T_{\rm b}/2}}$, where $\mathbf{F_{T_{\rm b}/2}}$ is the 
matrix obtained by integrating \eref{matrix2} over half a period and 
${\mathbf P_{1\leftrightarrow 2}}$ interchanges sites 1 and 2. Thus, 
only collisions of eigenvalues $\beta$ to 
${\mathbf P_{1\leftrightarrow 2}}\mathbf{F_{T_{\rm b}/2}}$ at 
$\beta= \pm 1$ can yield instabilities, and they must all correspond to 
eigenvalue collisions at $\alpha = +1$ for the total Floquet matrix 
$\mathbf{F_{T_{\rm b}}}$ since $\alpha=\beta^2$. Below, we report the 
numerical stability properties in terms of eigenvalues 
$\beta$ rather than $\alpha$. 

\subsection{The low-amplitude 'Type I' bifurcation}
\label{sec:TypeI}
\subsubsection{Small ${\mathcal N}/C $.}
\label{sec:Ismall}
We first discuss some properties of the two 
families of two-frequency solutions bifurcating from the stationary 
solution for $\Lambda/C < x_\ast$, i.e.\ ${\mathcal N}/C <  9.077...$, 
corresponding to the linear modes with
asymptotic oscillation patterns \eref{lin+} and \eref{lin-}, 
respectively, for small ${\mathcal N}/C$. Their continuation
versus $\omega_{\rm b}$ for fixed ${\mathcal N}/C=5$ is illustrated in 
figure \ref{fig_N=5}. 
\begin{figure}  
\begin{minipage}[l]{0.49\textwidth} 
\includegraphics[height=\textwidth,angle=270]{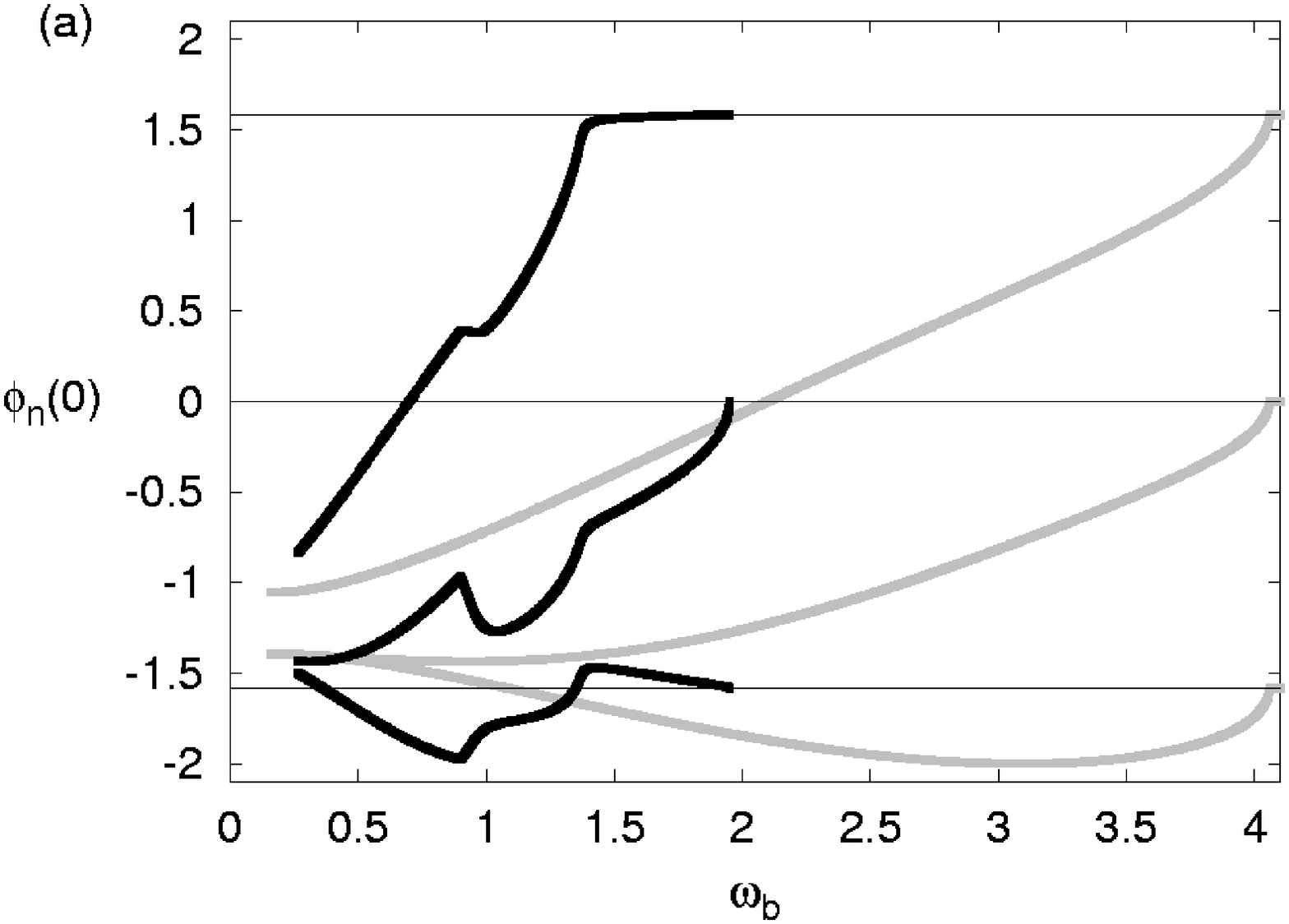} 
\end{minipage}%
\begin{minipage}[r]{0.49\textwidth} 
\includegraphics[height=\textwidth,angle=270]{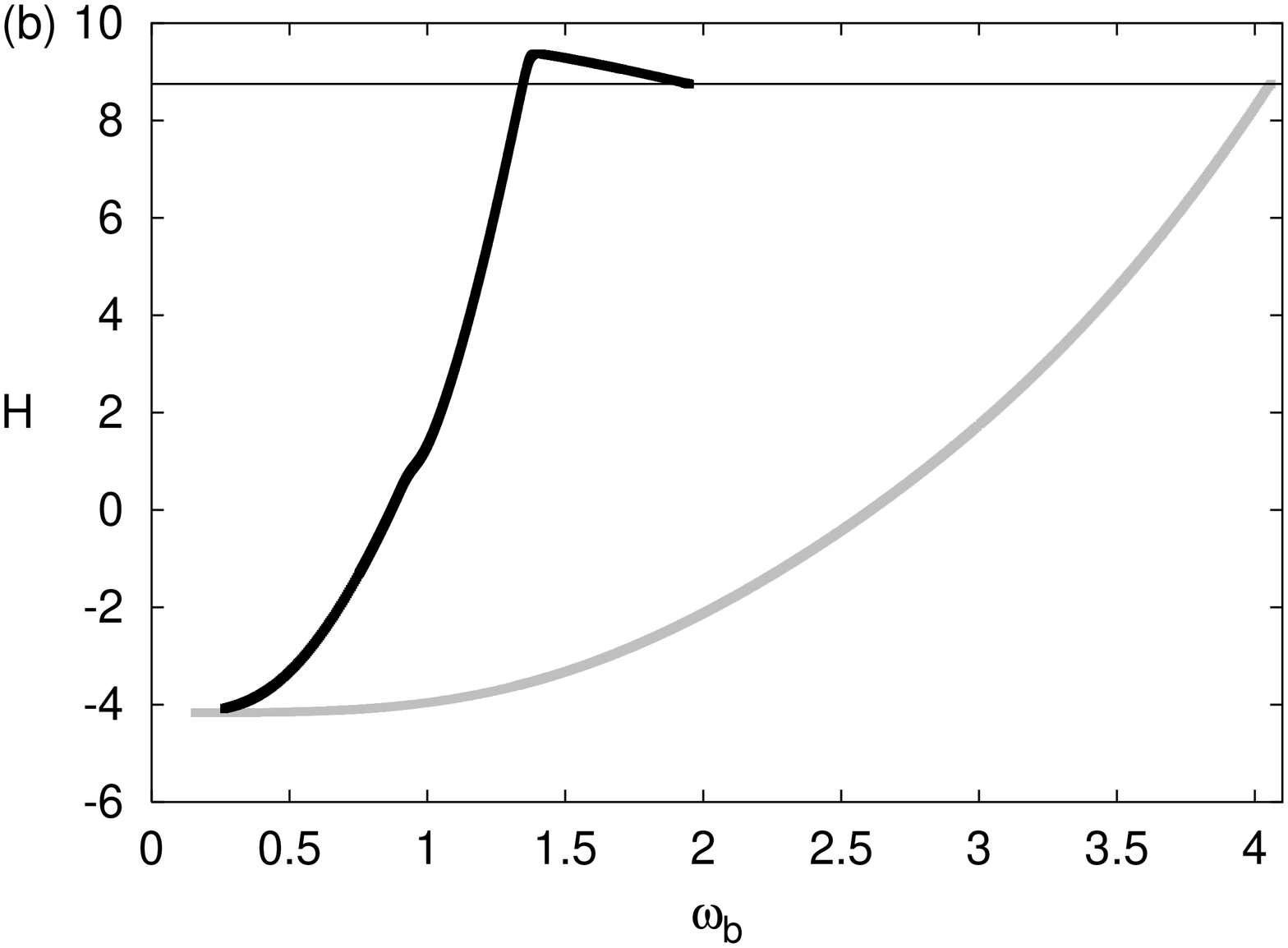} 
\end{minipage}\\
\begin{minipage}[l]{0.49\textwidth} 
\includegraphics[height=\textwidth,angle=270]{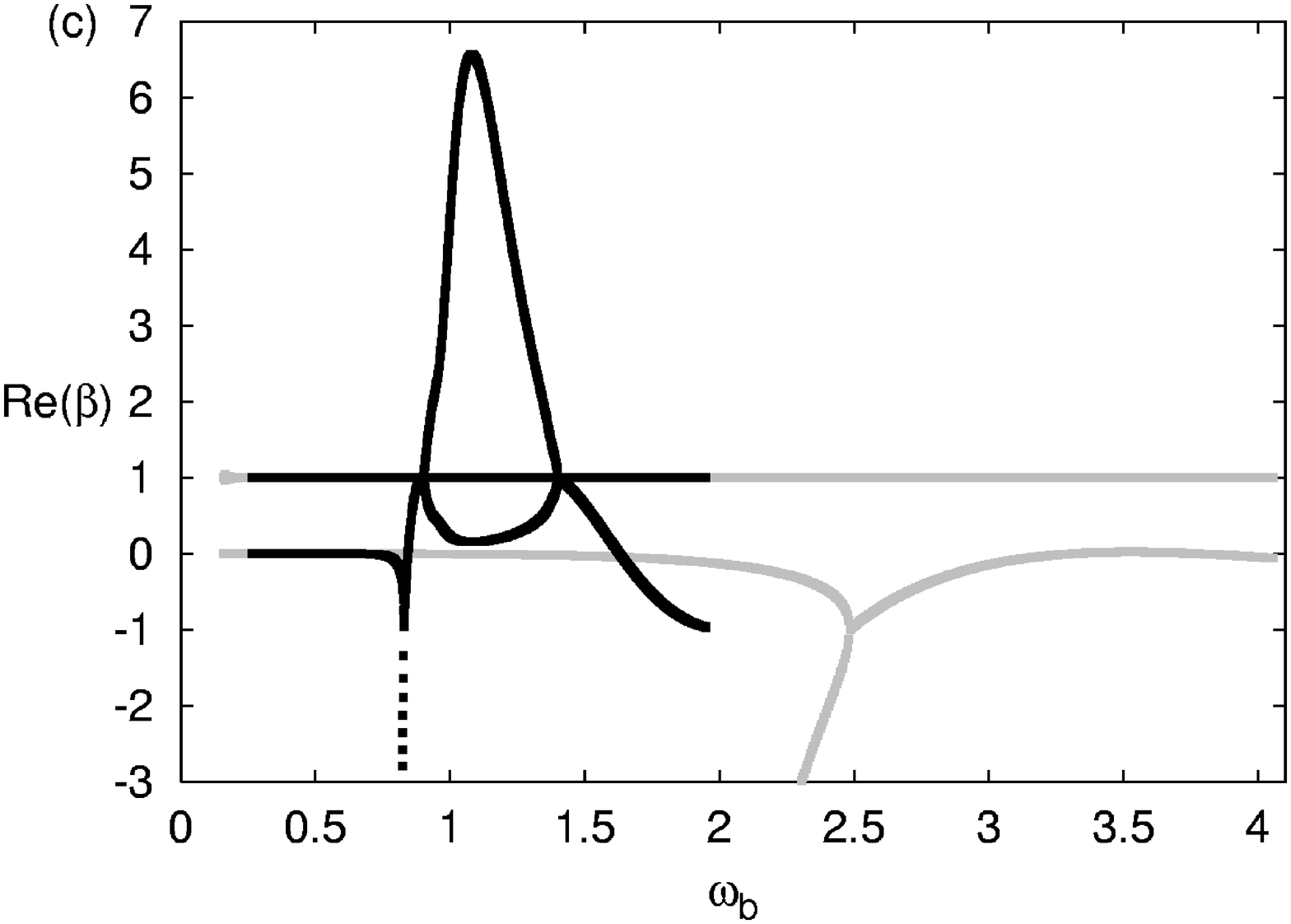}
\end{minipage}
\begin{minipage}[r]{0.49\textwidth} 
\includegraphics[height=\textwidth,angle=270]{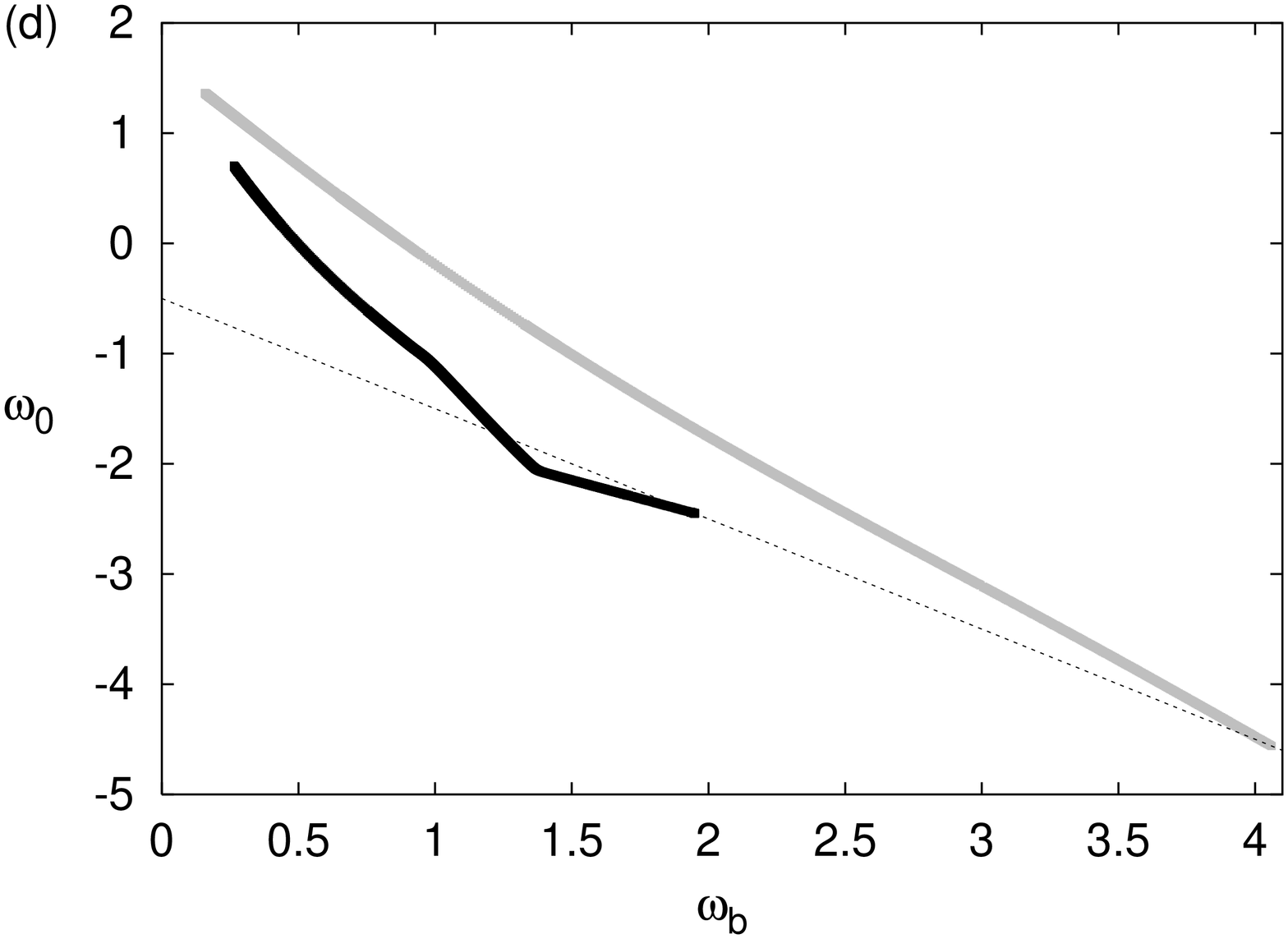}
\end{minipage}\caption{
Continuation versus $\omega_{\rm b}$ at fixed norm ${\mathcal N}=5$ ($C=1$) 
of the 
two-frequency solutions bifurcating 
from the stationary solution \eref{ALambda} with 
$\Lambda=-0.5$ at 
 $\omega_{\rm b}=\omega_{\rm l \pm}$ given by \eref{roots}; 
black [grey] curves correspond to $\omega_{\rm l -}$ [$\omega_{\rm l +}$]. 
(a) Initial amplitudes $\phi_n(0)$ 
for $n=1$ (upper), $n=2$ (lower) and $n=3$ 
(middle curves); (b) Hamiltonian $H$ \eref{hamil}; 
(c) real parts of stability eigenvalues $\beta$; (d) 
frequency $\omega_0$ of the rotating frame. 
The horizontal lines in (a) and (b) and the dashed line in (d) correspond to 
the stationary solution 
according to \eref{ALambda} and \eref{HLambda}. The large negative 
eigenvalues not shown in (c) for small $\omega_{\rm b}$ will continue to 
decrease rapidly (e.g.\ at $\omega_{\rm b}=0.2$,  
$\beta\approx-10^9$ for the grey solution). For $\omega_b\lesssim 0.90$, 
the black solution  is identical to 
the grey at frequency $3\omega_b$.
}
\label{fig_N=5} 
\end{figure} 
For both solutions, the continuation from the 
stationary solution  is monotonous 
in the direction of decreasing $\omega_{\rm b}$, and they are 
linearly stable 
for a rather large interval of frequencies 
$\omega_{\rm b}\leq \omega_{\rm l \pm}$, 
but become strongly unstable at small 
$\omega_{\rm b}$ (figure \ref{fig_N=5} (c)). 
For the (grey) solution corresponding to  $\omega_{\rm l +}$, 
the instability is caused by a collision of 
eigenvalues $\beta$ at -1 producing a rapid and monotonous growth of 
$|\beta|$ for small $\omega_{\rm b}$. 
This behaviour is typical for  ${\mathcal N}/C > 4.5$. 
For  ${\mathcal N}/C < 4.5$, 
 $\beta$ attains a minimum 
value and 
then returns towards the unit circle, and finally through a collision 
at $\beta=-1$ the grey solution bifurcates into the homogeneous 
stationary solution 
with $\phi_1=\phi_2=\phi_3=-\sqrt{{\mathcal N}/3}$, which is linearly stable 
in this 
regime \cite{CE85,ELS85}. (The continuation versus $\omega_b$ is typically 
non-monotonous close to this bifurcation point.) At ${\mathcal N}/C = 4.5$ 
the homogeneous solution becomes unstable \cite{CE85}, 
and by analyzing the dynamics of the grey solution in the 
limit $\omega_{\rm b}\rightarrow 0$
for ${\mathcal N}/C > 4.5$, we find that it asymptotically 
approaches a (strongly unstable) solution consisting of 
homoclinic connections of unstable
and stable manifolds of the homogeneous  solution 
(as further discussed in section \ref{sec:Ilarge}, see also figure 
\ref{fig_N=15dyn} (b) below).

The 
scenario for the (black)
solution corresponding to  $\omega_{\rm l -}$ 
is more complicated: 
it first becomes unstable through an eigenvalue collision at $\beta=+1$, 
associated with a change of nature of the solution 
as
seen in figure \ref{fig_N=5} (a). This is related to a 
third-harmonic resonance with the grey  solution, and 
the instability appears at $\omega_{\rm b}$ 
close to (but not exactly at) $\omega_{\rm l +}/3$. 
The instability first grows when
decreasing $\omega_{\rm b}$, but attains a 
maximum and finally,
through a second collision at $+1$, the black solution
 becomes identical to the (stable) grey 
solution 
at frequency $3\omega_{\rm b}$, which becomes 
unstable slightly after.  
As mentioned above, for symmetry reasons  higher-harmonic resonances can  
occur only 
for odd harmonics. Since  $\omega_{\rm l -}$ decreases 
towards zero when ${\mathcal N}/C$ decreases while $\omega_{\rm l +}$ 
remains finite, for smaller  ${\mathcal N}/C$ the resonances are 
of higher order. Thus, for example when ${\mathcal N}/C$ approaches
from above a critical value $\sim 3.53$ (where 
$\omega_{\rm l +}=3\omega_{\rm l -}$) the value of $\omega_{\rm b}$ where 
the black 
solution becomes unstable approaches 
the linear frequency $\omega_{\rm b}=\omega_{\rm l -}$, and for smaller 
 ${\mathcal N}/C$ the third-harmonic resonance 
disappears and the solution is stable until a similar fifth-harmonic 
resonance appears for smaller 
$\omega_{\rm b}$. It is interesting to note, that 
if the numerical continuation is performed 
'carelessly', i.e.\ with larger steps, one might 'jump' the regime yielding 
a $p$th harmonic 
resonance  and catch instead 
a solution in apparent (but non-smooth!) continuation of the original 
solution, as if the resonance had not existed. This solution can then be 
continued towards smaller $\omega_{\rm b}$ until a ($p+2$)th harmonic 
resonance is encountered, etc. Thus, a cascade of higher-order 
resonances will result, and the scenario is analogous to that previously 
found for higher-order resonances in Klein-Gordon chains in 
\cite{MJKA02} (e.g.\ figures 12 and 16 in this paper) and 
\cite{MJAK02}. The behaviour shown in figure \ref{fig_N=5} is typical for 
$3.6\lesssim{\mathcal N}/C\lesssim 5.5$, while for 
${\mathcal N}/C\approx 5.6$ two additional eigenvalue collisions at 
$\beta=+1$, producing a new pair of small stability/instability regimes, 
appear slightly before
the transformation of the black
solution into the grey 
$3\omega_{\rm b}$-solution.
A further 
increase of ${\mathcal N}/C$ yields one more pair of small regimes of 
stability/instability through a collision at $\beta=-1$, and for 
${\mathcal N}/C\gtrsim 6.8$ the continuation versus  $\omega_{\rm b}$ 
becomes nonmonotonous of 'Z-type',
where three different solutions of the family have the same frequency 
 $\omega_{\rm b}$, in some regime before the final bifurcation with the 
$3\omega_b$-solution (similar behaviour was also observed 
in \cite{MJKA02,MJAK02}). 

Note also (figure \ref{fig_N=5} (b)) that the Hamiltonian of the 
black solution changes 
slope so that ${\rm d}H/{\rm d}\omega_{\rm b}=0$ at the first eigenvalue 
collision at $\beta=+1$ (but not at the second where the solution 
transforms to the grey third-harmonic solution). Moreover, the 
additional eigenvalue collisions at $\beta=+1$ for 
${\mathcal N}/C\approx 5.6$ also produce a new pair of local min/max for
$H(\omega_{\rm b})$ at the collision points. 
Generally 
(see also several similar examples below), it appears 
that ${\rm d}H/{\rm d}\omega_{\rm b}=0$ for two-frequency 
solutions at fixed norm could be 
a sufficient (but not necessary) condition for a change in stability through 
an eigenvalue collision at +1, analogous to the wellknown  criteria 
${\rm d}{\mathcal N}/{\rm d}\Lambda=0$ for stationary solutions 
\cite{Laedke} and the more general 
${\rm d}{\mathcal I}/{\rm d}\omega=0$ for periodic solutions of 
Hamiltonian systems with frequency $\omega$ and action ${\mathcal I}$ 
\cite{Bridges91}. Indeed, this could be expected since a turning point of 
$H$ for relative periodic solutions at fixed $\mathcal N$ should correspond 
to a saddle-centre bifurcation. 

\subsubsection{Around the bifurcation point.}
\label{sec:Ibif}
\begin{figure}  
\begin{minipage}[l]{0.5\textwidth} 
\includegraphics[height=0.99\textwidth,angle=270]{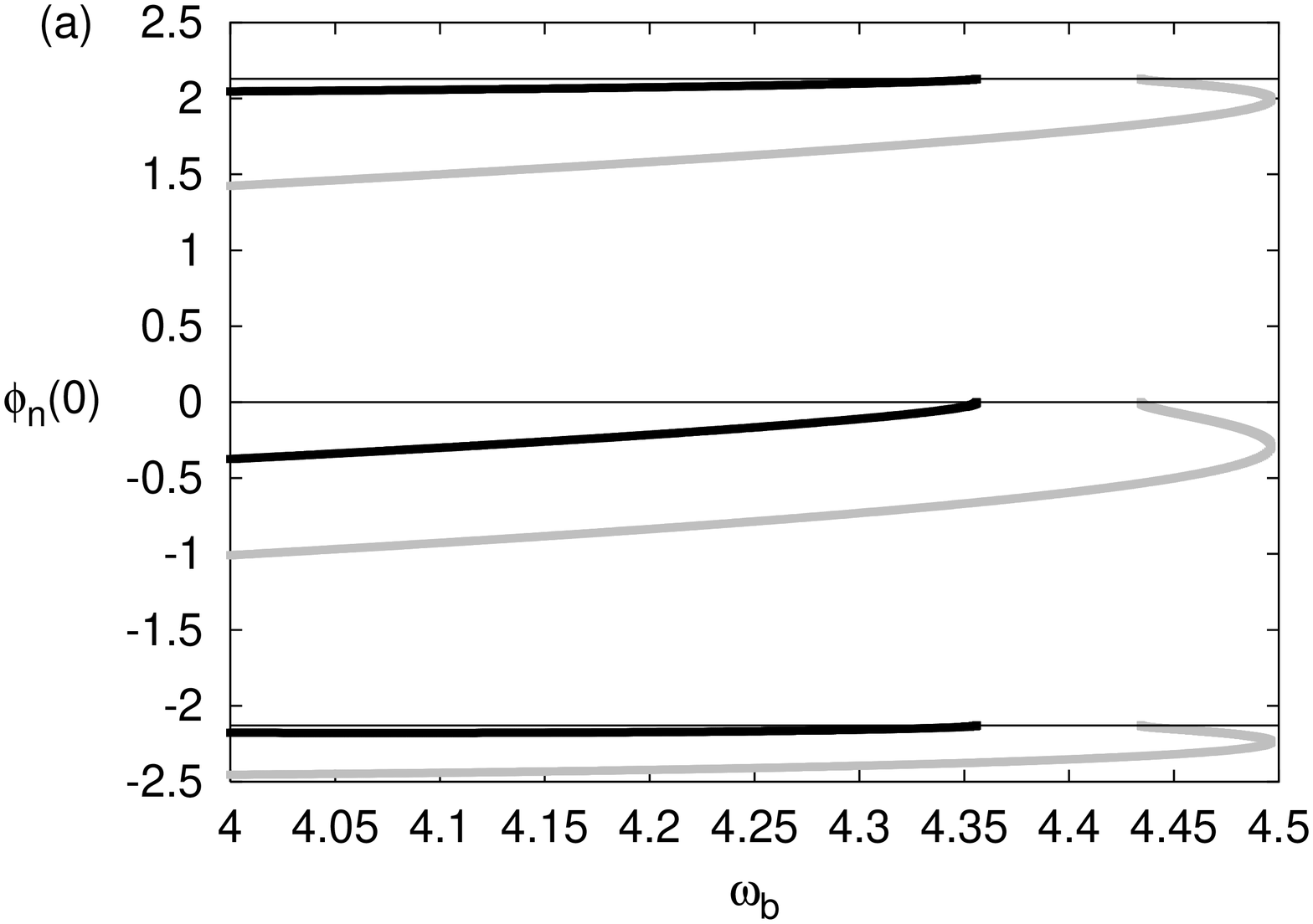} 
\end{minipage}%
\begin{minipage}[r]{0.5\textwidth} 
\includegraphics[height=0.99\textwidth,angle=270]{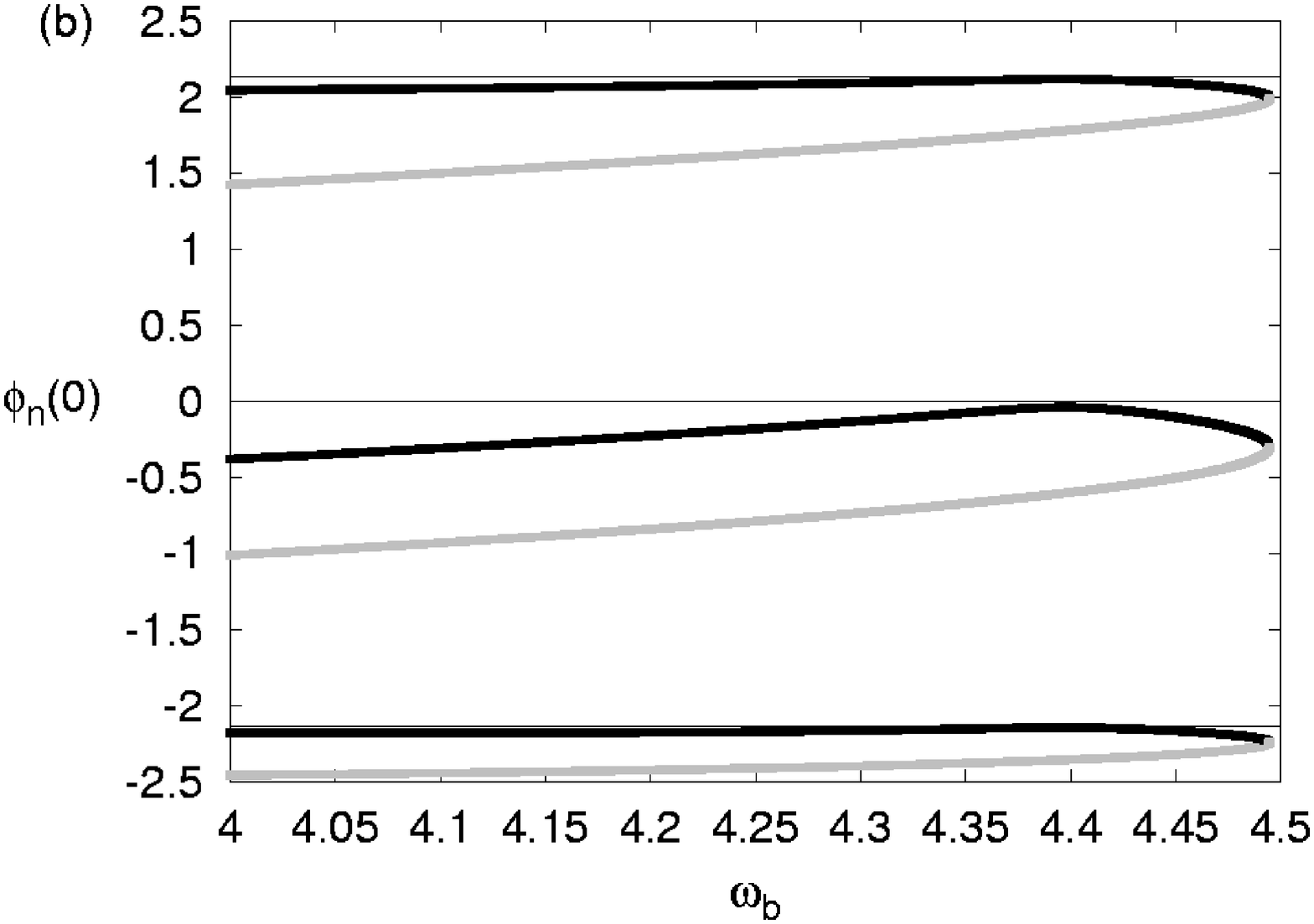} 
\end{minipage}\\
\begin{minipage}[l]{0.5\textwidth} 
\includegraphics[height=0.99\textwidth,angle=270]{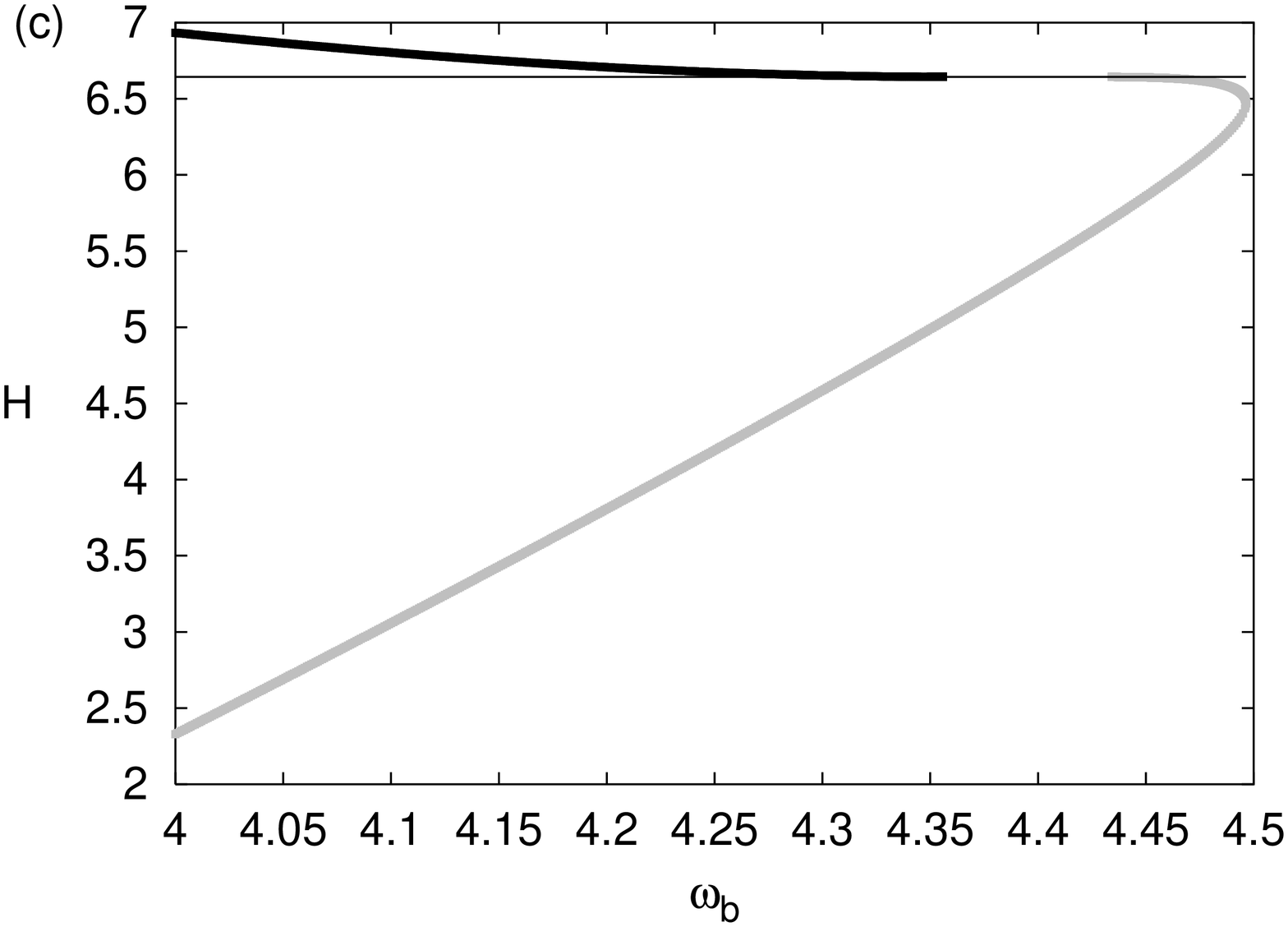} 
\end{minipage}%
\begin{minipage}[r]{0.5\textwidth} 
\includegraphics[height=0.99\textwidth,angle=270]{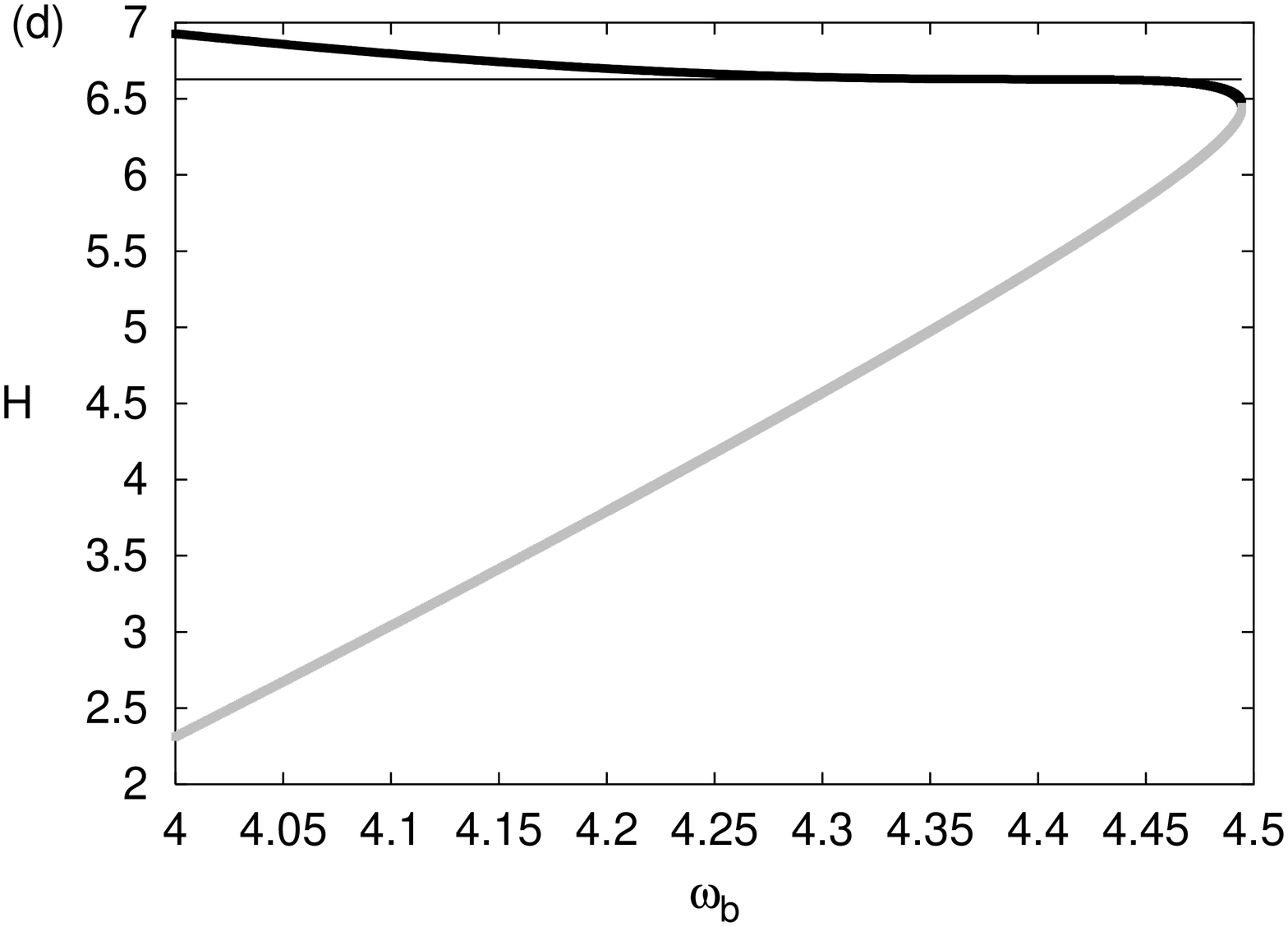} 
\end{minipage}\\
\begin{minipage}[l]{0.5\textwidth} 
\includegraphics[height=0.99\textwidth,angle=270]{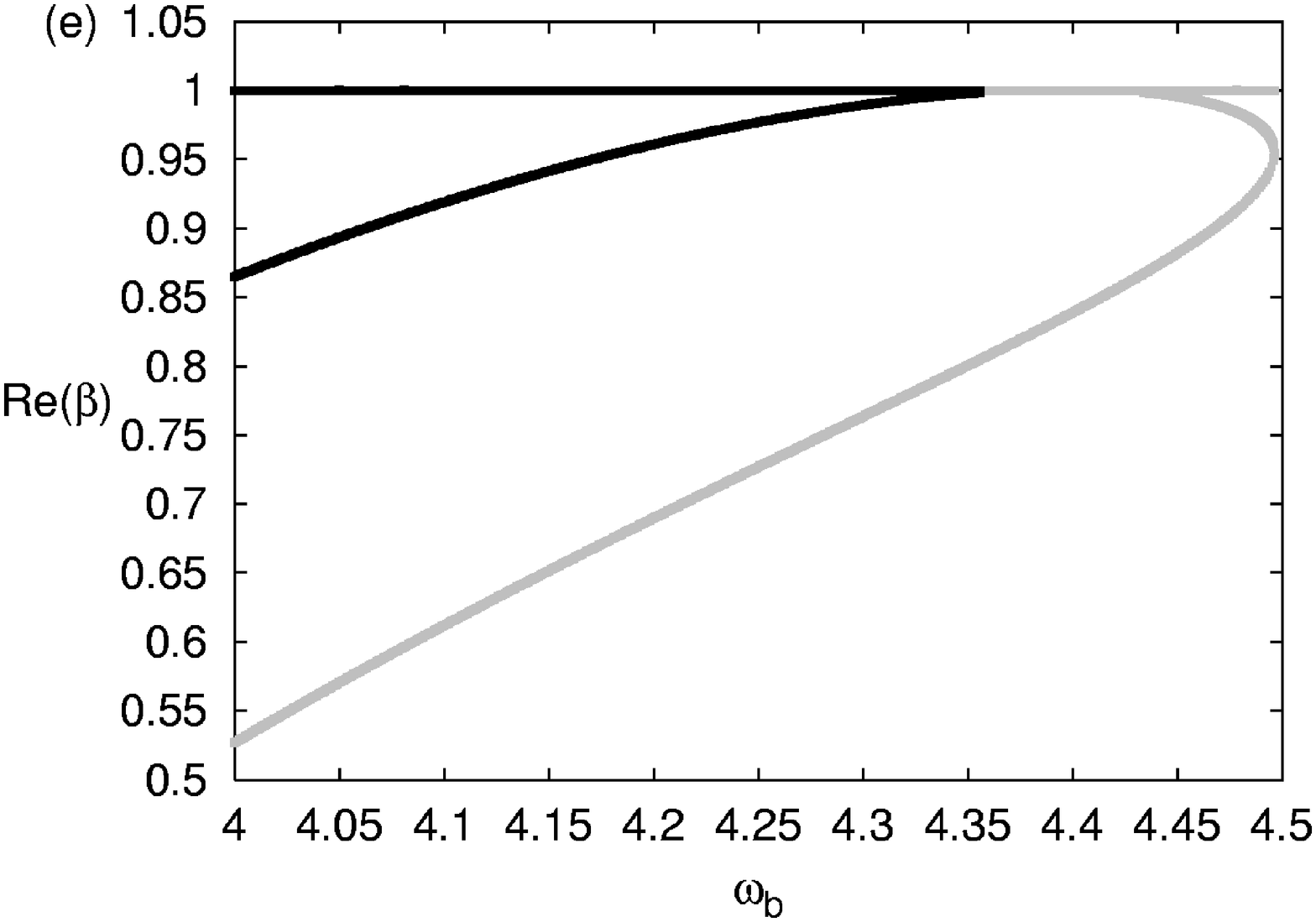} 
\end{minipage}%
\begin{minipage}[r]{0.5\textwidth} 
\includegraphics[height=0.99\textwidth,angle=270]{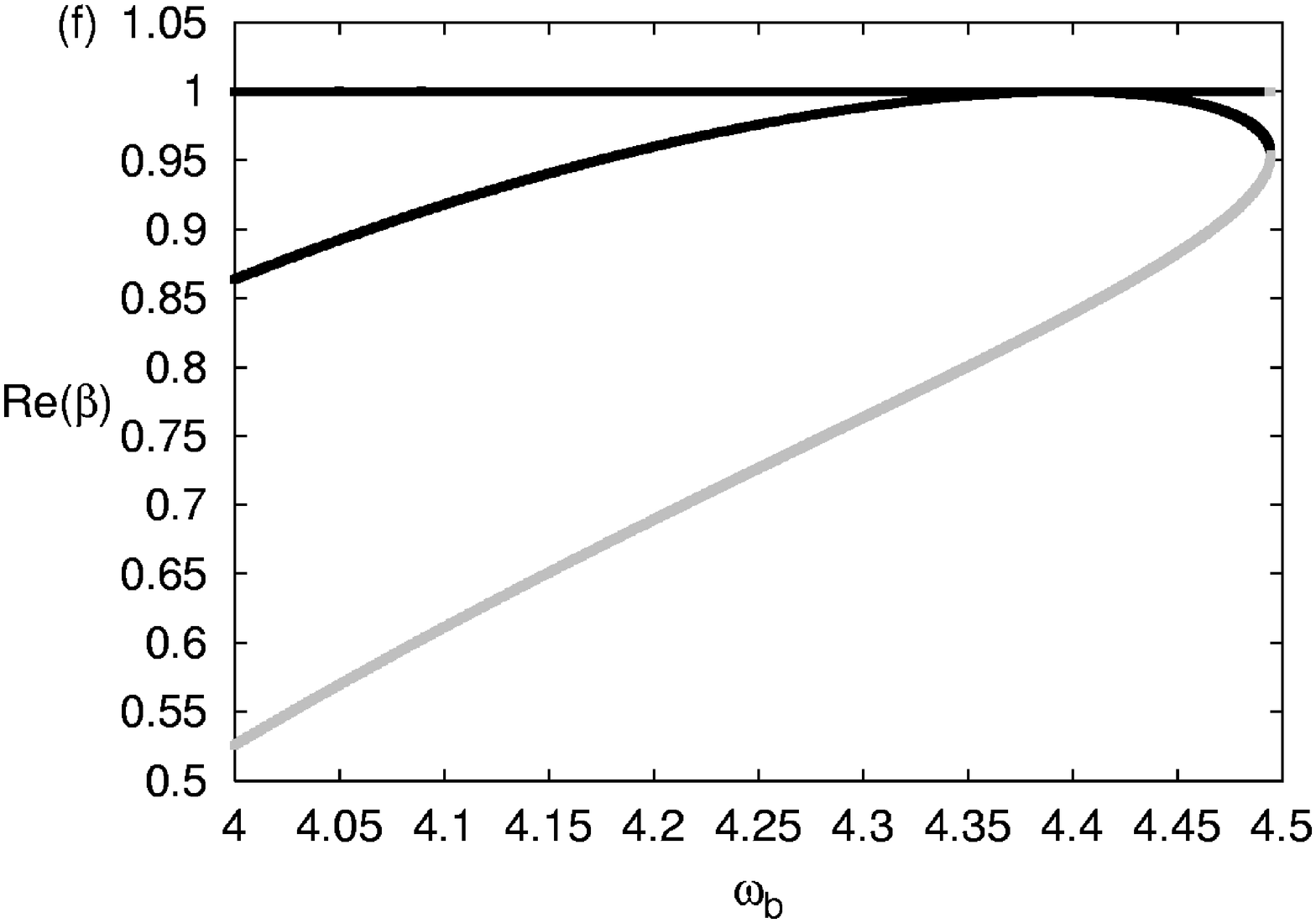} 
\end{minipage}\\
\caption{
Same as figure \ref{fig_N=5} (a)-(c) but for ${\mathcal N}=9.07$ 
(left figures) and 
${\mathcal N} =9.08$ (right figures). In (a), (c), (e) 
the meaning of the 
black and grey curves is the same as in figure \ref{fig_N=5}, while 
in (b), (d), (f) they 
illustrate the two different but connected branches of coexisting 
two-frequency solutions continued from the solutions in the left figures.
}
\label{fig_Ibif} 
\end{figure} 
Focusing on the regime of $\omega_{\rm b}$ close to 
$\omega_{\rm l\pm}$, we compare the 
solutions on both sides of the bifurcation point 
${\mathcal N}/C \approx  9.077$ in figure \ref{fig_Ibif}.
For 
$8.95 \lesssim {\mathcal N}/C \lesssim 9.077$, 
the continuation of the grey solution  
from the stationary solution
is no longer monotonous versus $\omega_{\rm b}$, but goes first 
in the direction of 
{\em increasing} $\omega_{\rm b}$ from 
$\omega_{\rm l +}$ up to a maximum value, 
and then turns and continues towards smaller 
$\omega_{\rm b}$ similarly as in figure \ref{fig_N=5}. 
As 
mentioned above, this 
multivaluedness causes numerical problems for the method of 
continuing for fixed norm by varying $\omega_0$ at one point 
(close to but different from the turning point), where 
${\rm d}{\mathcal N}/{\rm d}\omega_0$ for fixed $\omega_{\rm b}$ becomes 
infinite. 

At the bifurcation point ${\mathcal N}/C \approx  9.077$ the gap 
separating the black and grey families of solutions  in 
figure \ref{fig_Ibif} (a), (c), (e) closes 
(since $\omega_{\rm l -} = \omega_{\rm l +}$ at this point), and for 
larger ${\mathcal N}/C$ as in figure \ref{fig_Ibif} (b), (d), (f) they 
constitute two connected branches of a single family  of 
two-frequency solutions, 
now separated from the stationary solution.
Note that even though the two-frequency solutions are detached from 
the stationary solution on the superthreshold side of the bifurcation 
(where the latter is unstable and $\omega_{\rm l \pm}$ are complex), the 
distance between the two-frequency family and the stationary solution 
(which can be expressed in some suitable norm) decreases continuously 
to zero when approaching the bifurcation point from above. This is the 
characteristic property of a type I HH bifurcation 
\cite{LR01}, and  the scenario in figure 
\ref{fig_Ibif} is equivalent to that of figure 2 in \cite{LR01} (see also 
figure 2(b) in \cite{Bridges90}, \cite{Bridges91}). The importance of this 
property for the dynamics of the unstable stationary solution will be 
illustrated in section \ref{sec:dyn}. 

\subsubsection{Larger ${\mathcal N}/C$.}
\label{sec:Ilarge}
Increasing ${\mathcal N}/C$ on the superthreshold side of the type I 
bifurcation, the continued solutions persist as two connected branches 
of a single family detached from the stationary solution 
\eref{ALambda}, continuously increasing the distance to it. 
The typical behaviour for larger 
${\mathcal N}/C$ is illustrated in figure \ref{fig_N=15}.
\begin{figure}  
\begin{minipage}[l]{0.49\textwidth} 
\includegraphics[height=\textwidth,angle=270]{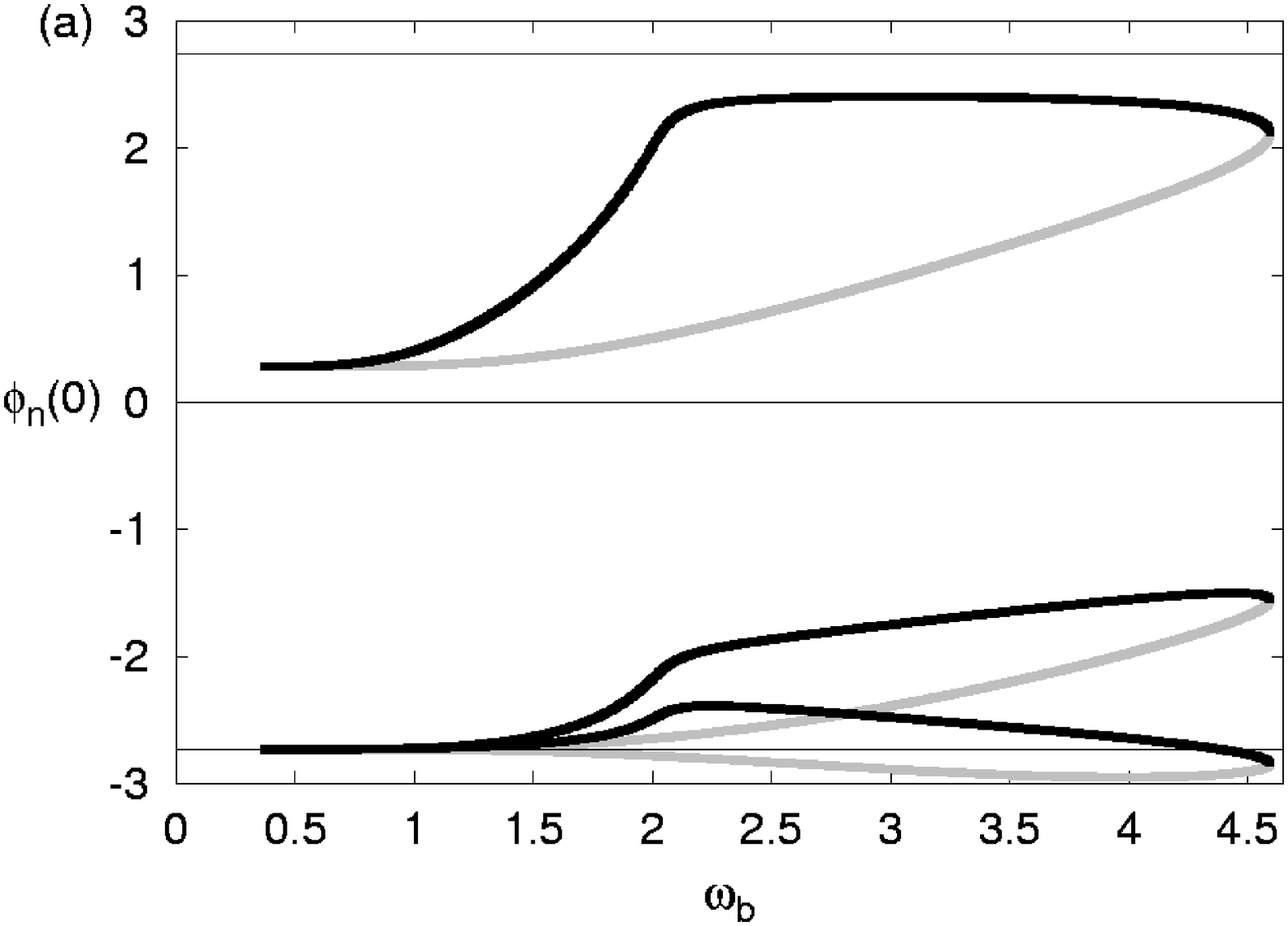} 
\end{minipage}%
\begin{minipage}[r]{0.49\textwidth} 
\includegraphics[height=\textwidth,angle=270]{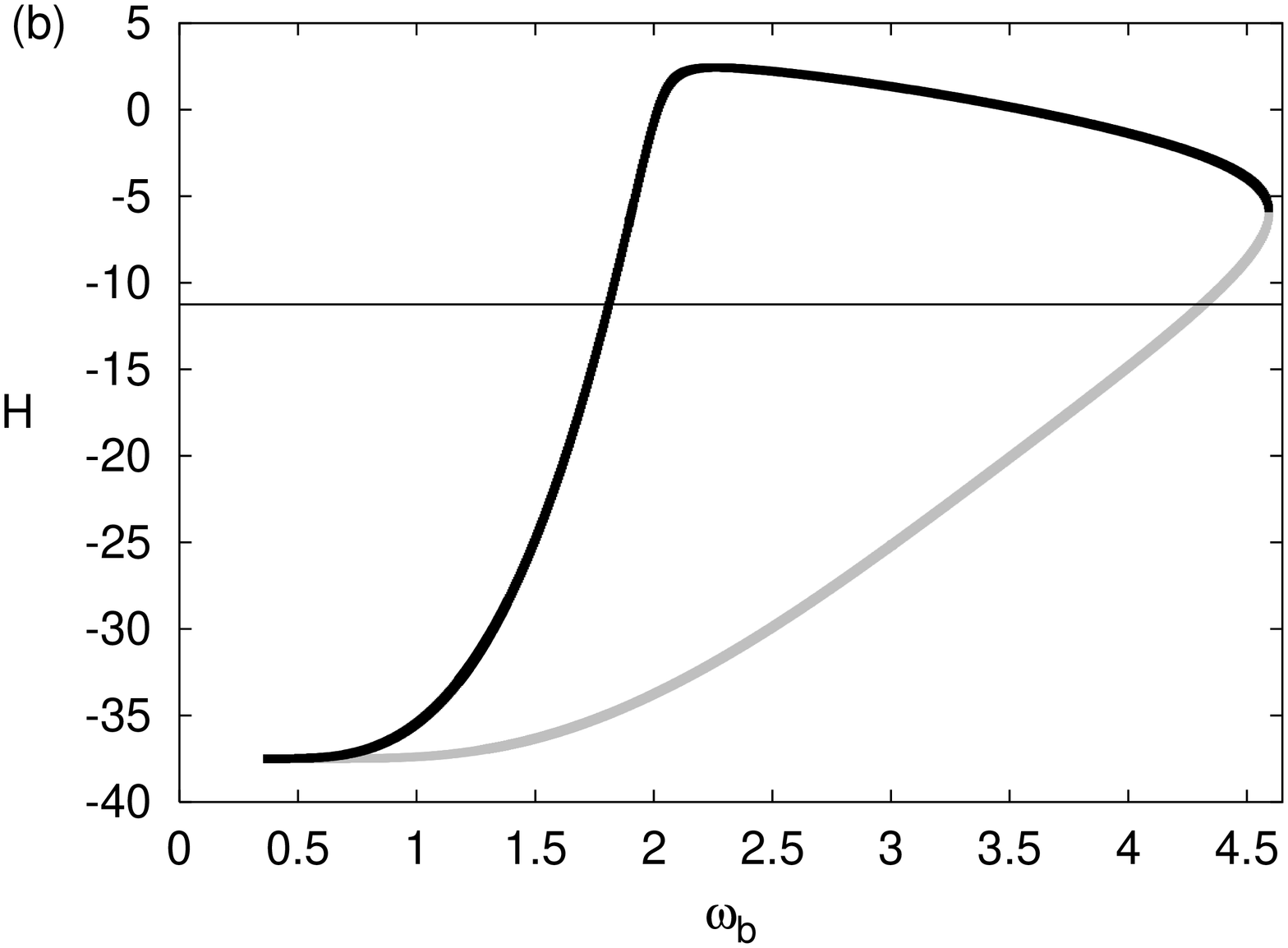} 
\end{minipage}\\
\begin{minipage}[l]{0.49\textwidth} 
\includegraphics[height=\textwidth,angle=270]{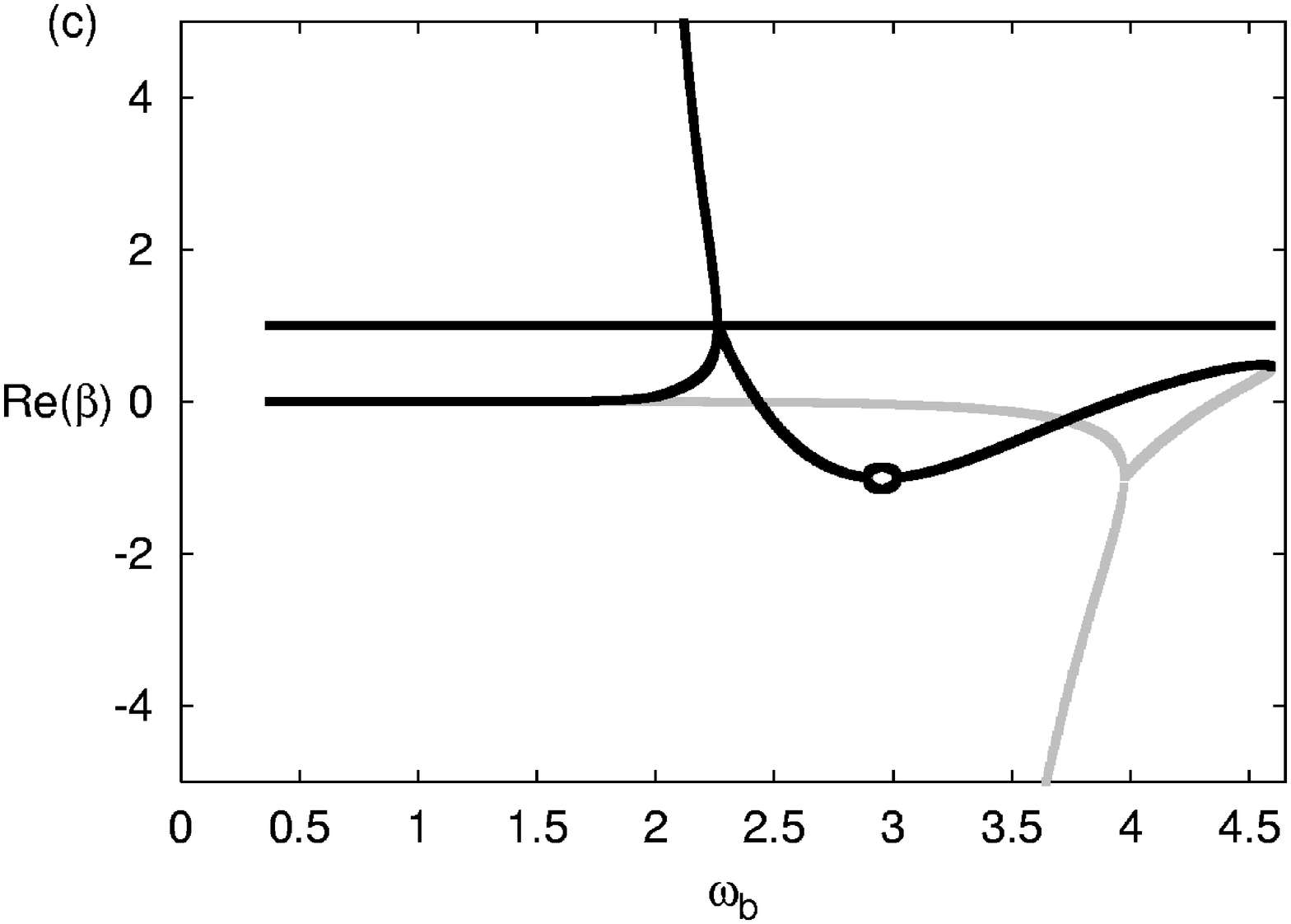}
\end{minipage}
\begin{minipage}[r]{0.49\textwidth} 
\includegraphics[height=\textwidth,angle=270]{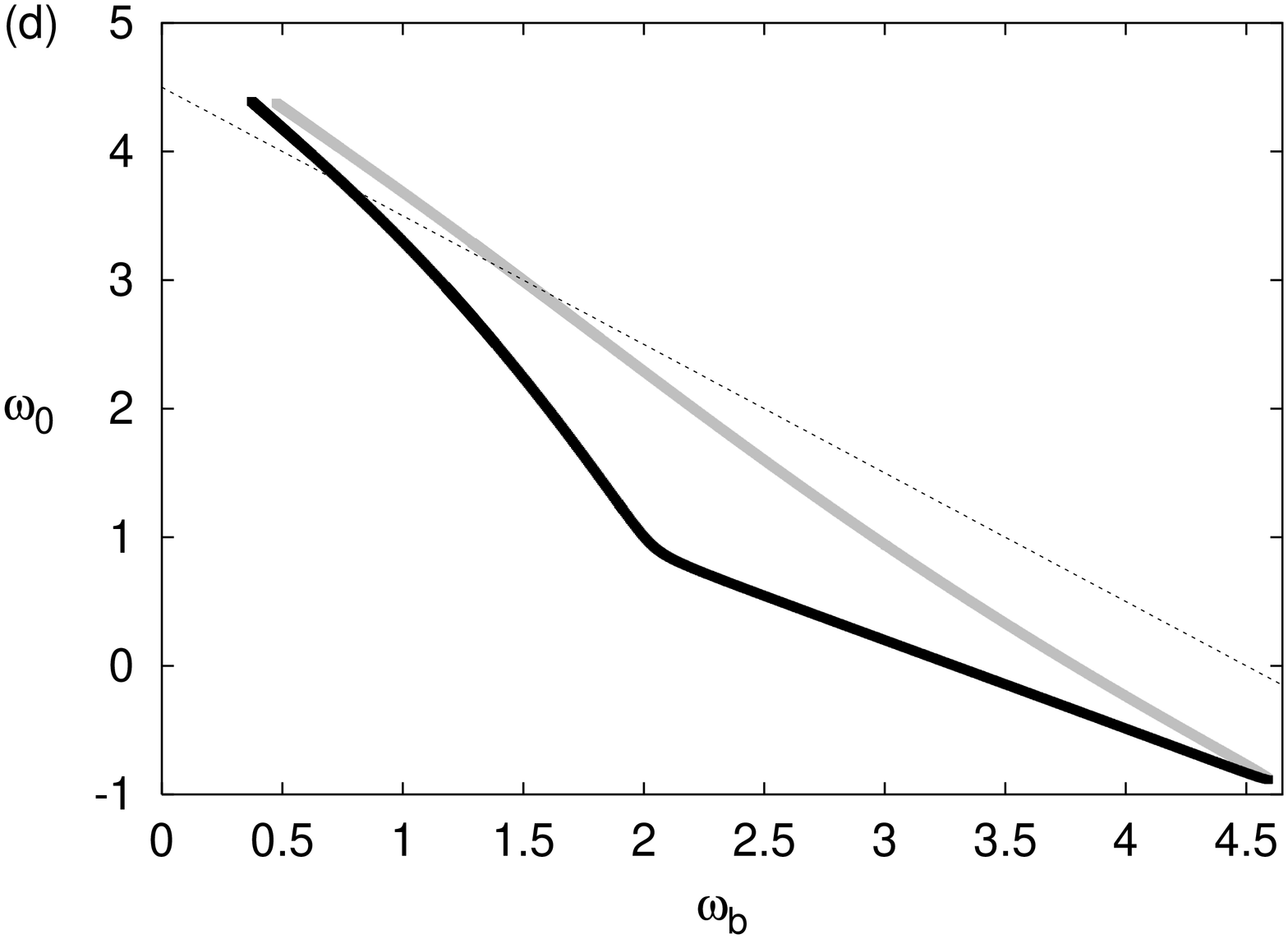}
\end{minipage}
\caption{
Continuation versus $\omega_{\rm b}$ at fixed norm ${\mathcal N}=15$ of the 
two-frequency solutions continued from those of figures 
\ref{fig_N=5}-\ref{fig_Ibif}. The 
meaning of the black and grey parts is the same as in figure \ref{fig_Ibif}
(b), (d), (f); other quantities have the same meaning as in figure 
\ref{fig_N=5}. 
The eigenvalues not shown in (c) for small $\omega_{\rm b}$ continue to 
decrease [increase] monotonously for the grey [black] solution; 
$\beta\approx-10^{12}$ [$\beta\approx 10^{11}$] at $\omega_{\rm b}=0.5$.
}
\label{fig_N=15} 
\end{figure}
Compared to the earlier discussed examples  
we note two main differences. First, there is a small 
bubble of instability through a collision at $\beta=-1$ 
for the black solution in  
figure \ref{fig_N=15} (c) for $\omega_{\rm b} \approx 2.9 - 3.0$. In fact, 
such an instability occurs in many regimes also for 
smaller 
${\mathcal N}/C$ (e.g.\ for ${\mathcal N}/C \gtrsim 5$), but it is then 
very weak and practically invisible numerically, while it grows
considerably stronger for larger ${\mathcal N}/C$. Second, 
for ${\mathcal N}/C \gtrsim 9.45$ the black solution no longer transforms 
into the third-harmonic of the grey solution for small $\omega_{\rm b}$, 
but approaches
 as $\omega_{\rm b}\rightarrow 0$ 
 a (strongly unstable) solution with the same values of 
$\phi_n(0)$, $H$ and 
$\omega_0$ as the asymptotic grey solution 
(figure \ref{fig_N=15} (a), (b), (d)), but with $\beta\rightarrow +\infty$ 
(figure \ref{fig_N=15} (c)). Thus, these asymptotic solutions are not 
identical, although they both consist of homoclinic connections 
of the unstable 
homogeneous stationary solution  $\phi_1=\phi_2=\phi_3$ 
(which has a doubly degenerate unstable eigenvalue 
\cite{CE85} so the corresponding stable and unstable manifolds 
can be connected in different ways). Analysis of the dynamics yields
that the grey solution in half a period contains two stable and two 
unstable manifolds from the homogeneous solution, and they are all 
associated with two-site localization with a third site of small 
$|\phi_n|^2$ 
as for $t=0$ in figure \ref{fig_N=15} (a) 
(see figure \ref{fig_N=15dyn} (b) below). The black solution 
on the other 
hand contains three stable and three unstable manifolds in one half-period, 
and of these two pairs are associated with one-site localization with 
two sites of smaller $|\phi_n|^2$, and only one pair with two-site 
localization (see  figure \ref{fig_N=15dyn} (a)).

We also note, by comparing figures \ref{fig_Ibif} (d) and 
\ref{fig_N=15} (b), that while close to the bifurcation point 
the Hamiltonian for the unstable stationary 
solution \eref{HLambda} 
cuts the two-frequency solution family on the black branch (figure 
\ref{fig_Ibif} (d)), the corresponding intersection point appears for larger 
 ${\mathcal N}/C$ on the grey branch (figure  \ref{fig_N=15} (b)). 
This change of branch occurs at  ${\mathcal N}/C \approx 10.6$, which, as 
will be discussed in section \ref{sec:dyn} below, is the value where the 
self-trapping transition of the unstable stationary solution is observed.

\subsection{The high-amplitude 'Type II' bifurcation}
\label{sec:TypeII}
Turning our attention to the two-frequency solutions associated 
with the high-amplitude Krein collision for the stationary solution at 
$\Lambda/C=6$, i.e.\ ${\mathcal N}/C =18$, we first describe some 
properties of these solutions (which  
are {\em not} identical to those involved 
in the low-amplitude bifurcation) on the stable side, 
i.e.\ for ${\mathcal N}/C > 18$.

\subsubsection{Large ${\mathcal N}/C $.}
\label{sec:IIlarge}
The continuation of the two-frequency solutions 
corresponding to linear modes with asymptotic oscillation patterns 
\eref{ac+} and \eref{ac-}, respectively, for large  
${\mathcal N}/C$, is shown in figure 
\ref{fig_N=20} for ${\mathcal N}/C = 20$. 
\begin{figure}  
\begin{minipage}[l]{0.49\textwidth} 
\includegraphics[height=\textwidth,angle=270]{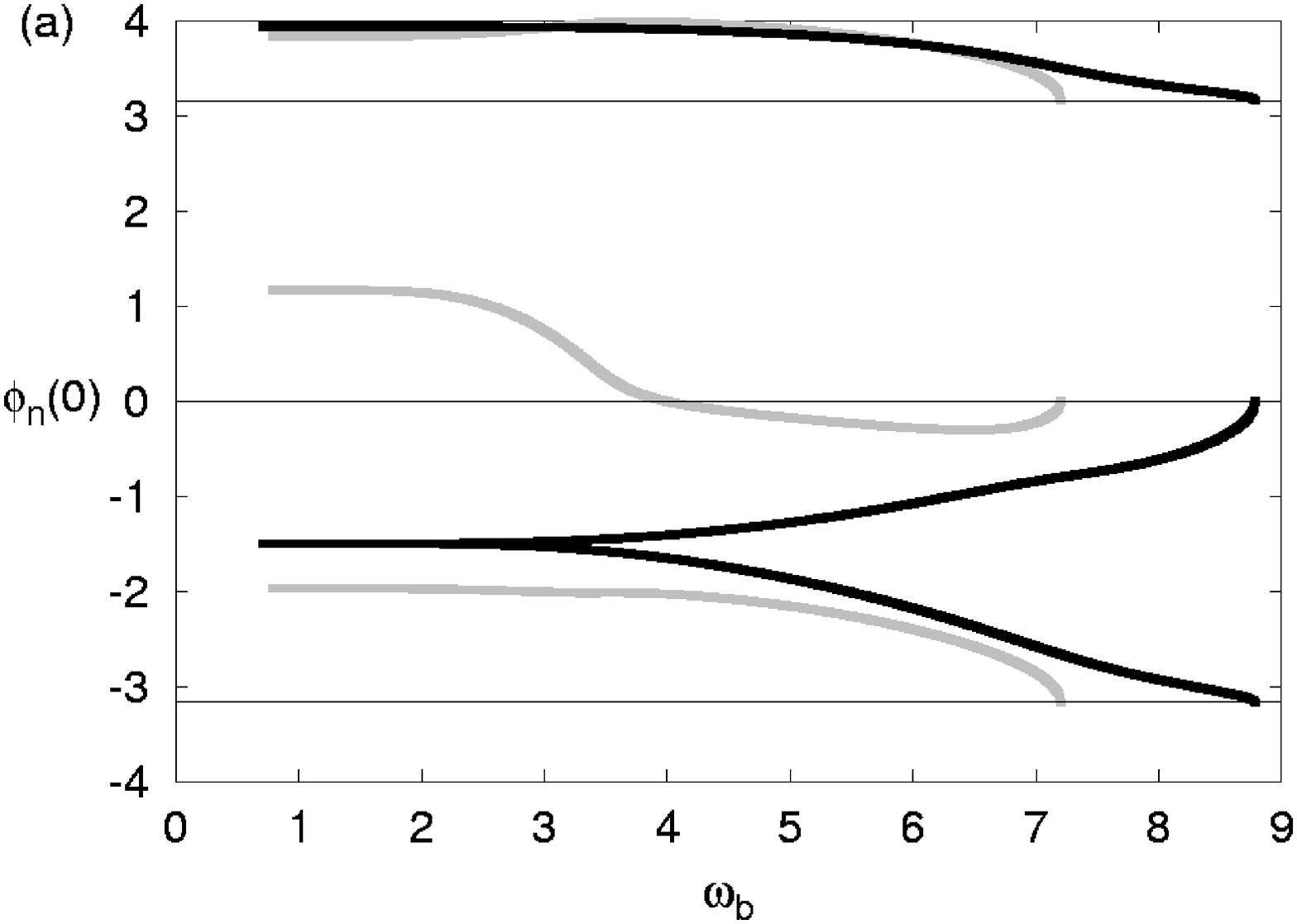} 
\end{minipage}%
\begin{minipage}[r]{0.49\textwidth} 
\includegraphics[height=\textwidth,angle=270]{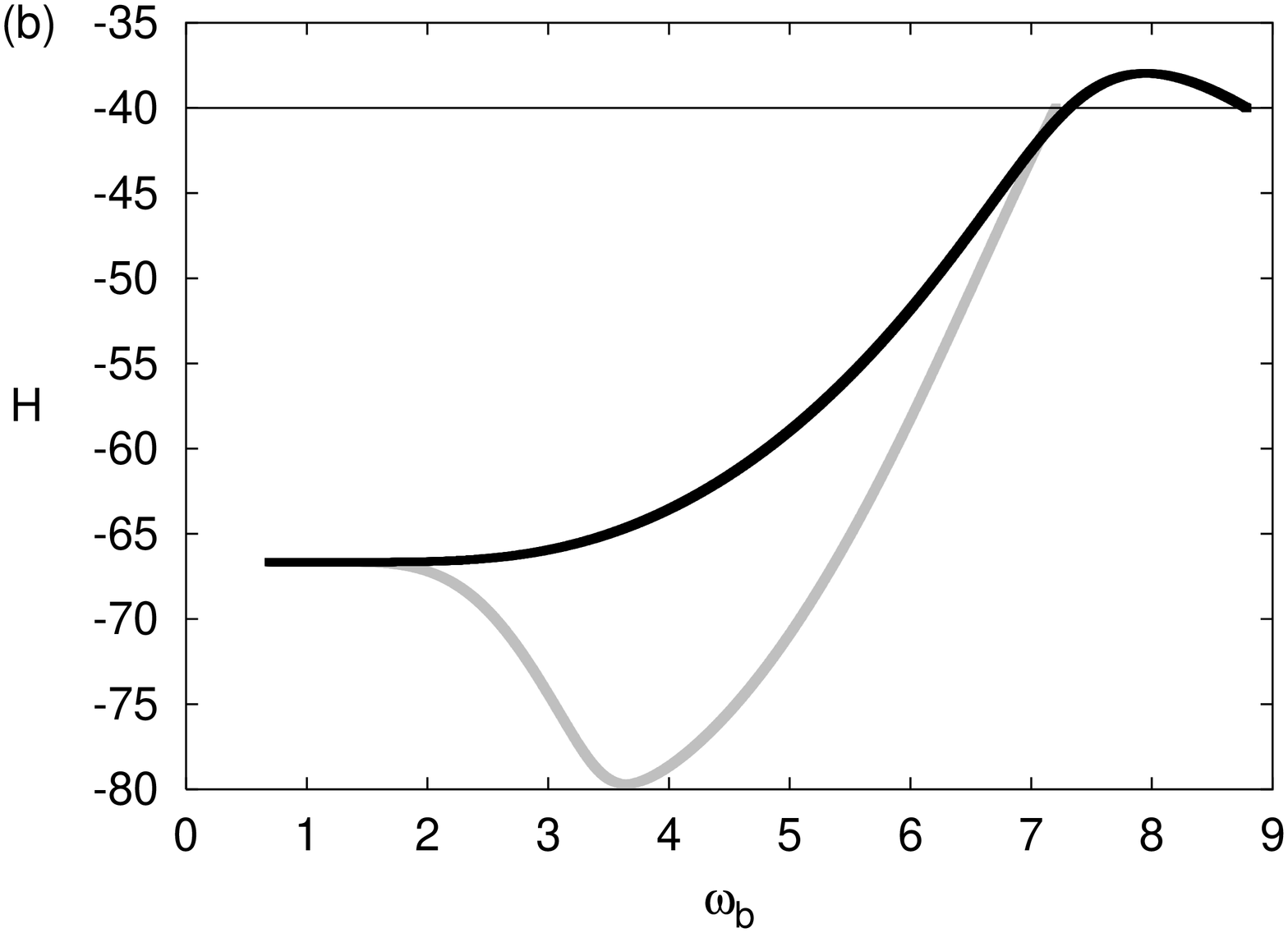} 
\end{minipage}\\
\begin{minipage}[l]{0.49\textwidth} 
\includegraphics[height=\textwidth,angle=270]{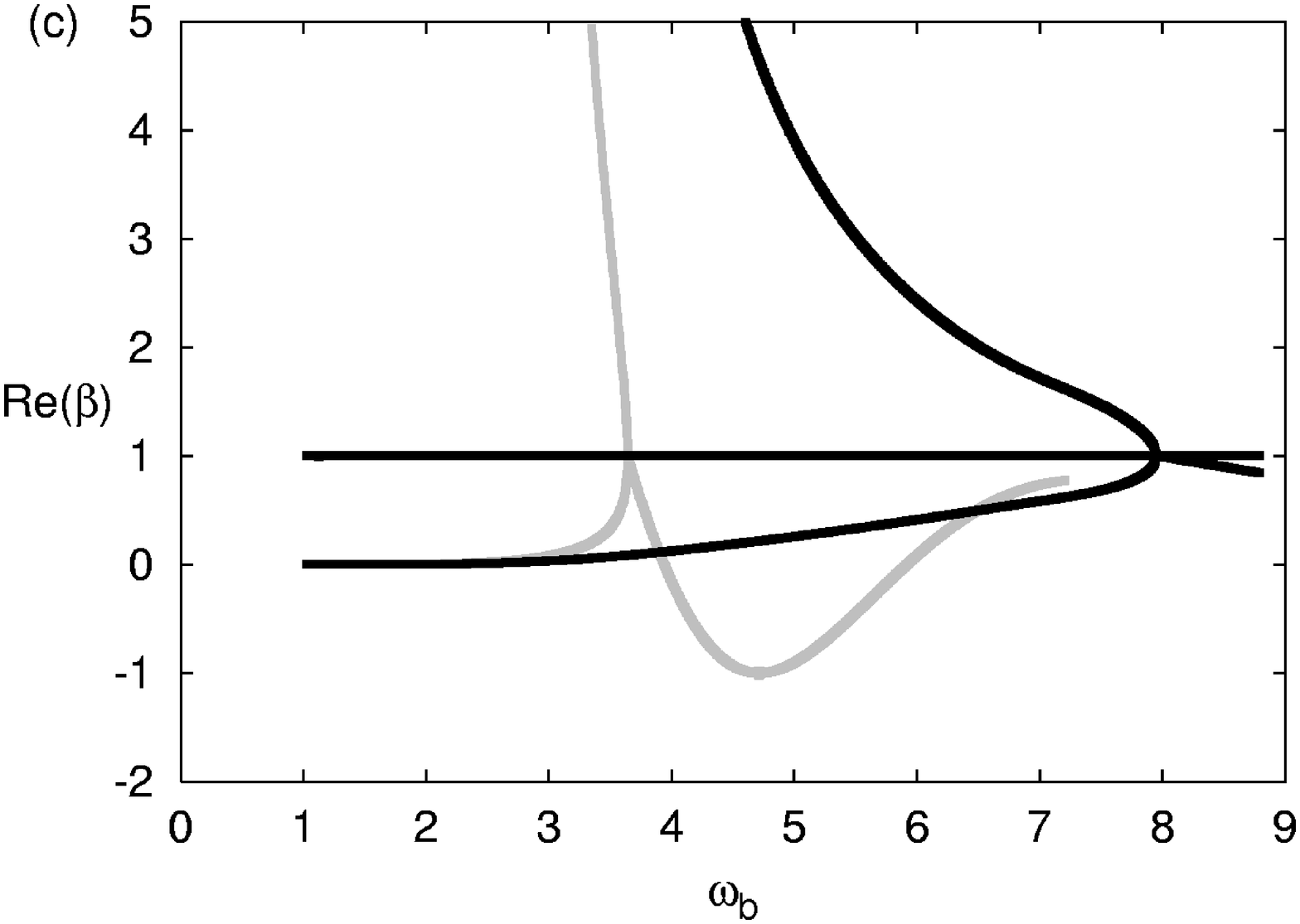}
\end{minipage}
\begin{minipage}[r]{0.49\textwidth} 
\includegraphics[height=\textwidth,angle=270]{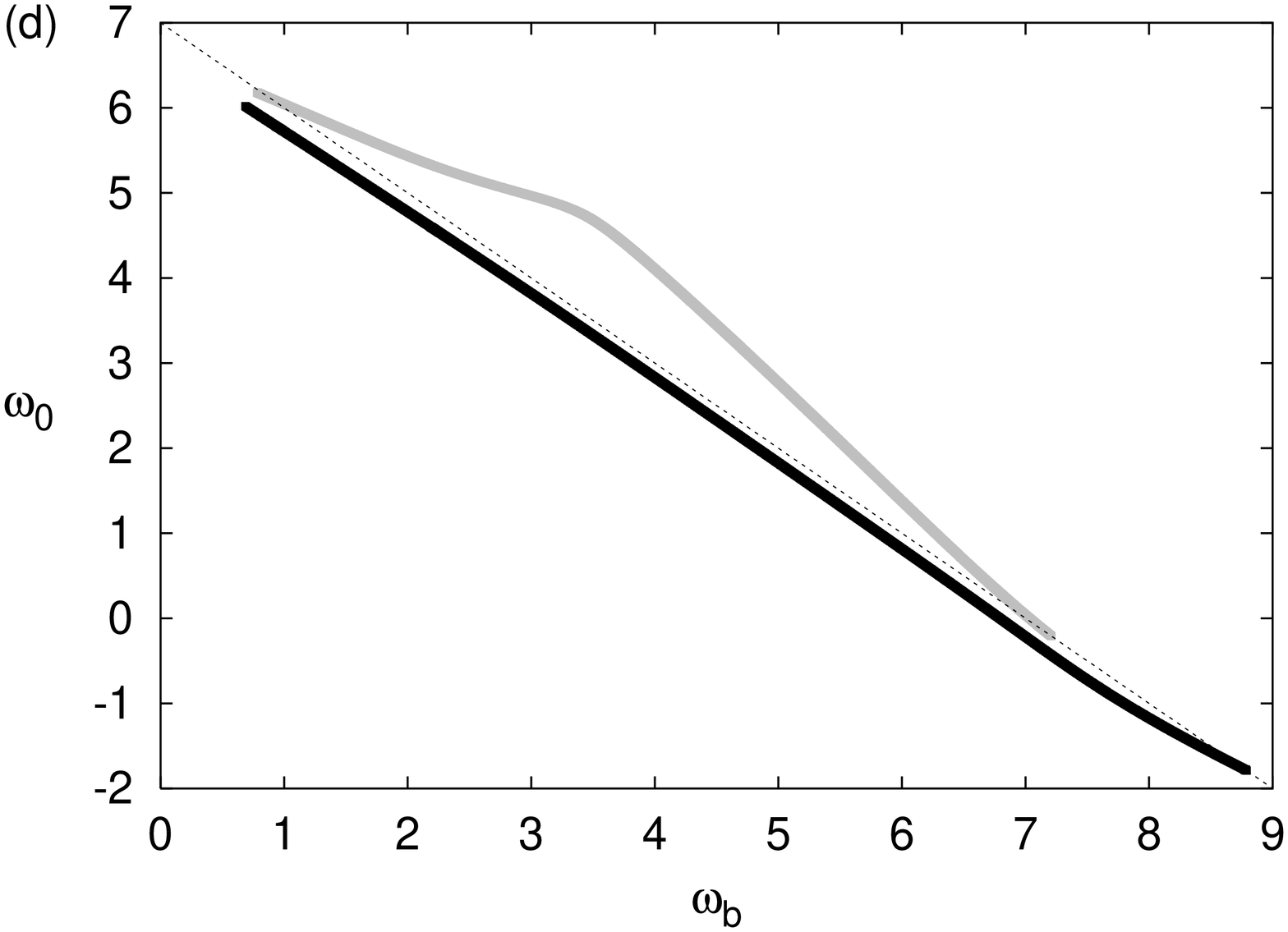}
\end{minipage}
\caption{
Same as figure \ref{fig_N=5} but for the two-frequency solutions 
bifurcating 
from the stationary solution with $\Lambda=7$ at  ${\mathcal N}=20$; 
grey [black] curves 
corresponding to solutions bifurcating at $\omega_{\rm b} = \omega_{\rm l -}$
[$\omega_{\rm b} = \omega_{\rm l +}$]. 
The eigenvalues not shown in (c) for small $\omega_{\rm b}$
increase monotonously for both solutions 
($\beta\sim10^{8}$ at $\omega_{\rm b}=0.8$).
}
\label{fig_N=20} 
\end{figure}
For both solutions the 
continuation from the linear modes is monotonous in the direction of 
decreasing  $\omega_{\rm b}$, and they are
linearly stable in some regime 
 $\omega_{\rm b} < \omega_{\rm l \pm}$ but become strongly 
unstable for smaller  $\omega_{\rm b}$  
(figure \ref{fig_N=20} (c)). The stability regime for the black solution 
is rather small, and it is 
unstable in the whole range where 
also the grey solution exists. Also here (figure  \ref{fig_N=20} (b))
${\rm d}H/{\rm d}\omega_{\rm b}=0$ at the instability thresholds 
occurring through collisions at $\beta=+1$ 
(there is also a tiny instability for the grey solution caused by 
collisions at $\beta=-1$ for $\omega_{\rm b} \approx 4.70-4.73$). 
As $\omega_{\rm b} \rightarrow 0$ also these 
solutions approach  homoclinic connections 
of stable and unstable manifolds of the homogeneous stationary solution 
$\phi_1=\phi_2=\phi_3$, which however are distinct both from each other and 
from those  described in section \ref{sec:TypeI}. 
They both  
contain one stable and one unstable manifold in each half-period, but 
while for the black solution the dynamics is of one-site-localization type 
with two sites of small equal $|\phi_n|^2$ 
(figure \ref{fig_N=15dyn} (c) below), the dynamics of the grey 
solution have all three sites of unequal amplitude as seen in figure 
\ref{fig_N=20} (a) (see figure \ref{fig_N=15dyn} (d)).

\subsubsection{Around the bifurcation point.}
\label{sec:IIbif}
Approaching from above the bifurcation point 
${\mathcal N}/C = 18$ the frequencies 
$\omega_{\rm l +}$ and $\omega_{\rm l -}$ approach each other, as 
do the corresponding families of two-frequency solutions
which at ${\mathcal N}/C \approx 18.7$ coincide at 
$\omega_{\rm b}\approx 6.8$ ($<\omega_{\rm l -} \approx 7.1$). At this 
point they bifurcate with each other, and compared to the 
scenario in figure 
\ref{fig_N=20} (c) an additional simultaneous collision at 
$\beta=+1$ appears for both solutions, in the regime where 
for larger ${\mathcal N}/C$ the black solution was unstable and the grey 
stable. For smaller  ${\mathcal N}/C$ the solutions split into a 
low-frequency 
and a high-frequency part separated by a forbidden gap 
in $\omega_{\rm b}$ as illustrated by 
figure \ref{fig_N=18.5}. 
\begin{figure}  
\begin{minipage}[l]{0.49\textwidth} 
\includegraphics[height=\textwidth,angle=270]{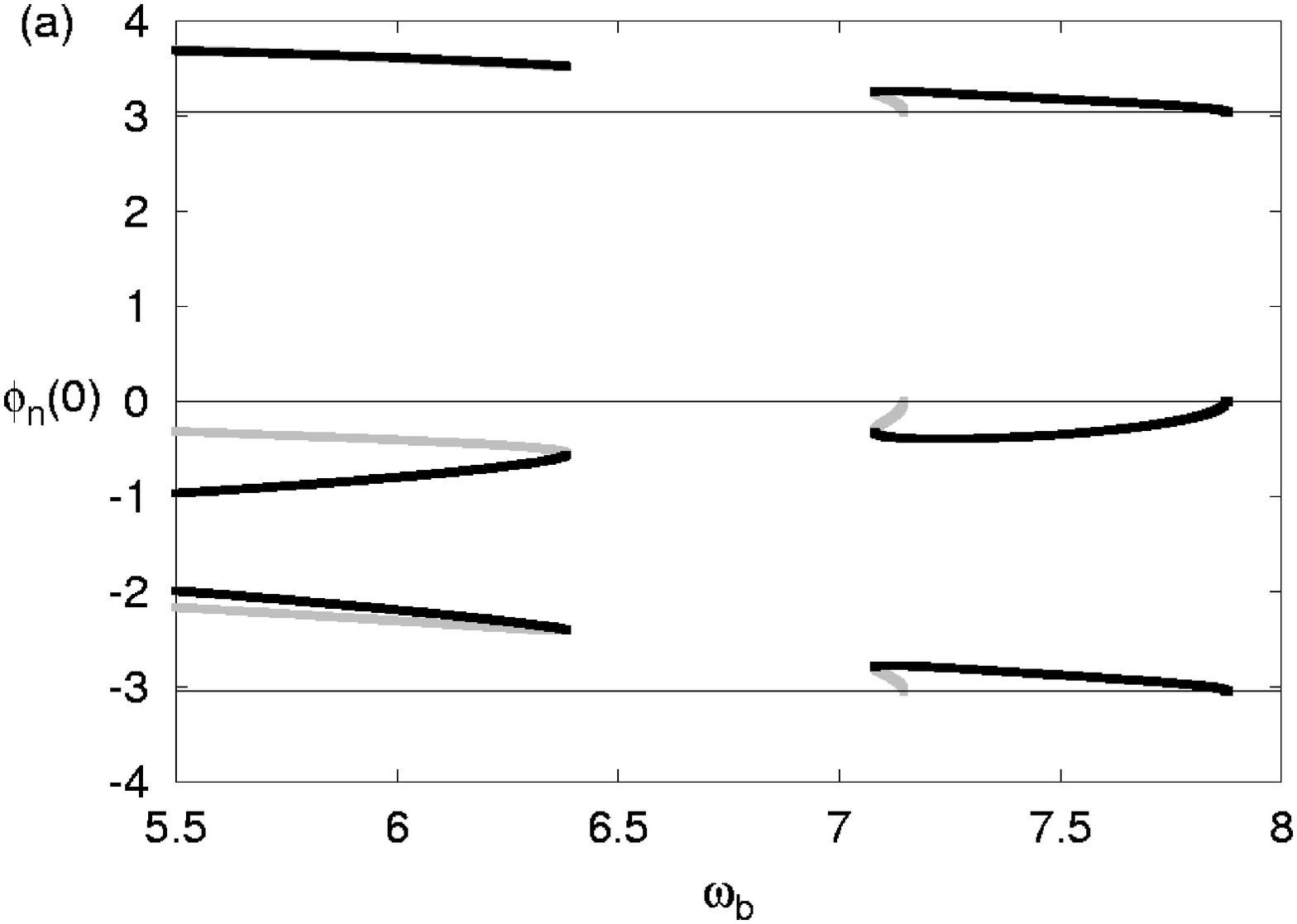} 
\end{minipage}%
\begin{minipage}[r]{0.49\textwidth} 
\includegraphics[height=\textwidth,angle=270]{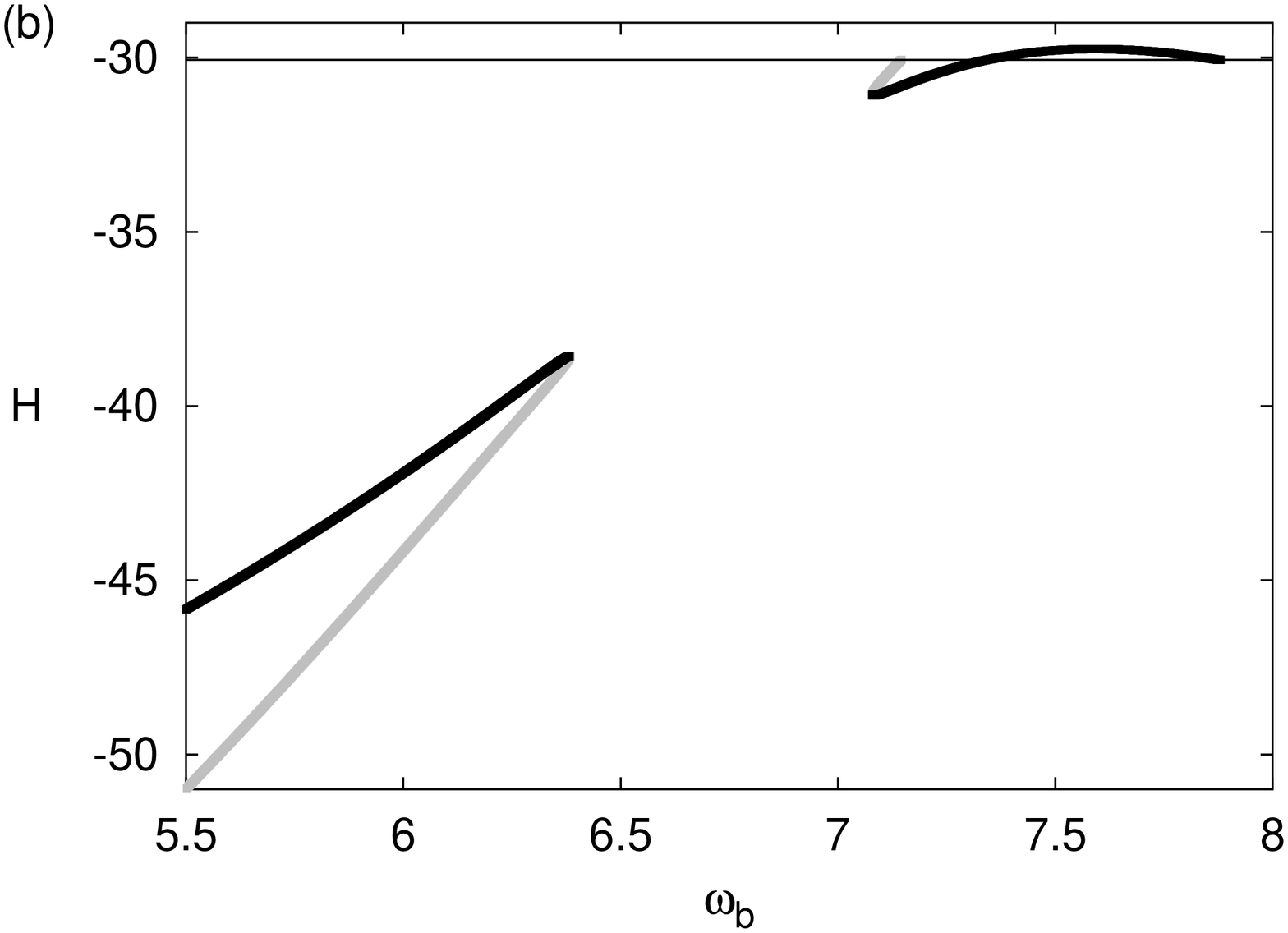} 
\end{minipage}\\
\begin{minipage}[l]{0.49\textwidth} 
\includegraphics[height=\textwidth,angle=270]{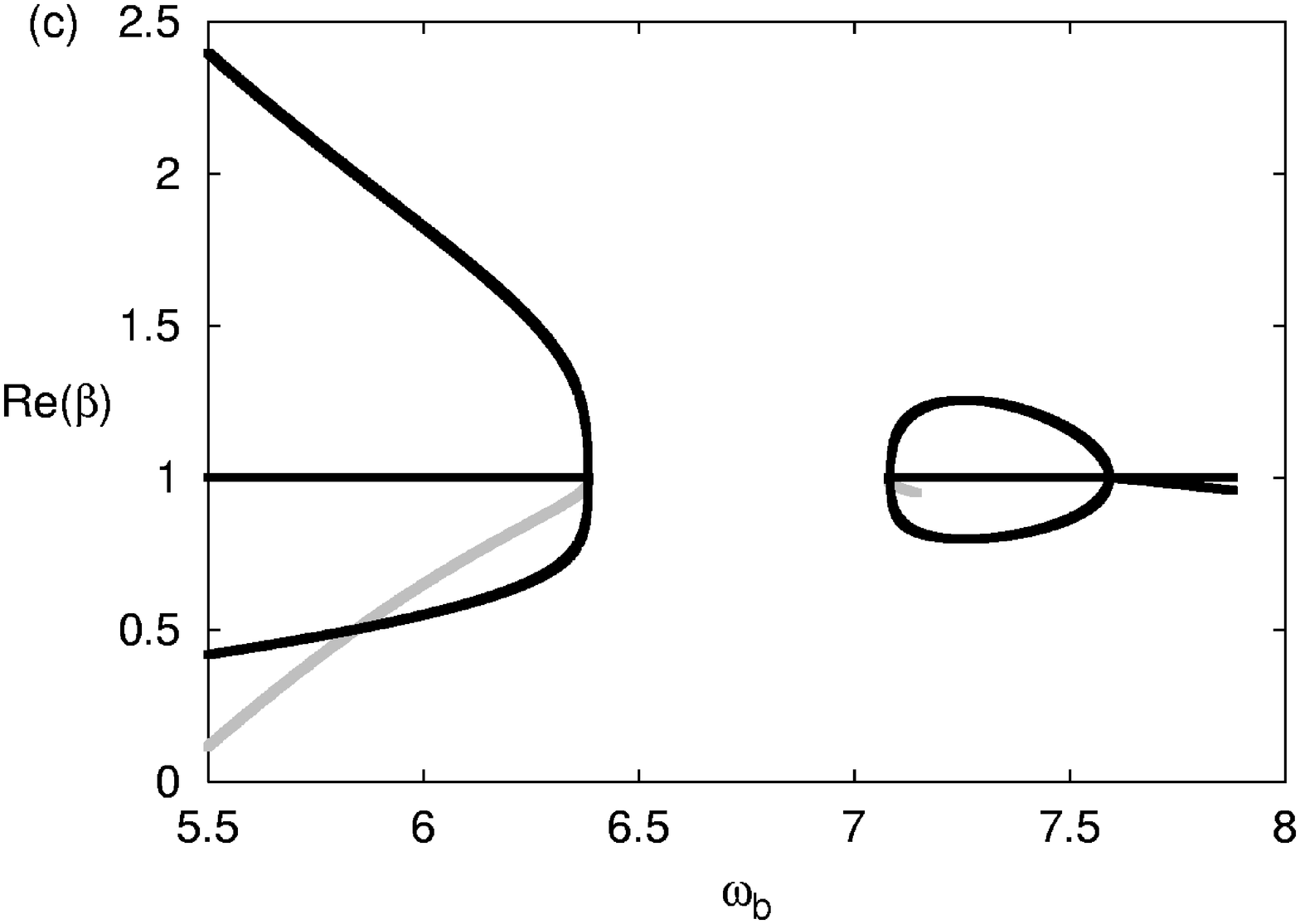}
\end{minipage}
\begin{minipage}[r]{0.49\textwidth} 
\includegraphics[height=\textwidth,angle=270]{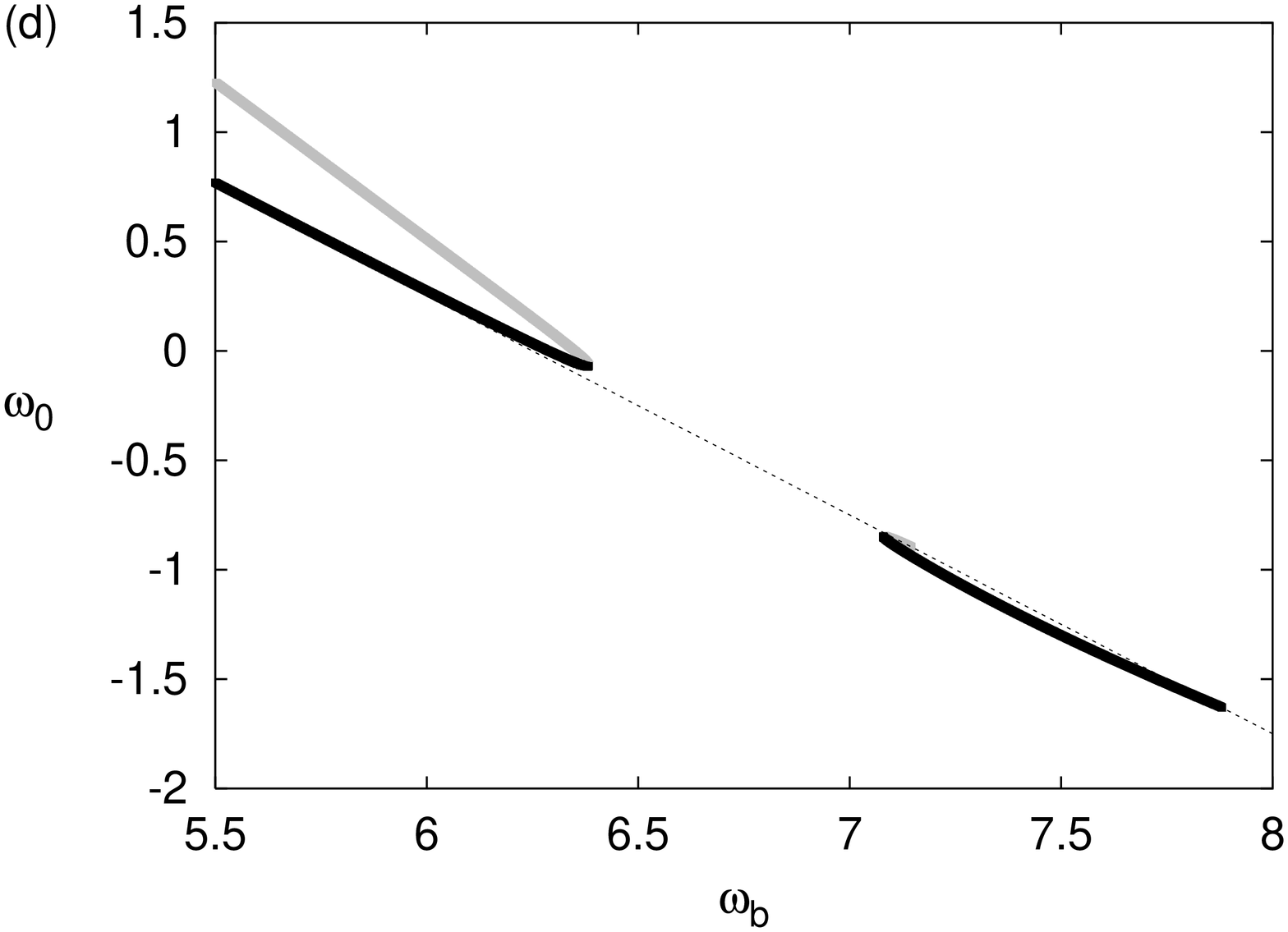}
\end{minipage}
\caption{
Continuation of the solutions in figure \ref{fig_N=20} for 
${\mathcal N}=18.5$. The behaviour for $\omega_{\rm b}< 5.5$ is 
qualitatively the same as in figure \ref{fig_N=20}.
}
\label{fig_N=18.5} 
\end{figure}
Thus, only the 
high-frequency part is now connected to the stationary solution. 

Decreasing further ${\mathcal N}/C$ the
high-frequency part shrinks as $\omega_{\rm l +}$ and $\omega_{\rm l -}$ 
approach each other, and furthermore at  ${\mathcal N}/C \approx 18.2$ 
(figure \ref{fig_IIbif} (a), (c), (e))
the continuation of the grey solution from 
$\omega_{\rm b}=\omega_{\rm l -}$ 
changes direction to occur towards larger $\omega_{\rm b}$.
\begin{figure}  
\begin{minipage}[l]{0.5\textwidth} 
\includegraphics[height=0.99\textwidth,angle=270]{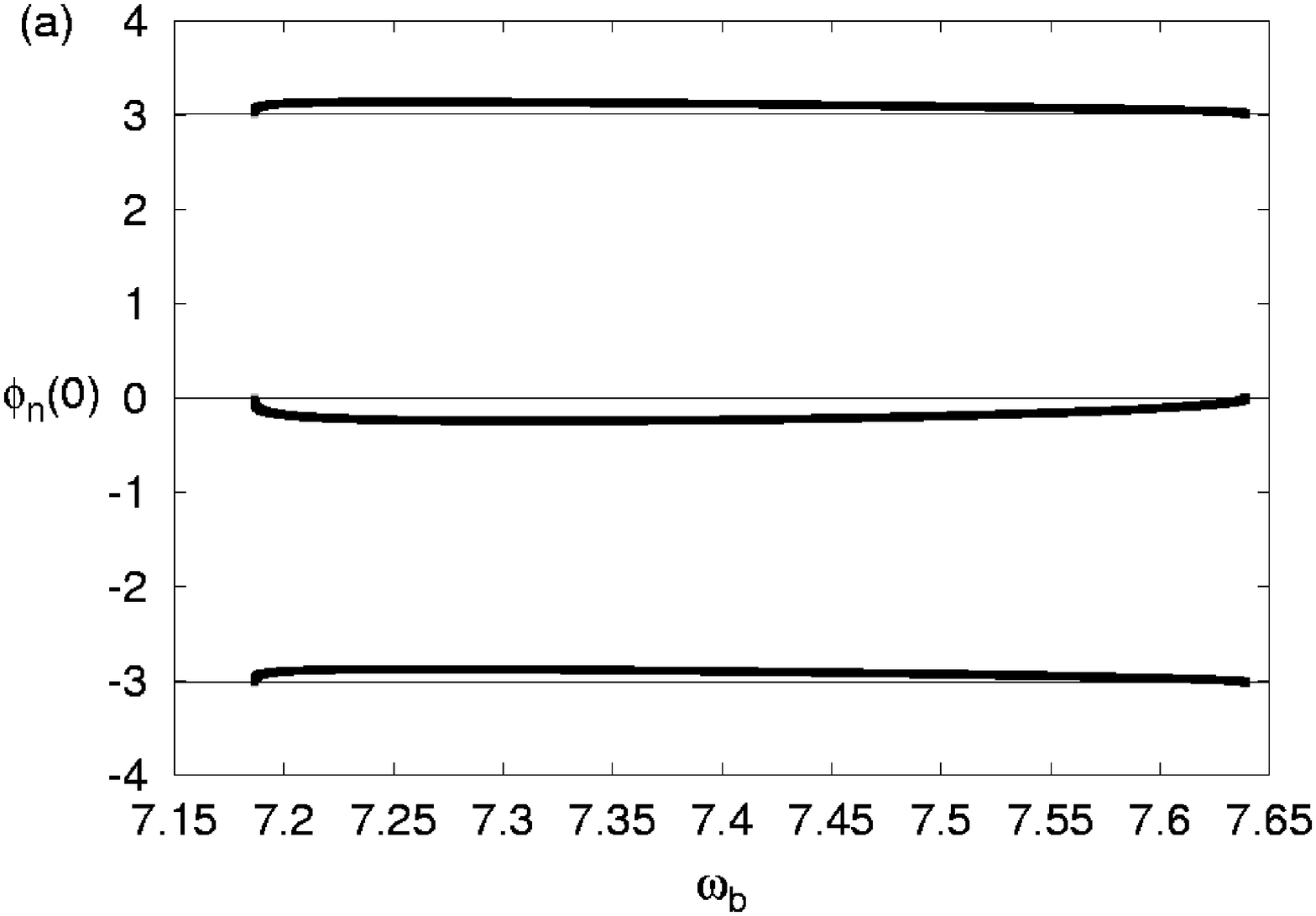} 
\end{minipage}%
\begin{minipage}[r]{0.5\textwidth} 
\includegraphics[height=0.99\textwidth,angle=270]{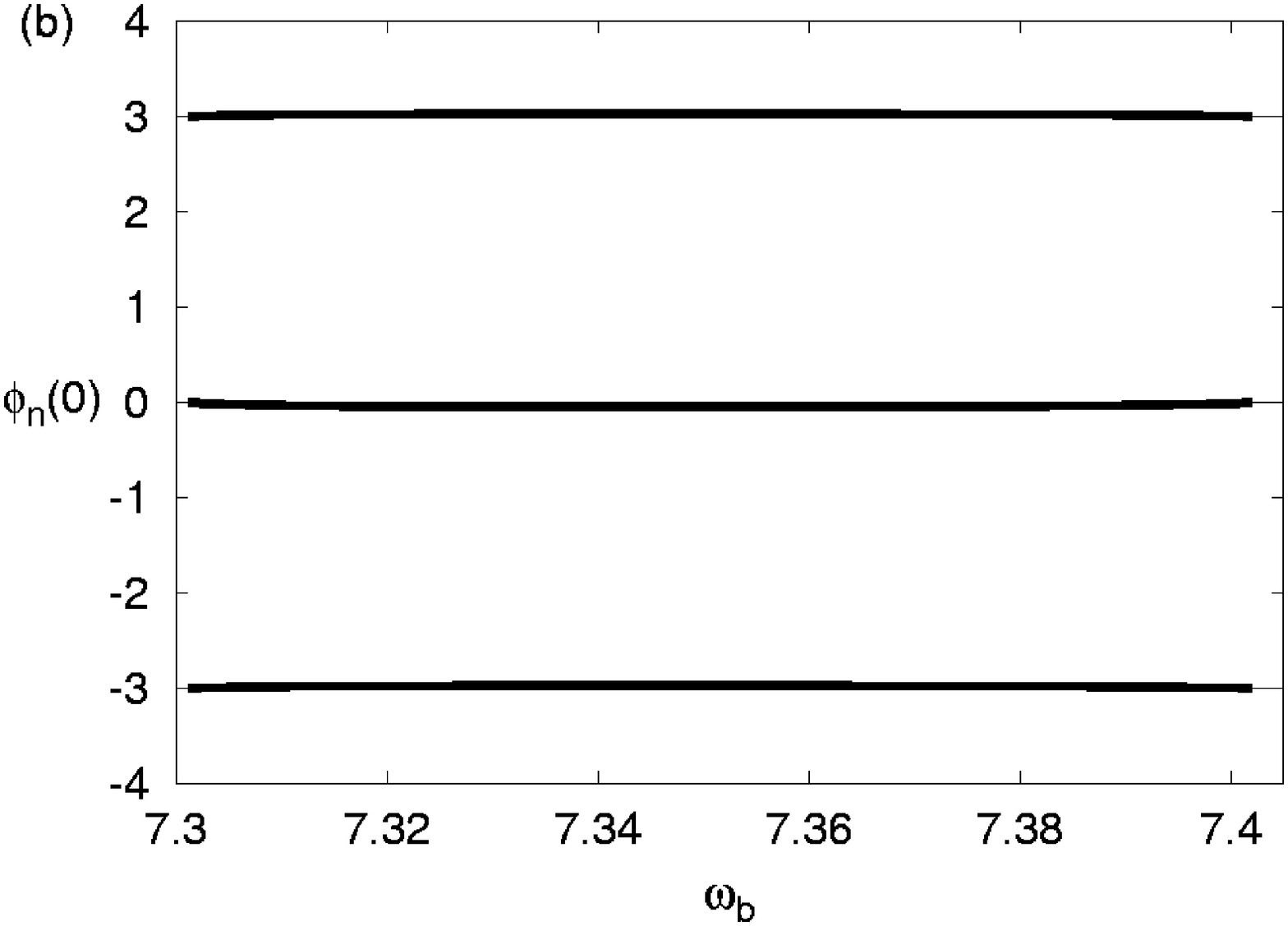} 
\end{minipage}\\
\begin{minipage}[l]{0.5\textwidth} 
\includegraphics[height=0.99\textwidth,angle=270]{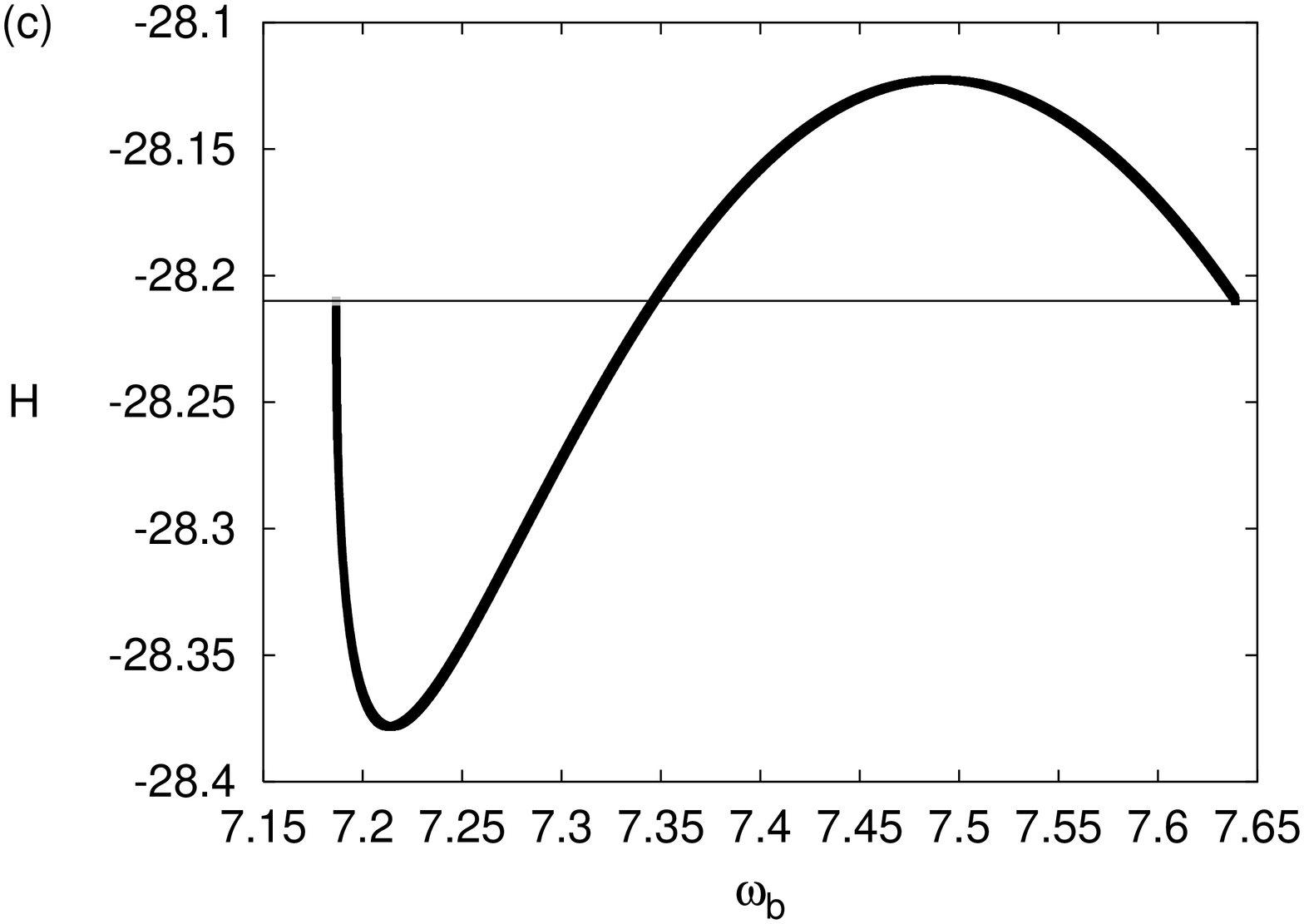} 
\end{minipage}%
\begin{minipage}[r]{0.5\textwidth} 
\includegraphics[height=0.99\textwidth,angle=270]{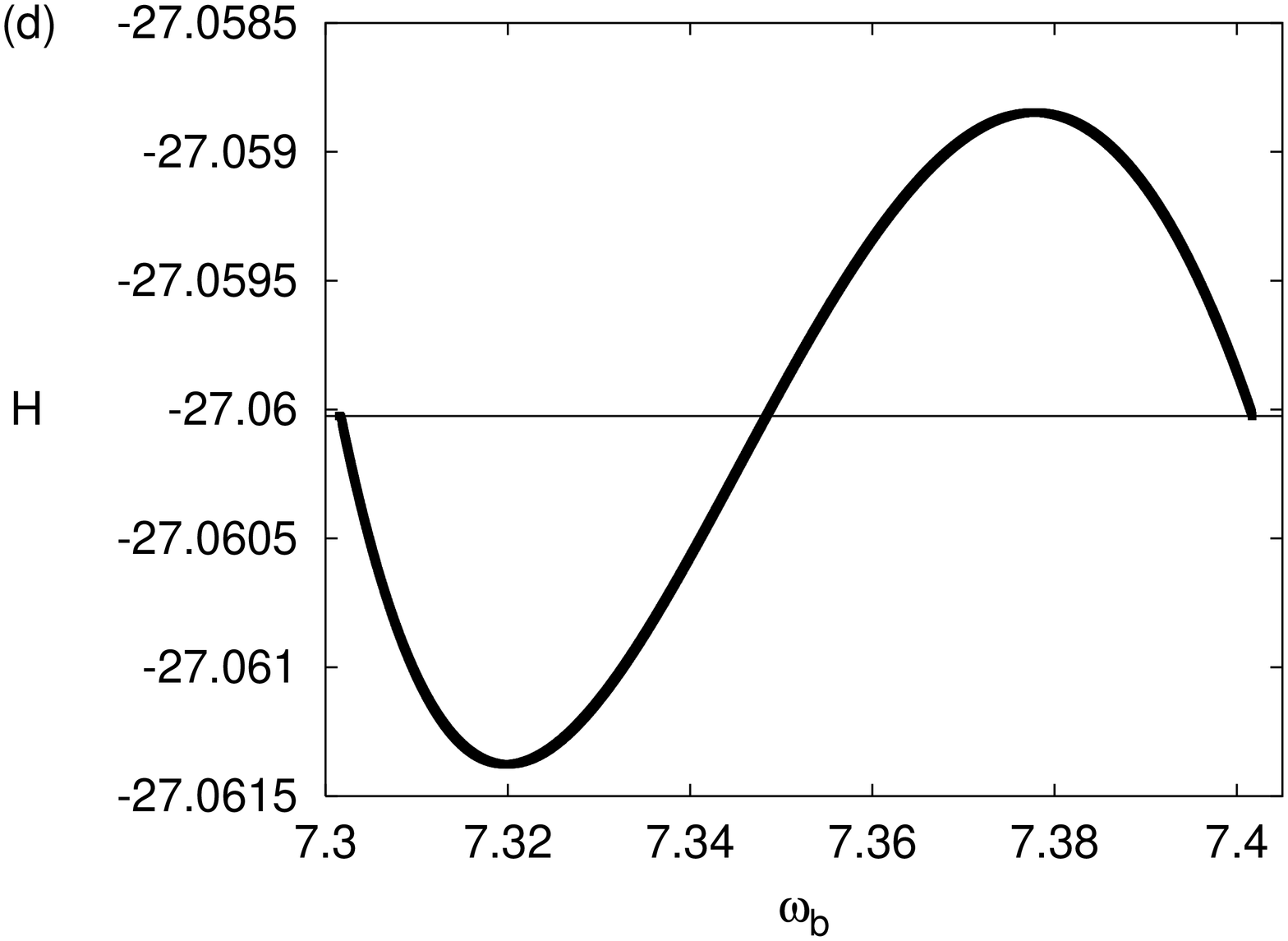} 
\end{minipage}\\
\begin{minipage}[l]{0.5\textwidth} 
\includegraphics[height=0.99\textwidth,angle=270]{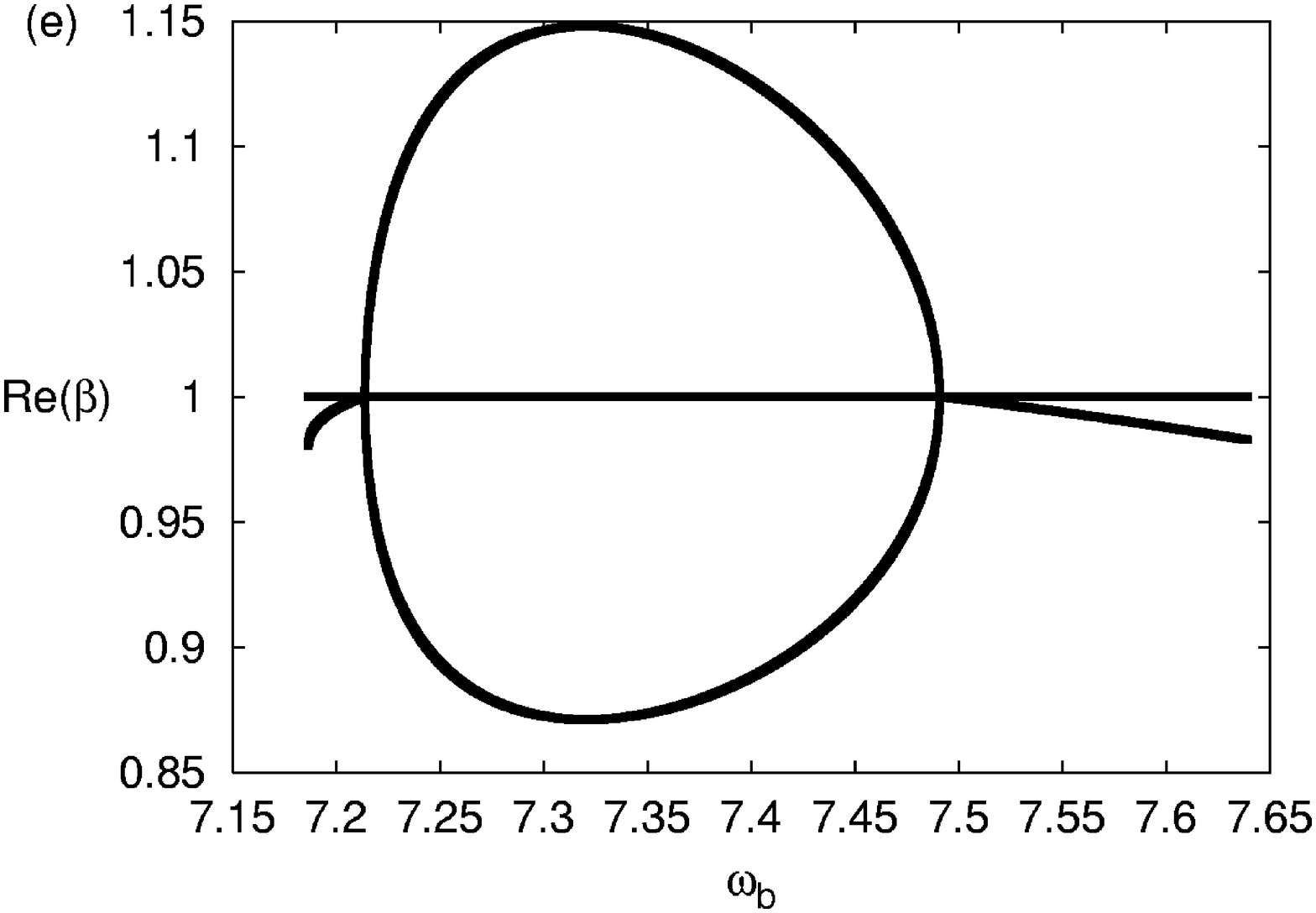} 
\end{minipage}%
\begin{minipage}[r]{0.5\textwidth} 
\includegraphics[height=0.99\textwidth,angle=270]{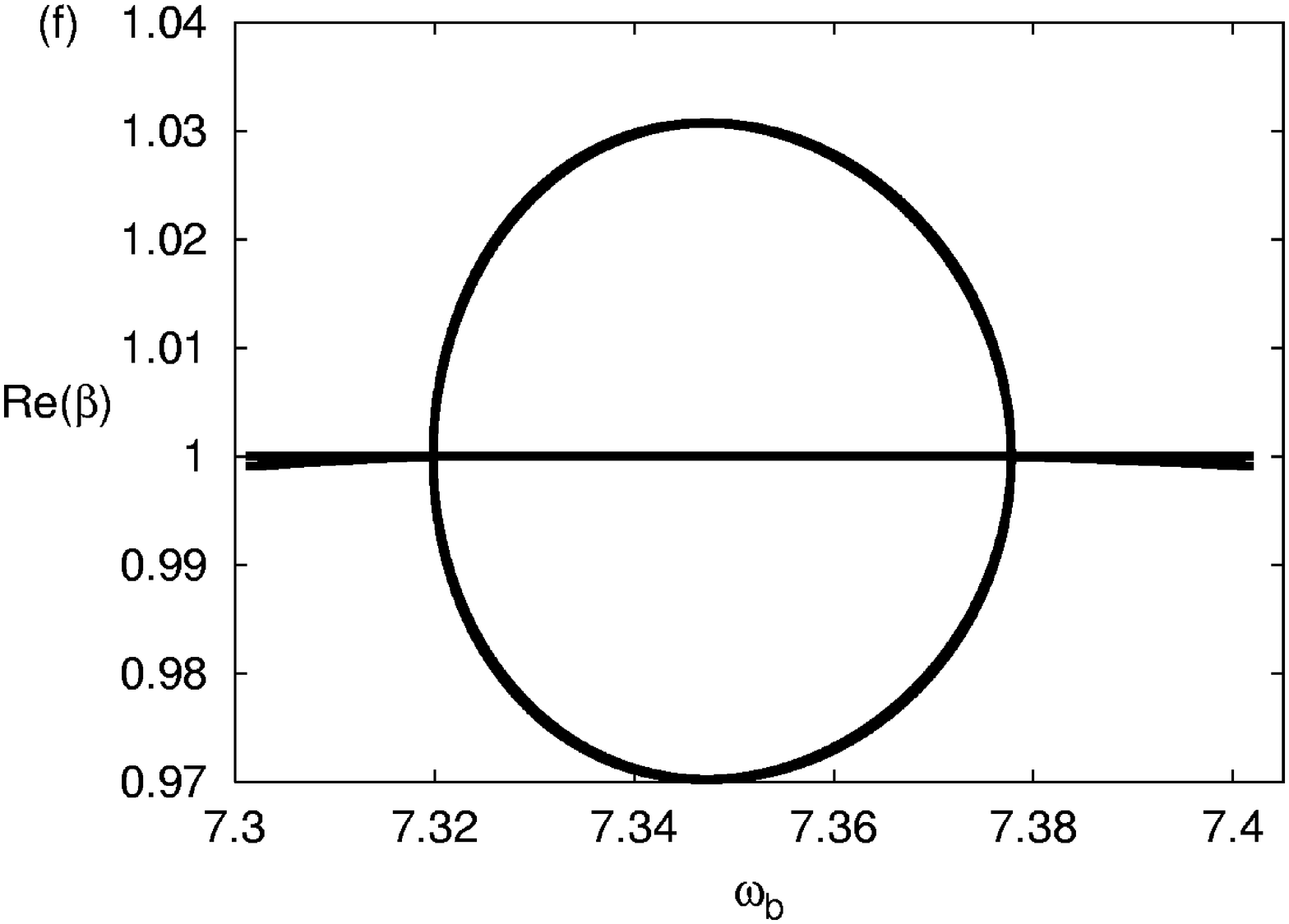} 
\end{minipage}\\
\caption{
Continuation of the high-frequency branches of  figure \ref{fig_N=18.5} 
(a)-(c) for ${\mathcal N}=18.2$ 
(left figures) and 
${\mathcal N} =18.01$ (right figures). At ${\mathcal N}=18.2$ there is still 
a tiny remainder of the grey part of previous figures which continues 
approximately $10^{-6}$ in the direction of negative $\omega_{\rm b}$ 
from the linear value 
$\omega_{\rm b} =\omega_{\rm l -} \approx 7.186\,6898$.
}
\label{fig_IIbif} 
\end{figure} 
Thus, 
for $18< {\mathcal N}/C \lesssim 18.2$ the high-frequency branch 
consists of one single family which through 
a monotonous continuation versus $\omega_{\rm b}$ connects the 
linear eigenmode at $\omega_{\rm l -}$ with that at $\omega_{\rm l +}$ 
(figure \ref{fig_IIbif} (b), (d), (f)). The solutions
are linearly stable close to
$\omega_{\rm b}=\omega_{\rm l \pm}$ but unstable for intermediate 
frequencies (figure \ref{fig_IIbif} (f)) (note also the 'inverse N-shape' 
of $H(\omega_{\rm b})$ in figure \ref{fig_IIbif} (d) consistent 
with 
the negative Krein signature of $\omega_{\rm l -}$ and the positive of 
 $\omega_{\rm l +}$). At the bifurcation point  ${\mathcal N}/C =18$, 
$\omega_{\rm l -} = \omega_{\rm l +} $ and the loop of surrounding 
two-frequency solutions shrinks to zero. Consequently, 
for ${\mathcal N}/C < 18$ where the stationary solution is unstable, no 
nearby two-frequency solutions exist, and the scenario is that of a type 
II  HH bifurcation \cite{LR01} (cf.\ figure 3 in \cite{LR01}  
and figure 2(a) in \cite{Bridges90}, \cite{Bridges91}). However, the 
low-frequency branches in figure \ref{fig_N=18.5} exist also 
for ${\mathcal N}/C < 18$, but at a 
considerable distance from the stationary solution (e.g.\ at 
${\mathcal N} = 18$ and $C=1$ the Hamiltonian \eref{HLambda} 
for the stationary solution is $H=-27$, while its maximum value on the 
disconnected branch of two-frequency solutions is $H\approx -39.4$). 
We illustrate the consequences for the dynamics of the unstable 
stationary solution in section \ref{sec:dyn}. 

\subsubsection{Smaller ${\mathcal N}/C $.}
\label{sec:IIsmall}
Continuing the low-frequency branches of figure \ref{fig_N=18.5}
towards smaller  ${\mathcal N}/C $, their features remain qualitatively the 
same.  The example for ${\mathcal N}/C = 15$ in figure \ref{fig_N=15II} 
shows
that they are distinct from 
those originating from the low-amplitude bifurcation  
(cf.\ figure \ref{fig_N=15}). 
\begin{figure}  
\begin{minipage}[l]{0.49\textwidth} 
\includegraphics[height=\textwidth,angle=270]{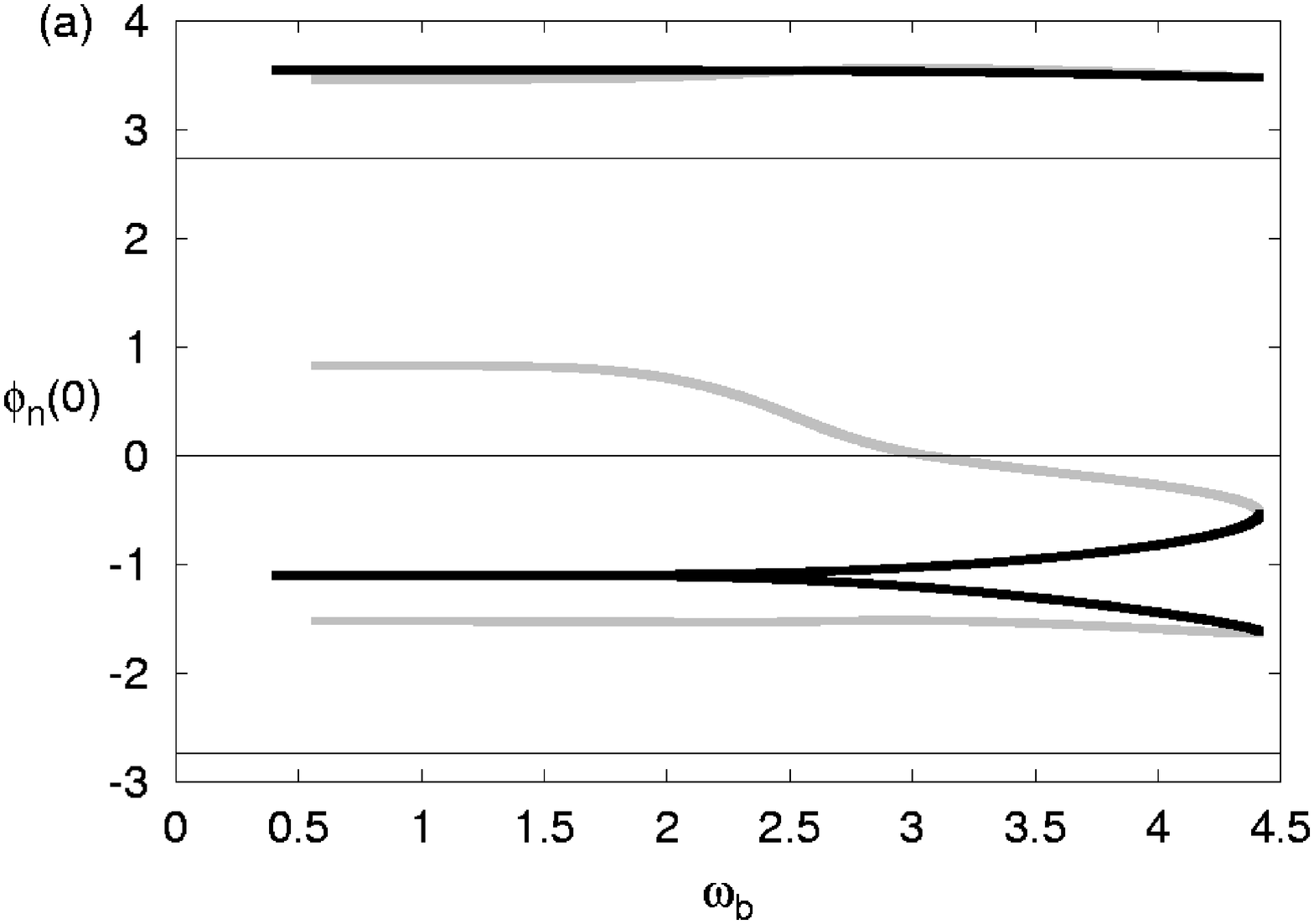} 
\end{minipage}%
\begin{minipage}[r]{0.49\textwidth} 
\includegraphics[height=\textwidth,angle=270]{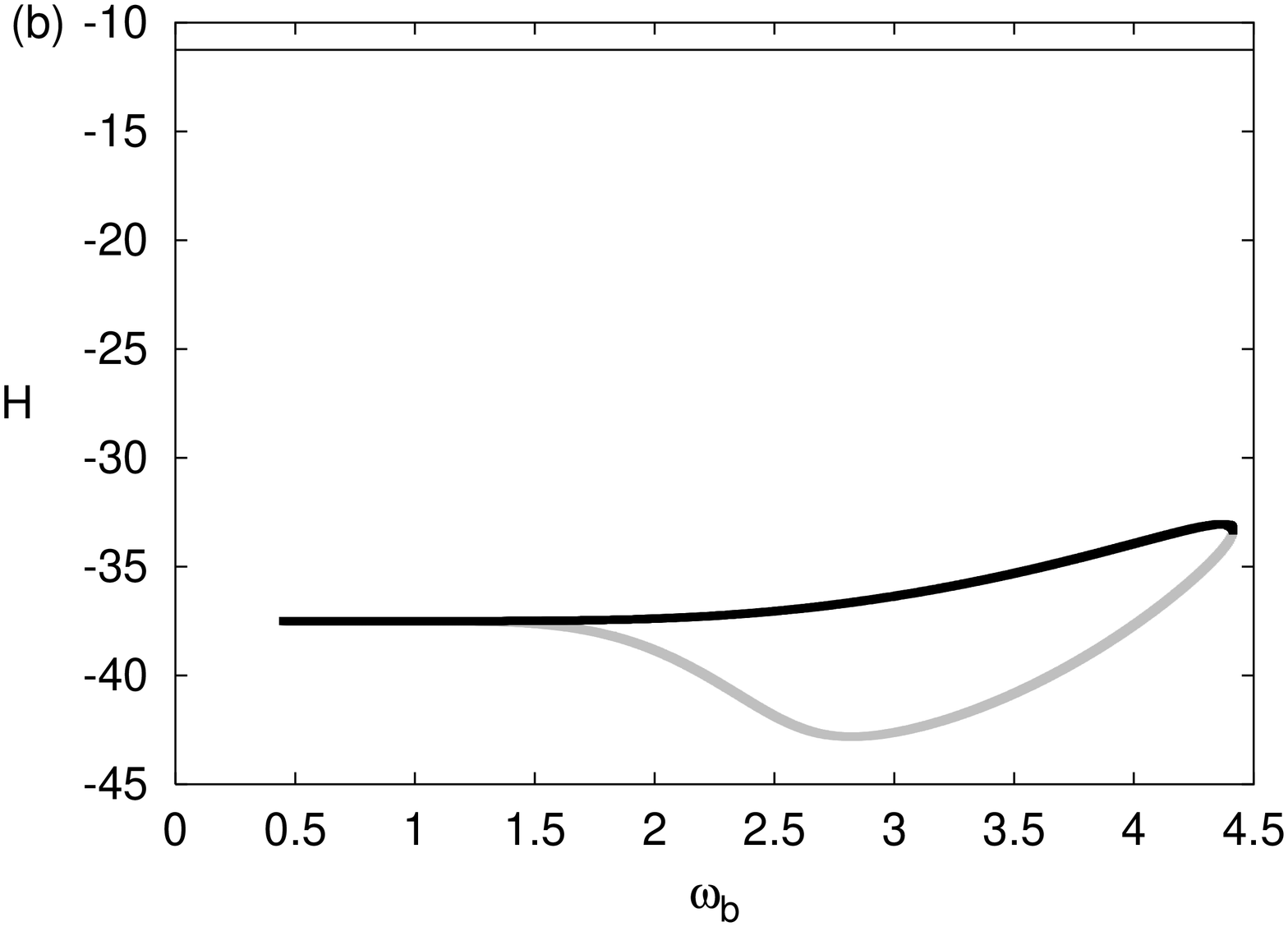} 
\end{minipage}\\
\begin{minipage}[l]{0.49\textwidth} 
\includegraphics[height=\textwidth,angle=270]{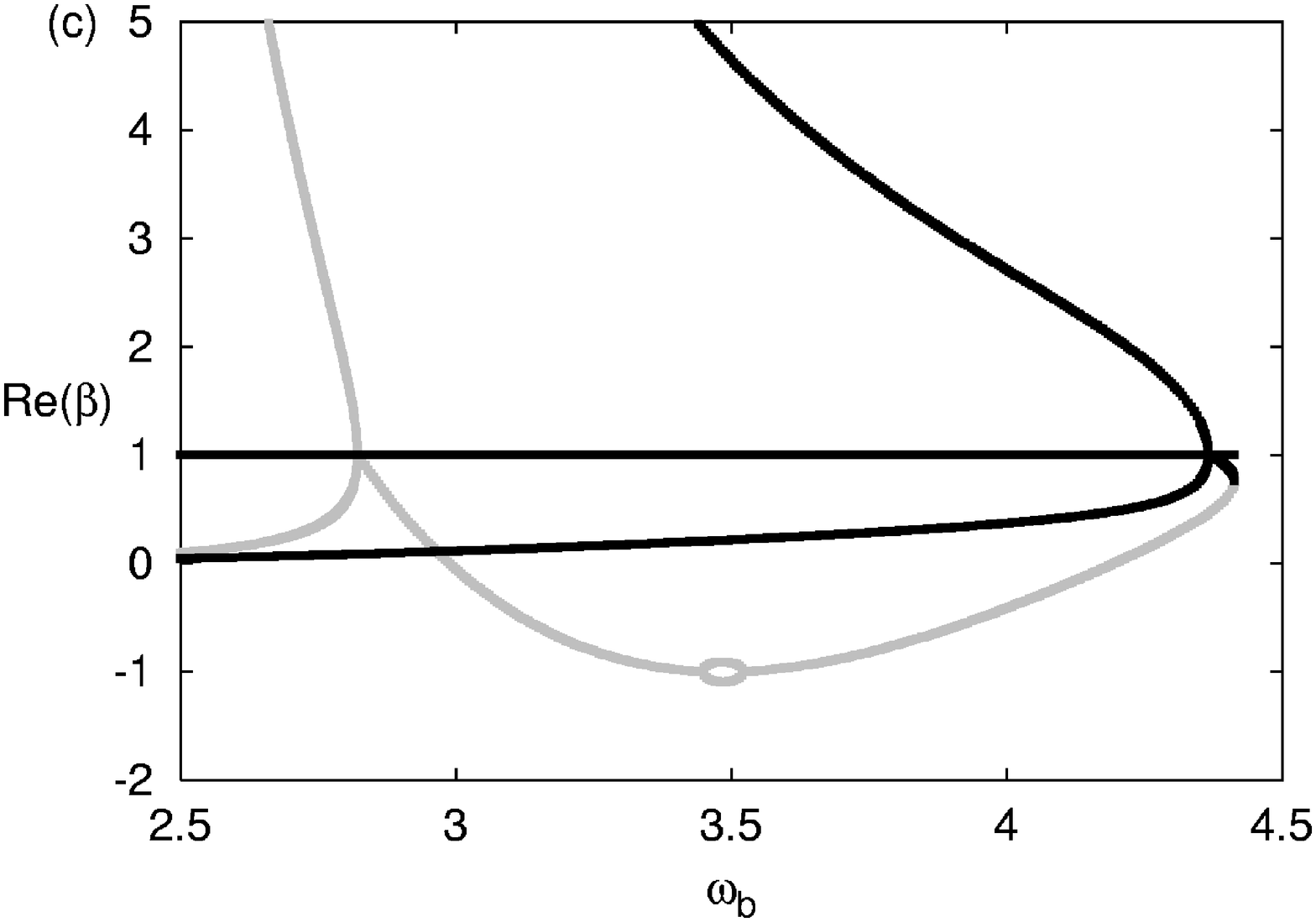}
\end{minipage}
\begin{minipage}[r]{0.49\textwidth} 
\includegraphics[height=\textwidth,angle=270]{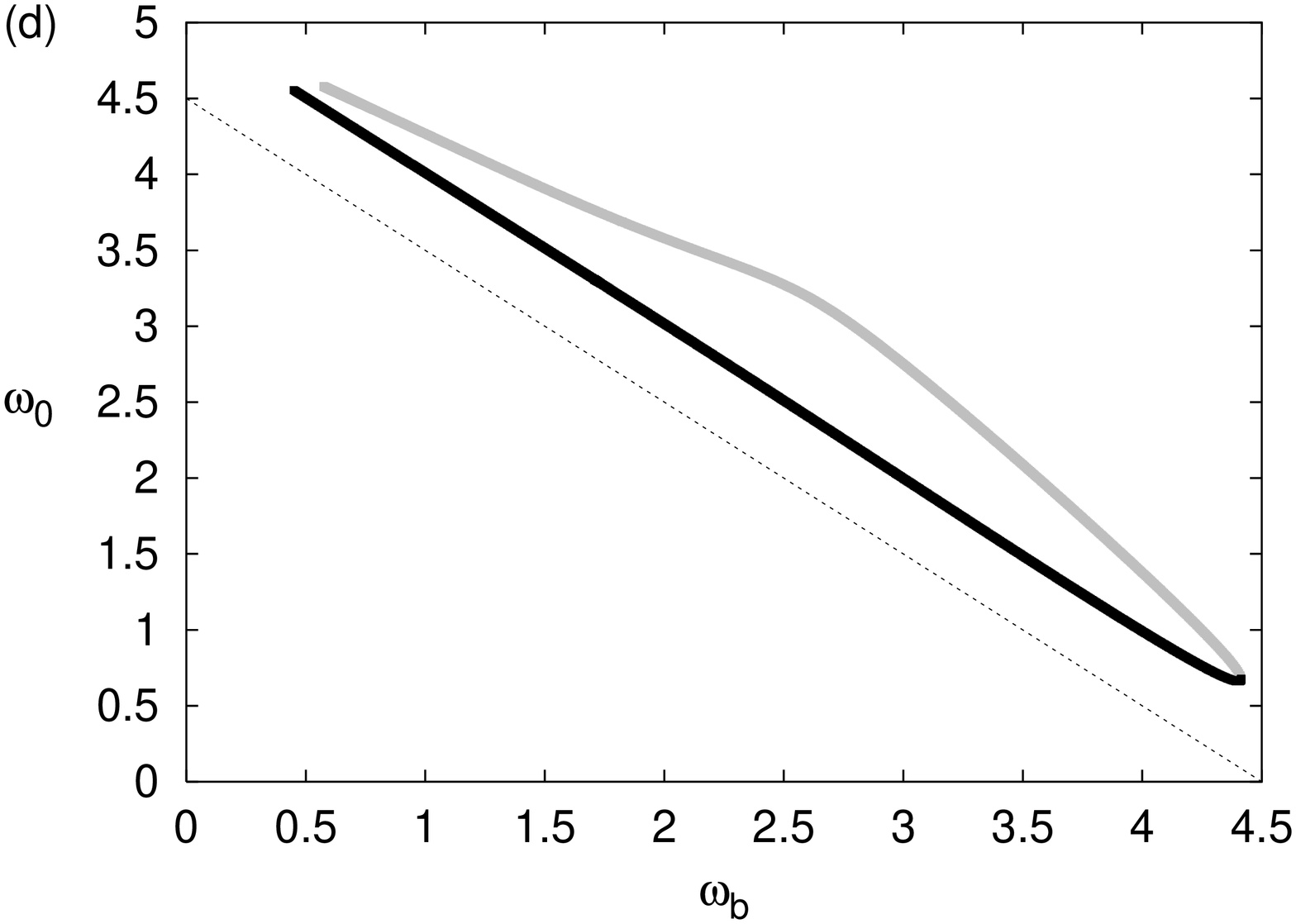}
\end{minipage}
\caption{
Continuation of the low-frequency branches of figure \ref{fig_N=18.5} 
for ${\mathcal N}=15$.
The eigenvalues not shown in (c) for small $\omega_{\rm b}$ 
increase monotonously for both solutions 
($\beta\sim 5\cdot 10^{6}$ at $\omega_{\rm b}=0.8$). As before, 
horizontal lines in (a), (b) and dashed line in (d) show 
corresponding quantities for the (unstable) stationary solution 
\eref{ALambda} ($C=1$).
}
\label{fig_N=15II} 
\end{figure}
Moreover, they remain 
at a large distance from the unstable stationary solution 
\eref{ALambda} also when  
${\mathcal N}/C $ decreases (see e.g.\ the large difference in $H$ in 
figure \ref{fig_N=15II} (b)), and their maximum value of $\omega_{\rm b}$ 
(i.e.\ where the black and grey branches meet) continues to 
decrease (e.g.\ at ${\mathcal N}=10$ and $C=1$ they meet at 
$\omega_{\rm b}\approx 2.33$). 

\begin{figure}  
\begin{minipage}[l]{0.49\textwidth} 
\includegraphics[height=\textwidth,angle=270]{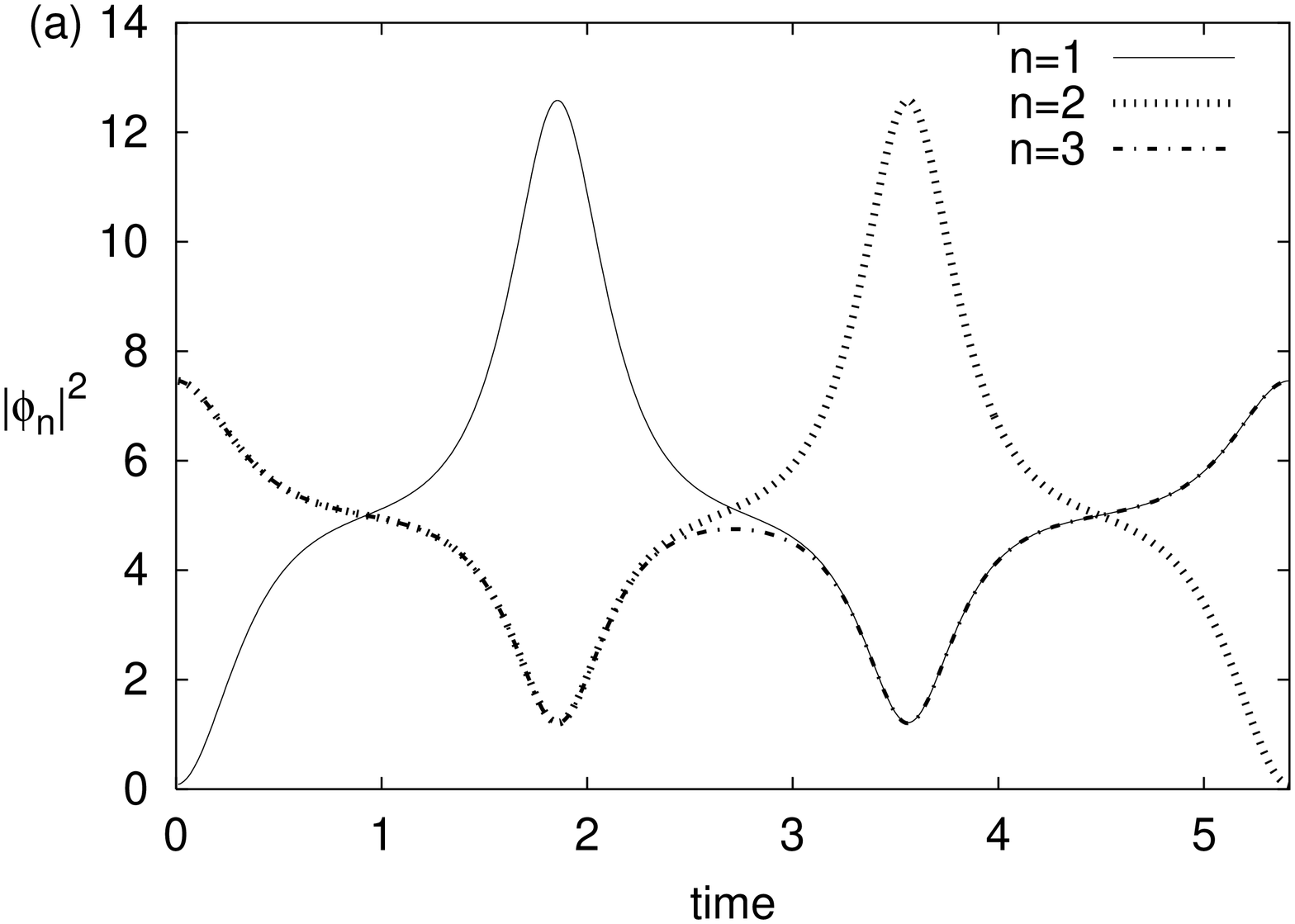} 
\end{minipage}%
\begin{minipage}[r]{0.49\textwidth} 
\includegraphics[height=\textwidth,angle=270]{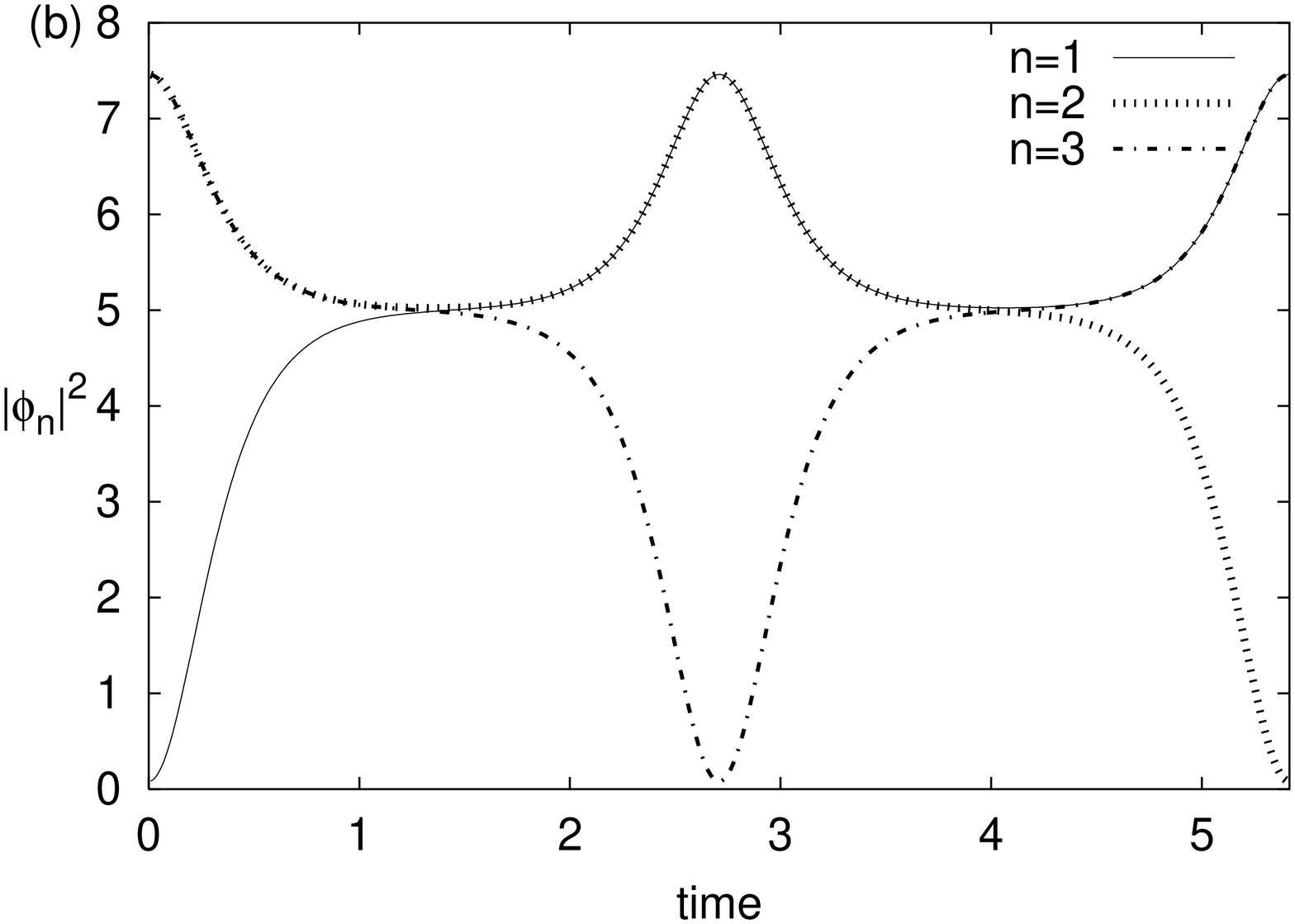} 
\end{minipage}\\
\begin{minipage}[l]{0.49\textwidth} 
\includegraphics[height=\textwidth,angle=270]{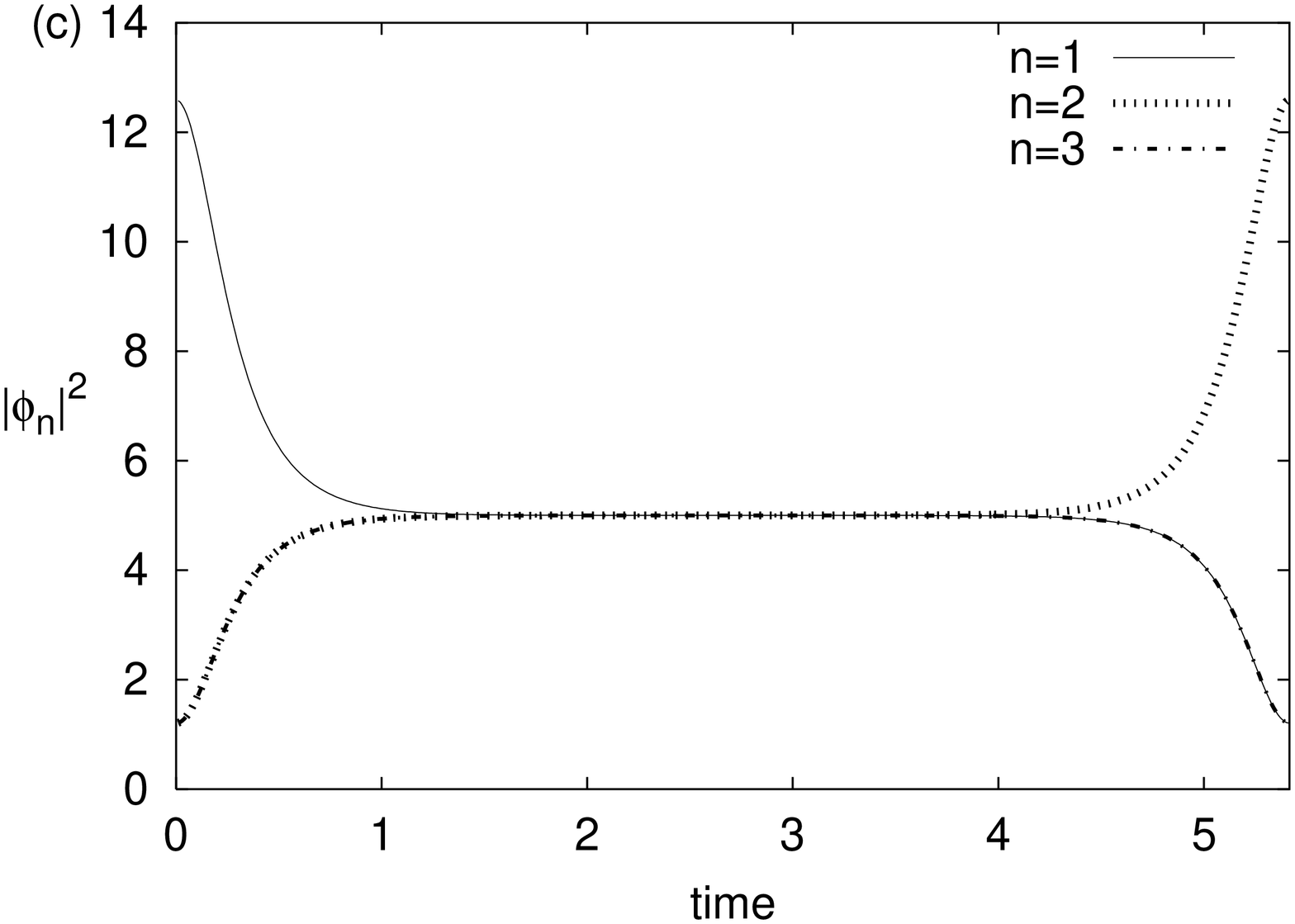}
\end{minipage}
\begin{minipage}[r]{0.49\textwidth} 
\includegraphics[height=\textwidth,angle=270]{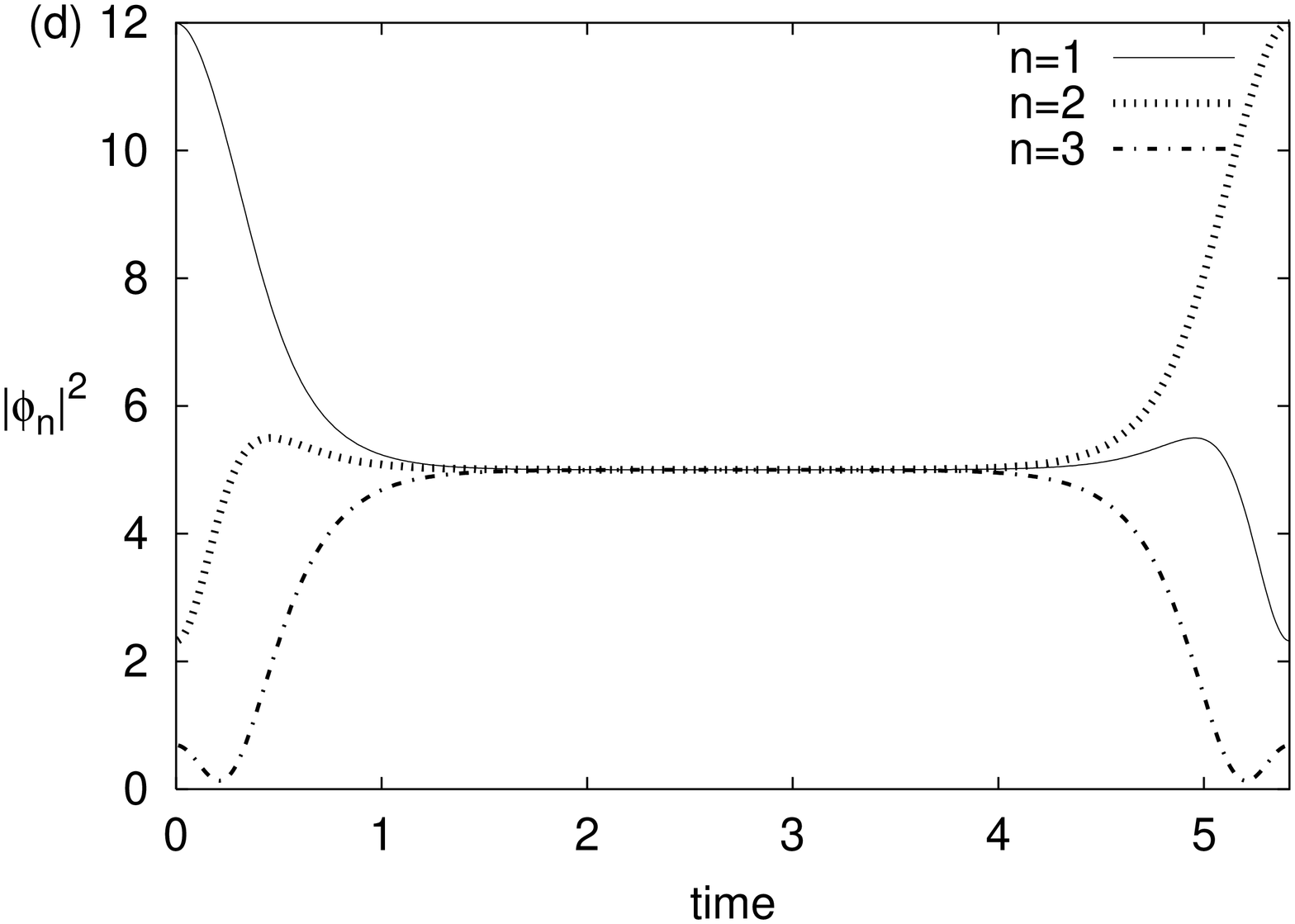}
\end{minipage}
\caption{
Dynamics for one half-period of the four different solutions in figures 
\ref{fig_N=15} and \ref{fig_N=15II} at  ${\mathcal N}=15$, 
$\omega_{\rm b}=0.58$ and $C=1$. (a) and (c) [(b) and (d)] correspond 
to the black [grey] solutions in figures \ref{fig_N=15} and 
\ref{fig_N=15II}, respectively.
}
\label{fig_N=15dyn} 
\end{figure}

To illustrate the different nature of the solutions in 
figures \ref{fig_N=15} and \ref{fig_N=15II} for 
small $\omega_{\rm b}$, we plot in figure \ref{fig_N=15dyn} the 
resulting dynamics for one half-period $T_{\rm b}/2$ (remembering that
$\phi_1(t+T_{\rm b}/2)=\phi_2(t), \phi_3(t+T_{\rm b}/2)=\phi_3(t)$ yields
the complete dynamics) at $\omega_{\rm b}=0.58$.
As $\omega_{\rm b}$ decreases the flat parts of the curves will flatten 
and stretch out in time, so that as $\omega_{\rm b} \rightarrow 0$ the 
previously described asymptotic solutions are approached. 
(Since these solutions are strongly unstable, their accurate 
numerical determination even at $\omega_{\rm b} = 0.58$ and for only half a 
period requires working in quadruple precision.)

\section{Instability-generated dynamics} 
\label{sec:dyn}
\begin{figure}  
\begin{minipage}[l]{0.49\textwidth} 
\includegraphics[height=\textwidth,angle=270]{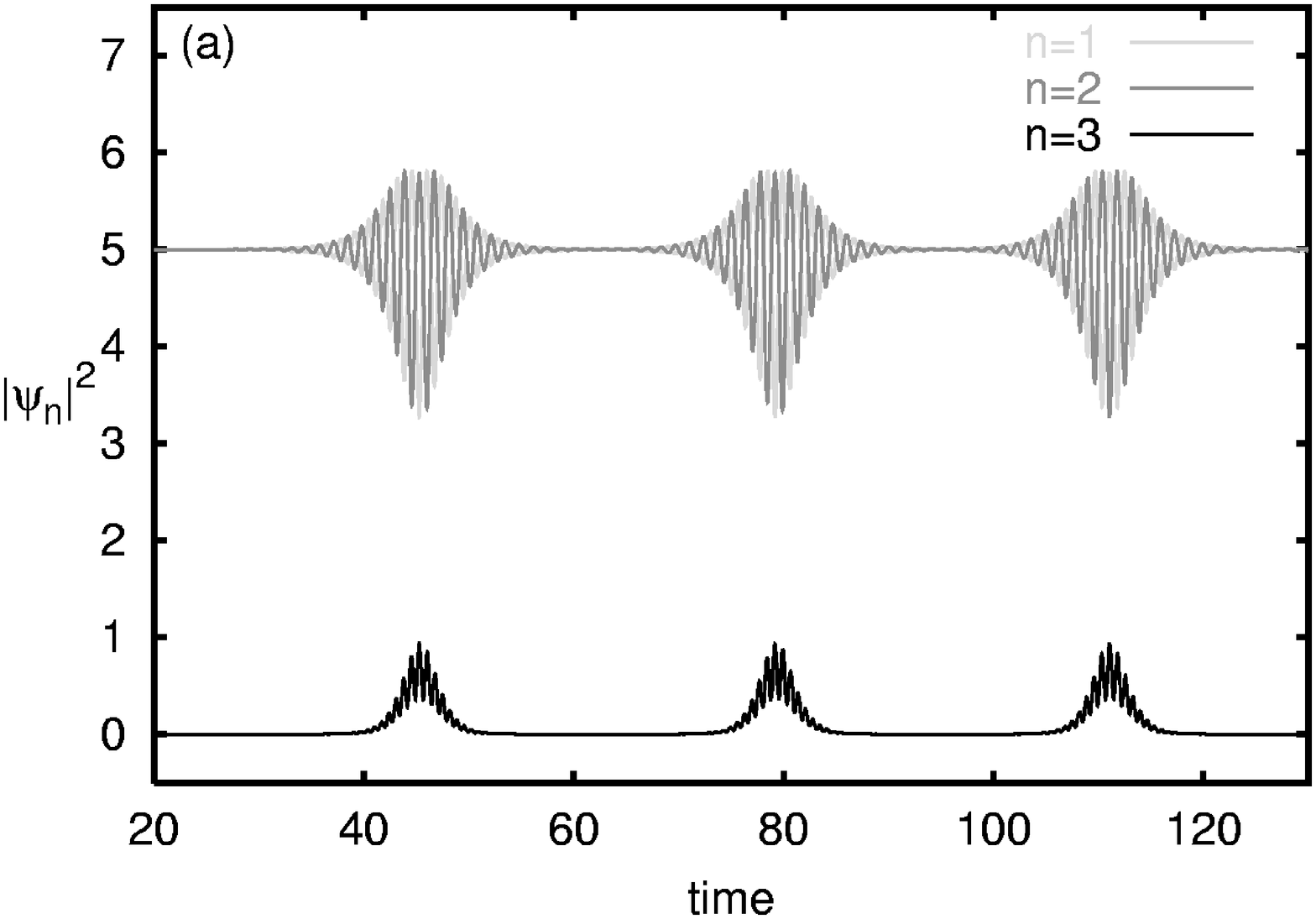} 
\\
\end{minipage}%
\begin{minipage}[r]{0.49\textwidth} 
\includegraphics[height=\textwidth,angle=270]{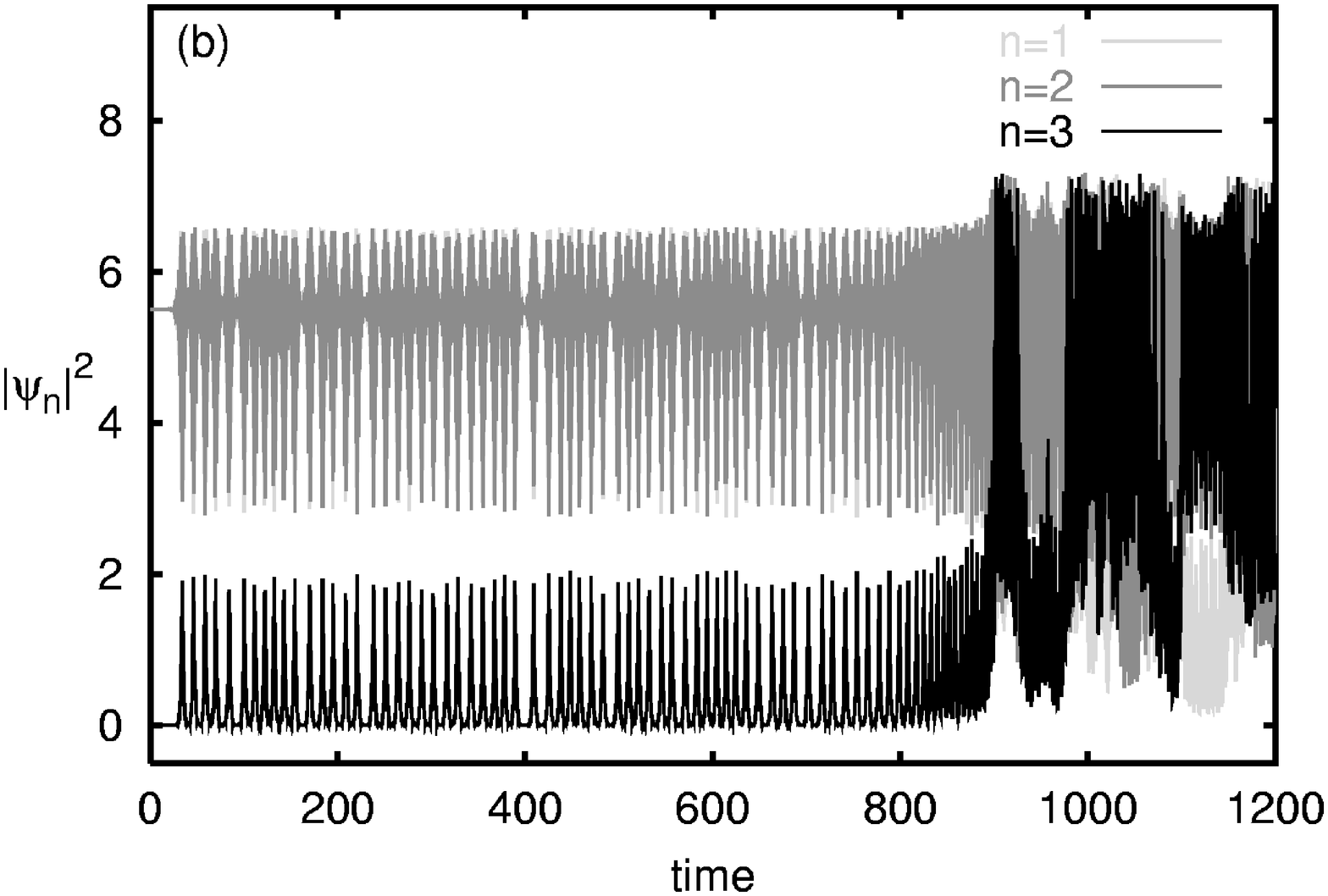} 
\\
\end{minipage}\\
\begin{minipage}[c]{0.49\textwidth} 
\includegraphics[height=\textwidth,angle=270]{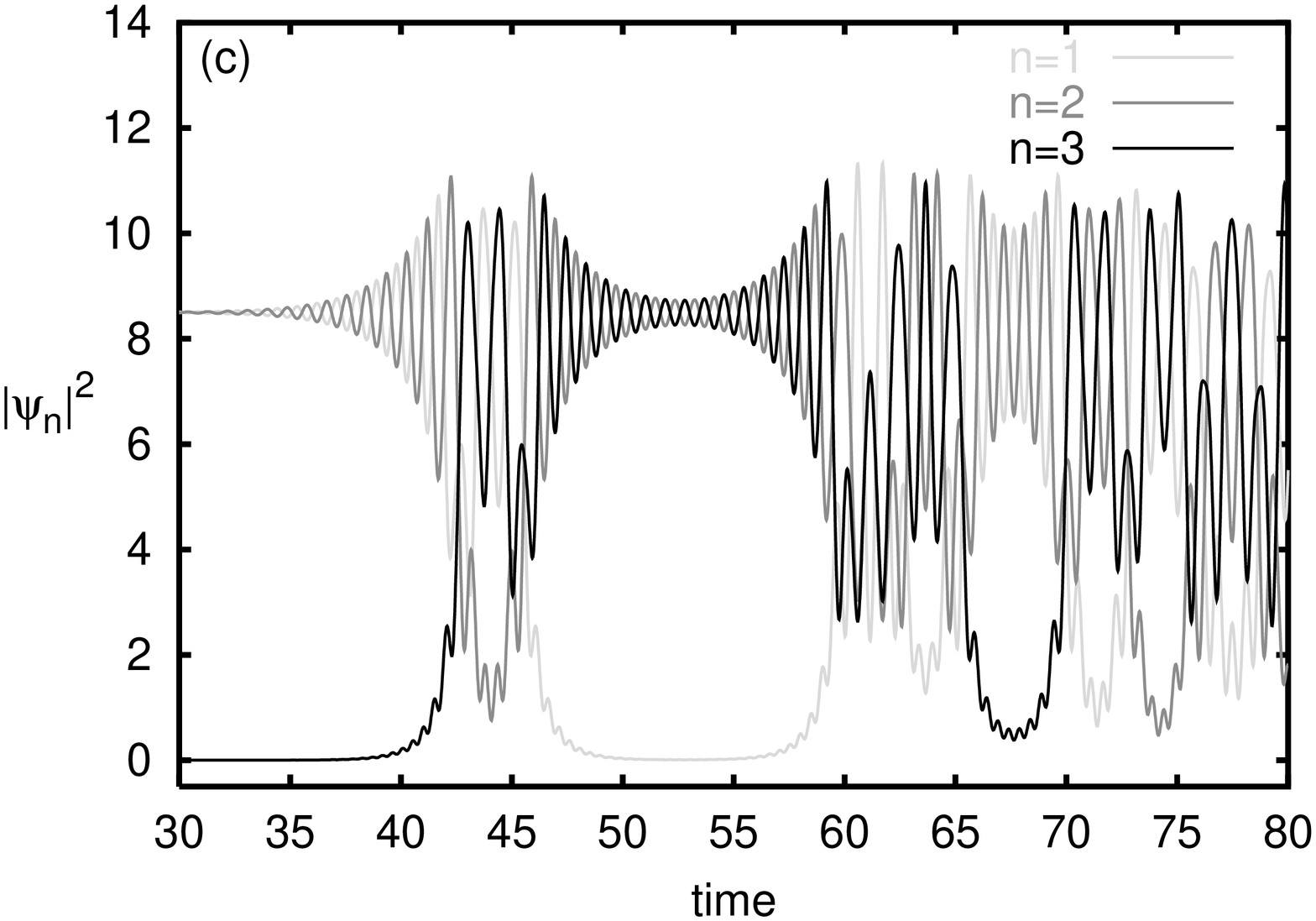}
\end{minipage}
\caption{
Typical instability-generated dynamics for randomly perturbed unstable
stationary solutions \eref{ALambda} with $C=1$ and (a) 
${\mathcal N}=10$, (b) ${\mathcal N}=11$, and (c) ${\mathcal N}=17$, 
respectively. The size of the 
initial perturbation is of the order of $10^{-8}$.
}
\label{fig_dyn} 
\end{figure}

Let us now discuss the dynamics resulting from initial conditions 
taken as a slightly perturbed stationary solution
\eref{ALambda}
in various parts of its Krein instability 
regime, with three examples shown in 
figure \ref{fig_dyn}.
Close to the low-amplitude instability threshold (figure \ref{fig_dyn} (a)), 
the dynamics remains trapped forever in 
a state of bounded oscillations around the initial 
solution. In particular, the amplitude of the initially 
unexcited site ($n=3$) always remains small compared to the 
other two sites. As ${\mathcal N}/C$ increases the amplitude of these 
oscillations increases, and above a certain value 
${\mathcal N}/C \approx 10.6$ the self-trapping is destroyed and 
the dynamics of all three sites becomes qualitatively the same. 
Slightly above this value  the dynamics is still initially trapped 
(figure \ref{fig_dyn} (b)), 
but finally the trapping is destroyed and the amplitude of the third 
site increases drastically. The length of the initial transient
approaches infinity as ${\mathcal N}/C \rightarrow \sim 10.6$, but decreases 
for increasing ${\mathcal N}/C$ 
so that close to the high-amplitude instability threshold
${\mathcal N}/C = 18$ no transient is observed 
(figure \ref{fig_dyn} (c)). In all cases, the 
dynamics in the untrapped regime can be characterized as 
'intermittent population inversion', which was observed also 
for certain initial conditions of the (nonequivalent) trimer 
studied in \cite{FP02}. This is most clearly seen in figure 
\ref{fig_dyn} (c). Here, one can during certain time intervals unambiguously 
identify two sites of large amplitude and a single small-amplitude site, 
which changes from initially being $n=3$ to $n=2$, and then, during a rather 
long time, to $n=1$. After that follows a regime of more 
complicated oscillations, which then again is followed by a regime with 
a single small-amplitude site changing from $n=3$ to $n=1$, etc. 

To understand more clearly the origin of this self-trapping transition, we 
analyze the dynamics close to the unstable stationary solution 
in terms of Poincar{\'e} sections. They can be introduced 
in many different ways; here we apply similar ideas as in \cite{Cruzeiro90} 
making use of the transformation into action-angle variables 
$\{P_n,\theta_n\}$ defined by 
$\psi_n=\sqrt{P_n} \rme^{-\rmi \theta_n}$ (a slightly different approach 
was used in \cite{FP02}). It is then convenient to replace one of the action 
variables, which we here choose as $P_1$, with the conserved 
quantity ${\mathcal N}$. Then, the angle-variables conjugated to 
the set of generalized momenta $\{{\mathcal N}, P_2, P_3\}$ are 
\cite{Cruzeiro90} $\{\theta_1,\theta_2-\theta_1,\theta_3-\theta_1\}$, so 
that $\theta_1$ describing an overall phase becomes an ignorable coordinate. 
Thus, the essential dynamics takes place in a four-dimensional space where 
the surface of constant energy $H$ is three-dimensional, so that a proper 
Poincar{\'e} section through it becomes two-dimensional. Consequently, 
although chaotic trajectories may fill a large portion of the available 
phase space \cite{ELS85,Cruzeiro90}, Arnold diffusion is prohibited 
\cite{FFS89,FCS91} since 
the presence of any regular KAM tori will disconnect the phase space. 
(Arnold diffusion does however appear for the four-site DNLS 
model \cite{FCS91,HG95}.)

\begin{figure}  
\begin{minipage}[l]{0.49\textwidth} 
\includegraphics[height=\textwidth,angle=270]{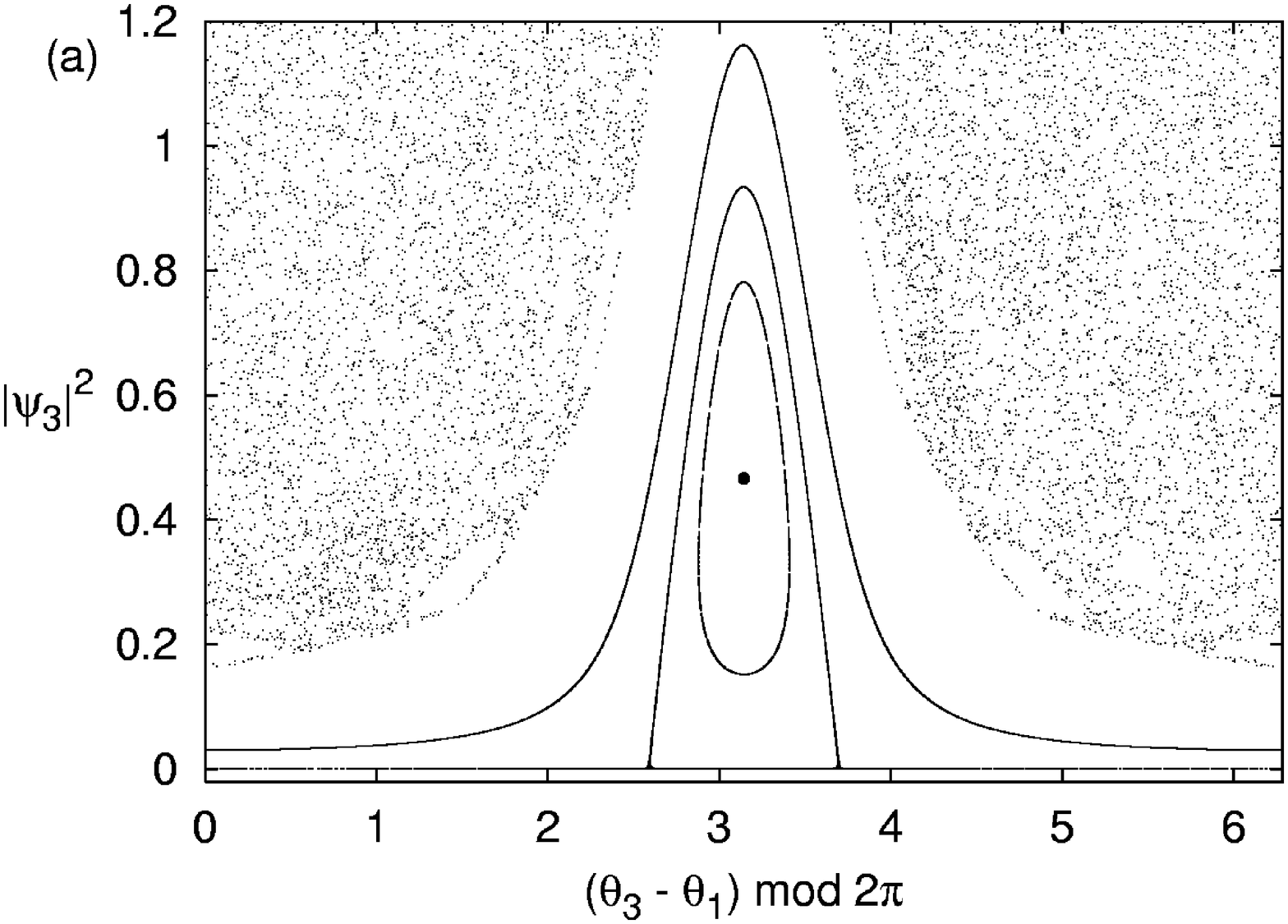} 
\\
\end{minipage}%
\begin{minipage}[r]{0.49\textwidth} 
\includegraphics[height=\textwidth,angle=270]{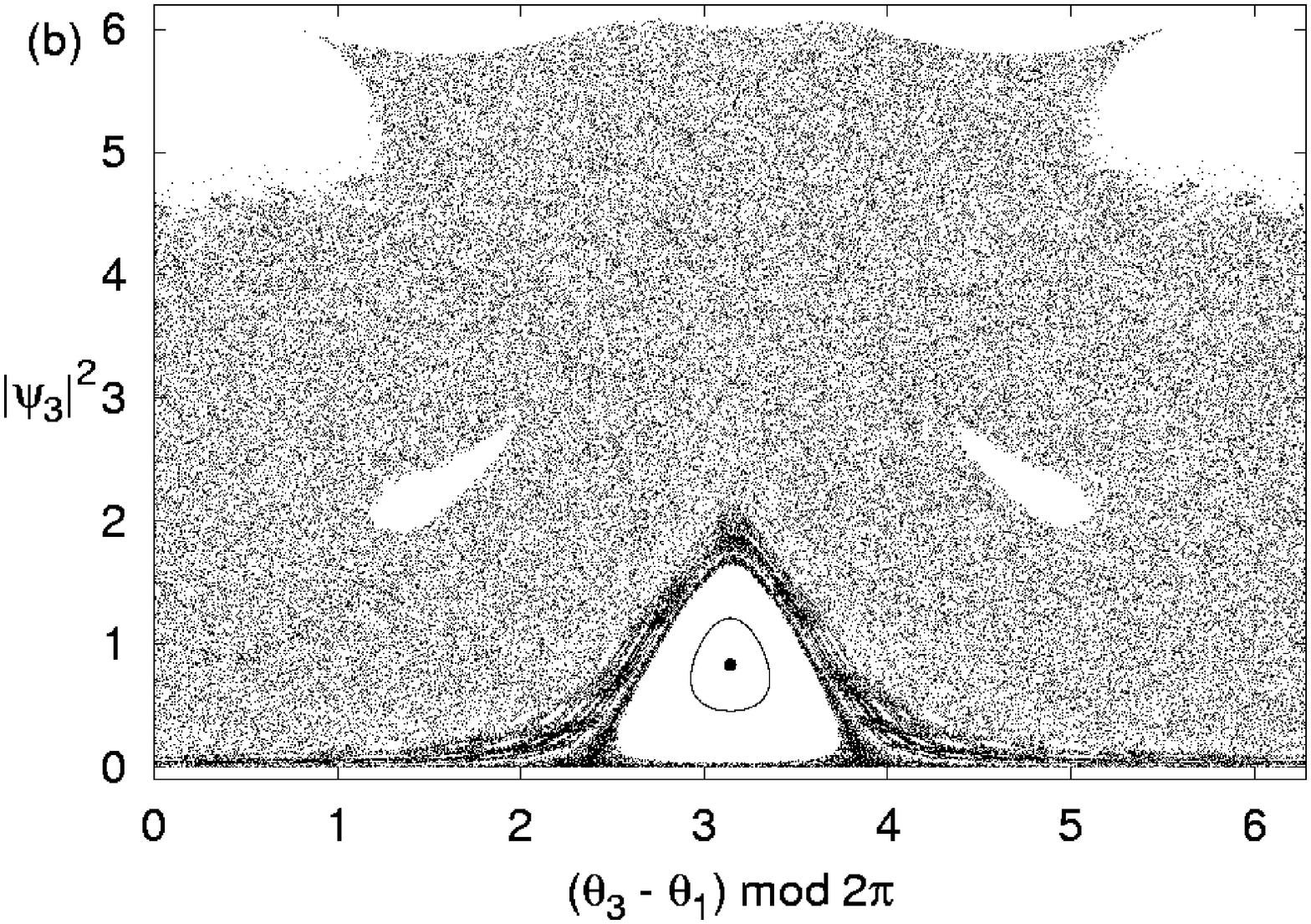} 
\\
\end{minipage}\\
\begin{minipage}[l]{0.49\textwidth} 
\includegraphics[height=\textwidth,angle=270]{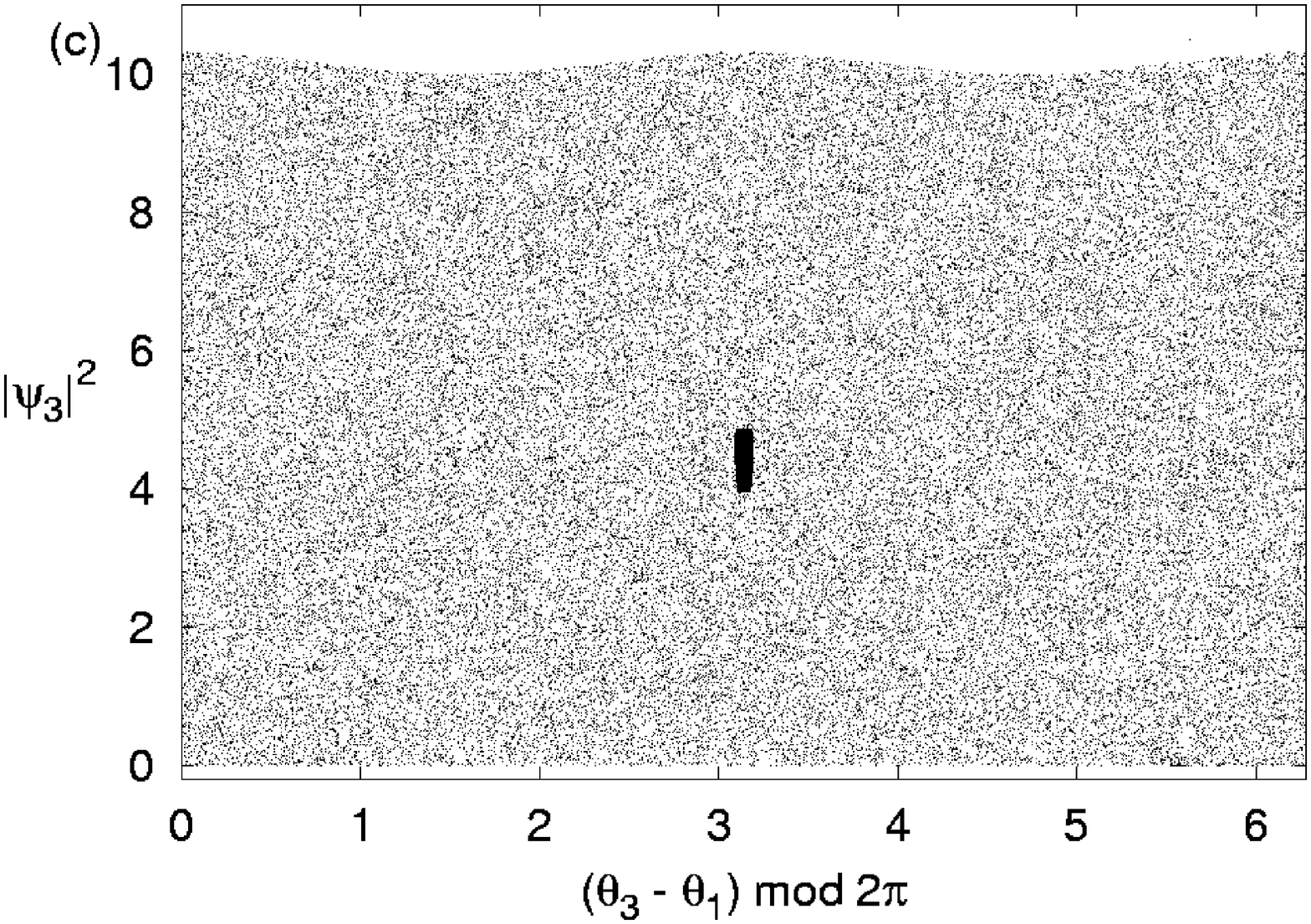}
\end{minipage}
\begin{minipage}[r]{0.49\textwidth} 
\includegraphics[height=\textwidth,angle=270]{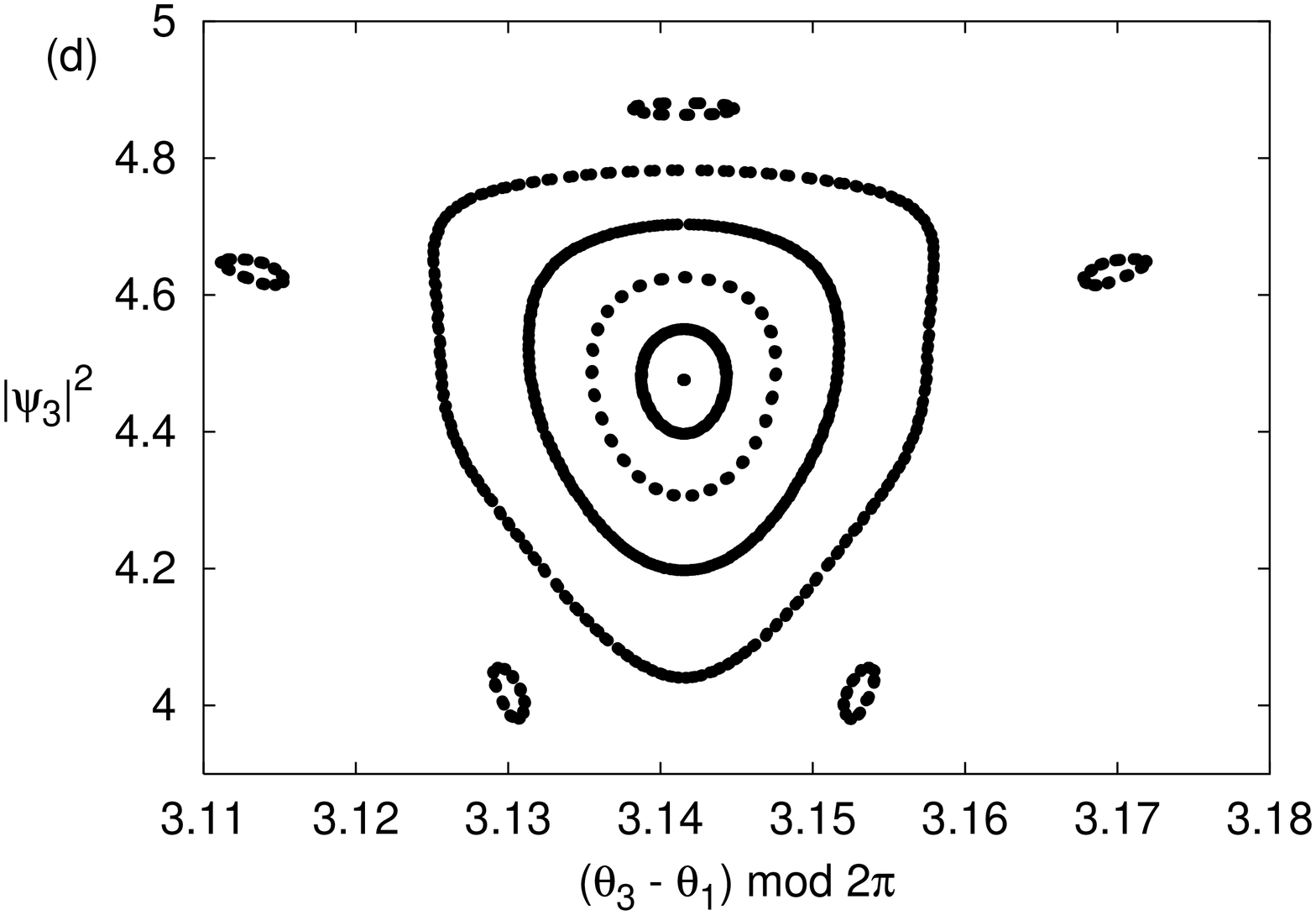}
\end{minipage}
\caption{
Poincar{\'e} sections $\theta_2-\theta_1 = \pi$, 
$\frac{\rmd}{\rmd t} (\theta_2 -\theta_1)<0 $ for the dynamics close to 
unstable
stationary solutions \eref{ALambda} with $C=1$ and (a) 
${\mathcal N}=10$ ($H=5$), (b) ${\mathcal N}=10.7$ ($H=3.4775$), and 
(c), (d) ${\mathcal N}=17$ ($H=-21.25$), respectively. 
The stationary solution is represented by horizontal lines
$|\psi_3|^2 = 0$. The elliptic fixed point (filled circle)
represents continuations of the stable branch of two-frequency solutions 
in figures 
\ref{fig_Ibif}-\ref{fig_N=15}. The window of regular orbits in (d) 
is marked by a short vertical black line in (c).
}
\label{fig_Poin} 
\end{figure}

As a particular choice of Poincar{\'e} section giving a clear illustration 
of the self-trapping transition, we  plot 
in figure \ref{fig_Poin} $P_3 = |\psi_3|^2$ 
versus $\theta_3-\theta_1$ at each time instant when 
$\theta_2-\theta_1 = \pi$ and $\frac{\rmd}{\rmd t} (\theta_2 -\theta_1)<0$. 
Since the stationary solution has zero amplitude at $n=3$ its phase 
$\theta_3$ is undefined, and thus it is represented by a horizontal line 
$|\psi_3|^2 = 0$. The scenario in the self-trapped 
regime is illustrated by  figure \ref{fig_Poin} (a), corresponding to the 
dynamics in figure \ref{fig_dyn} (a). Here the elliptic fixed point at 
$\theta_3-\theta_1=\pi$ represents the stable two-frequency 
solution of section \ref{sec:TypeI} with the same value of $H$ as the 
stationary solution. As in figure \ref{fig_Ibif} (d)
this solution belongs to 
the black branch.
This elliptic fixed point appears at $|\psi_3|^2 = 0$ 
at the type I HH bifurcation point
${\mathcal N}/C \approx 9.077$, and moves vertically in the direction of 
increasing $|\psi_3|^2$ for increasing ${\mathcal N}/C$ 
(cf.\ figure \ref{fig_Ibif} (b)). It 
is surrounded by two different kinds of 
regular 
periodic or quasiperiodic orbits, where  the latter constitute KAM 
tori corresponding generically to quasiperiodic {\em three}-frequency  
solutions in the original DNLS dynamics. With a pendulum 
analogy, these orbits can be classified as 'rotating' and 'vibrating', 
respectively, where the former extend for all values of 
$\theta_3-\theta_1$ while the latter only exist in a bounded 
region close to $\theta_3-\theta_1 = \pi$. Then, the unstable stationary 
solution becomes the separatrix between these different kinds of solutions. 
Although the separatrix is chaotic (which can be seen from a careful 
look at figure \ref{fig_Poin} (a)) the chaos is confined between KAM-tori, 
and in particular the existence of confining 'rotating' tori makes it 
impossible for $|\psi_3|^2$ to exceed some upper limit value 
($|\psi_3|^2 \approx 1$ at $\theta_3-\theta_1 = \pi$ in figure 
\ref{fig_Poin} (a)). Thus, self-trapping results. 

Below  ${\mathcal N}/C \approx 9.077$, 
where the stationary solution is stable and the elliptic fixed point 
with nonzero $|\psi_3|^2$ 
not yet born, all surrounding KAM tori are of the 'rotating' kind. As 
${\mathcal N}/C $ is increased and the elliptic fixed point moves upwards 
towards larger $|\psi_3|^2$, more and more of the 'rotating' KAM tori get 
destroyed, and finally at ${\mathcal N}/C \approx 10.6$ the last 'rotating' 
KAM torus breaks up, and the self-trapping is destroyed. The 
Poincar{\'e} plot for ${\mathcal N}/C =10.7$, i.e.\ just above the 
transition point, is shown in figure \ref{fig_Poin} (b) (the corresponding 
dynamics is qualitatively similar to figure  \ref{fig_dyn} (b) 
but with 
a longer transient $t\sim 17000$). 
It can be seen, that although the dynamics finally spreads 
to a large part 
of the available phase space, the darker parts signifies regions 
where it will be almost trapped 
for long times. This should be expected close to the transition 
point, as destroyed KAM tori should transform into cantori by a 
transition by breaking of analyticity \cite{Aub78}. Close to the transition 
the cantori should have almost full measure, and can thus trap the dynamics 
for very long times. (The white regions in the upper part 
around $\theta_3-\theta_1 = 0$ mod $2\pi$ correspond to 
dynamics trapped around the equivalent two-frequency solution 
with main amplitudes at sites $n=2$ and $n=3$.)

At the self-trapping transition 
${\mathcal N}/C \approx 10.6$, the solution family 
represented by the elliptic fixed point in figure \ref{fig_Poin} changes 
from the black (larger $H$) to the grey (smaller $H$) branch 
(cf.\ section \ref{sec:Ilarge} and figure \ref{fig_N=15} (b)).
Increasing further   
${\mathcal N}/C$, this fixed point continues to move towards 
larger $|\psi_3|^2$ (cf.\ figure \ref{fig_N=15} (a)), and the surrounding 
island of KAM tori shrinks so that the unstable stationary solution 
invades almost all the available phase space 
(figure \ref{fig_Poin} (c), (d), corresponding 
to the dynamics of figure \ref{fig_dyn} (c)). Here, there are no
visible traces of destroyed rotating KAM tori (if cantori exist they 
should have very small measure), so the dynamics 
shows no transient trapping. This behaviour 
persists until the stationary solution 
regains stability through the type II HH bifurcation 
at ${\mathcal N}/C = 18$. 
Before, at ${\mathcal N}/C \approx 17.2$, the 
grey two-frequency solution becomes unstable through a period-doubling type
bifurcation (cf.\ collision at $\beta=-1$
in figure  \ref{fig_N=15} (c)), so that the elliptic fixed point 
in figure \ref{fig_Poin} (d) becomes 
hyperbolic and splits into two new elliptic points surrounded by 
tiny KAM tori.

\section{Conclusions}
\label{Conclusions}
In conclusion, we have investigated  the Hamiltonian Hopf 
bifurcations in the three-site DNLS model, 
resulting in a regime of oscillatory instability for the 
stationary solution with two sites of anti-phased oscillations and the 
third site of zero amplitude ('single depleted well' \cite{FP02}). 
Using numerical continuation techniques to calculate
the (generally quasiperiodic) two-frequency solutions involved in the two 
bifurcations, we found them to be of two different types. 
At the low-amplitude instability threshold the bifurcation is of 'type I', 
with 
stable two-frequency solutions close to the unstable stationary solution 
near the bifurcation point. These two-frequency solutions are themselves 
surrounded by KAM tori, representing generally quasiperiodic three-frequency 
solutions of the full DNLS model. At the high-amplitude instability 
boundary, the HH bifurcation is of 'type II', 
and the stationary solution has 
no surrounding two-frequency solutions on the unstable side of the 
bifurcation. This reflects itself as a self-trapping transition 
in the instability-induced dynamics of the stationary solution, so that 
in the low-amplitude instability regime the dynamics remains trapped close 
to the initial state with small amplitude on the third site, while in the 
high-amplitude instability regime an 'intermittent population inversion' 
dynamics is observed with the small-amplitude oscillation moving chaotically 
between the sites. This self-trapping transition, which we 
believe has not been described earlier in the literature, 
was found to correspond 
to the destruction of phase-space dividing KAM tori. 

As discussed in the introduction, 
our original motivation was to obtain an increased 
understanding for the dynamics resulting from oscillatory instabilities of 
certain multibreather configurations in infinite lattices, which have been 
discovered recently in many contexts. Based on the results obtained here, 
we will address these issues in a forthcoming publication. However, the 
three-site DNLS model is of interest in itself, in particular in view 
of the recent experimental progress in the fields of coupled optical 
waveguides and Bose-Einstein condensates. It seems likely, that the 
existence of stable families of two-frequency solutions, corresponding to 
intensities periodically oscillating around their mean values, as well 
as the existence of a new type of self-trapping transition in the unstable 
dynamics of the stationary solution, should be experimentally verifiable 
within these contexts.

\ack
This work was initiated at the Seminar and Workshop on Nonlinear 
Lattice Structure and Dynamics, Dresden, Germany, September 03-30,
2001, and I thank the Max-Planck-Institut f\"ur Physik komplexer Systeme, 
and in particular Sergej Flach, for their hospitality. I also thank 
Chris Eilbeck for some helpful remarks, and for sending reprints of 
\cite{Eilbeck87}. Parts of this work were inspired by earlier 
discussions with Serge Aubry. 
Financial support from the Swedish Research Council is also gratefully 
acknowledged.

\end{document}